\documentclass[aps,prb,twocolumn,showpacs,superscriptaddress,floatfix]{revtex4-1}
\usepackage{graphicx,graphics}
\usepackage{dcolumn}
\usepackage{amsmath,amssymb,amsfonts}
\usepackage{latexsym,verbatim}
\usepackage{bm}
\usepackage{color}
\usepackage{ulem}
\usepackage[breaklinks=true,colorlinks,citecolor=blue,linkcolor=blue,urlcolor=blue]{hyperref}

\def\be{\begin{eqnarray}}
\def\ee{\end{eqnarray}}

\begin{document}
\title{Bulk and shear viscosities of the 2D electron liquid in a doped graphene sheet}
\author{Alessandro Principi}
\email{aprincipi@science.ru.nl}
\affiliation{Department of Physics and Astronomy, University of Missouri, Columbia, Missouri 65211,~USA}
\affiliation{Institute for Molecules and Materials, Radboud University Nijmegen, Heijndaalseweg 135, 6525 AJ, Nijmegen, The Netherlands}
\author{Giovanni Vignale}
\affiliation{Department of Physics and Astronomy, University of Missouri, Columbia, Missouri 65211,~USA}
\author{Matteo Carrega}
\affiliation{CNR-SPIN, Via Dodecaneso 33, 16146 Genova,~Italy}
\author{Marco Polini}
\affiliation{NEST, Istituto Nanoscienze-CNR and Scuola Normale Superiore, I-56126 Pisa,~Italy}
\affiliation{Istituto Italiano di Tecnologia, Graphene Labs, Via Morego 30, I-16163 Genova,~Italy}
\begin{abstract}
Hydrodynamic flow occurs in an electron liquid when the mean free path for electron-electron collisions is the shortest length scale in the problem. In this regime, transport is described by the Navier-Stokes equation, which contains two fundamental parameters, the bulk and shear viscosities. In this Article we present extensive results for these transport coefficients in the case of the two-dimensional massless Dirac fermion liquid in a doped graphene sheet. Our approach relies on microscopic calculations of the viscosities up to second order in the strength of electron-electron interactions and in the high-frequency limit, where perturbation theory is applicable. We then use simple interpolation formulae that allow to reach the low-frequency hydrodynamic regime where perturbation theory is no longer directly applicable. The key ingredient for the interpolation formulae is the ``viscosity transport time'' $\tau_{\rm v}$, which we calculate in this Article. 
The transverse nature of the excitations contributing to $\tau_{\rm v}$ leads to the suppression of scattering events with small momentum transfer, which are inherently longitudinal. Therefore, contrary to the quasiparticle lifetime, which goes as $-1/[T^2 \ln(T/T_{\rm F})]$, in the low temperature limit we find $\tau_{\rm v} \sim 1/T^2$. 
\end{abstract}
\maketitle

\section{Introduction}

Hydrodynamics~\cite{Landau_6,batchelor,NPF} is a powerful non-perturbative theory to deal with transport properties of strongly interacting many-particle systems. 

\begin{figure}[t]
\begin{center}
\begin{tabular}{c}
\includegraphics[width=0.99\columnwidth]{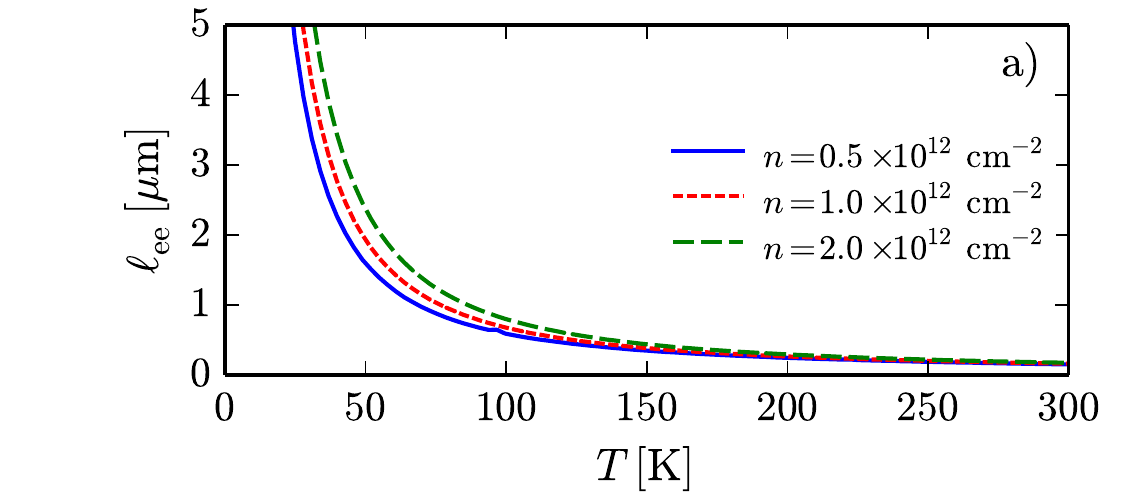}
\\
\includegraphics[width=0.99\columnwidth]{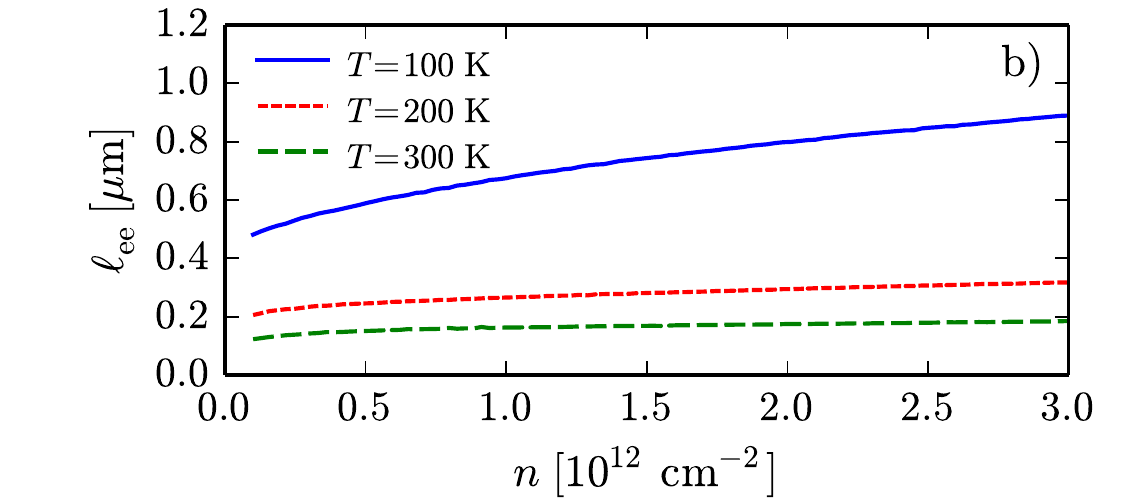}
\end{tabular}
\end{center}
\caption{(Color online) Panel a) The e-e mean free path $\ell_{\rm ee}$ (in $\mu{\rm m}$) in a 2D MDF liquid is plotted as a function of temperature $T$ (in ${\rm K}$). Different curves refer to different values of the excess carrier density, $n = 0.5\times 10^{12}~{\rm cm}^{-2}$, $n = 1.0\times 10^{12}~{\rm cm}^{-2}$, and $n = 2.0\times 10^{12}~{\rm cm}^{-2}$. Panel b) The e-e mean free path $\ell_{\rm ee}$ (in $\mu{\rm m}$) in a 2D MDF liquid is plotted as a function of the excess carrier density $n$ (in units of $10^{12}~{\rm cm}^{-2}$), for three values of $T$, i.e.~$T = 100$, $200$, and $300~{\rm K}$. In both panels the e-e dimensionless coupling constant has been set to $\alpha_{\rm ee} = 0.5$.
\label{fig:one}}
\end{figure}

In the solid state, interactions need to be sufficiently strong to ensure that the mean free path $\ell_{\rm ee} = v_{\rm F} \tau_{\rm ee}$ for electron-electron (e-e) collisions is the {\it shortest} length scale in the problem, i.e. $\ell_{\rm ee} \ll \ell_{\rm p}, L, v_{\rm F}/\omega$. Here, $v_{\rm F}$ is the Fermi velocity, $\tau_{\rm ee}$ is the quasiparticle lifetime due to electron-electron collisions~\cite{Giuliani_and_Vignale}, $\ell_{\rm p}$ is the mean free path for momentum-non-conserving collisions, $L$ is the sample size, and $\omega$ is the frequency of the external perturbation. In Fig.~\ref{fig:one} we show the result of 
microscopic calculations of the e-e mean free path 
for the two-dimensional (2D) massless Dirac fermion (MDF) liquid in a doped graphene sheet~\cite{castroneto_rmp_2009,kotov_rmp_2012,roadmap} embedded between two semi-infinite uniform and isotropic media with dielectric constants $\epsilon_1$ and $\epsilon_2$. The two-dimensional Fourier transform of the e-e interaction in this case is $v_{\bm q} = 2\pi e^2/(\epsilon q)$, where $\epsilon \equiv (\epsilon_1 + \epsilon_2)/2$. Technical details on these many-body diagrammatic perturbation theory calculations can be found, e.g., in Refs.~\onlinecite{polini_arxiv_2014,Li_prb_2013}. We clearly see that, for sufficiently large temperatures, there is a wide range of carrier concentrations in which $\ell_{\rm ee}$ becomes much shorter than the typical device size ($L \sim 10~{\rm \mu m}$).

 The regime defined by the above inequalities is named in what follows ``hydrodynamic'', ``low-frequency'' or ``collisional''.  In the range of temperatures and carrier densities in which this regime is attained, e-e interactions drive the system towards a local quasi-equilibrium state characterized by slowly-varying time-dependent density $n({\bm r}, t)$ and drift velocity ${\bm v}({\bm r}, t)$, which obey the continuity and Navier-Stokes equations. The latter are controlled by two transport coefficients, the shear viscosity, $\eta_{\omega \to 0}$, which describes the friction between adjacent layers of fluid moving with different velocities, and the bulk viscosity, $\zeta_{\omega \to 0}$, which describes the dissipation arising in the liquid when it undergoes a homogeneous compression-like deformation~\cite{Landau_6}. 

When, on the contrary,  the frequency $\omega$ of the external perturbation is much larger than the quasiparticle collision rate (i.e.~$\omega \tau_{\rm ee} \gg 1$)---but still much smaller than the characteristic free-particle frequencies epitomized by the Fermi energy---e-e interactions fail to drive the system towards local quasi-equilibrium~\cite{Giuliani_and_Vignale}. Nonetheless, it is still possible to describe the system by hydrodynamic equations of motion~\cite{Conti_prb_1999}, provided that the low-frequency bulk and shear viscosities are replaced by their high-frequency counterparts~\cite{Giuliani_and_Vignale} ($\zeta_\infty$ and $\eta_\infty$, respectively) and that a finite value of the shear modulus (${\cal S}_\infty$) is allowed. This regime is named in what follows ``high-frequency'' or ``collisionless''.  We emphasize that it is a ``high-frequency" regime only on the scale of e-e collisions, but not at all on the scale of the Fermi energy.

In this Article we calculate the frequency-dependent viscosities $\zeta_\omega$ and $\eta_\omega$ for the 2D MDF liquid in a doped graphene sheet~\cite{kotov_rmp_2012}. Doping, which creates a Fermi liquid of electrons or holes in the upper or lower Dirac band, is of essence here:  we note that the zero-frequency shear viscosity of thermally excited electron-hole pairs in an {\it undoped} graphene sheet was previously calculated in Ref.~\onlinecite{Muller_prl_2009}. 
The low-frequency bulk and shear viscosities of an ordinary three-dimensional (3D) parabolic-band electron gas in the Fermi liquid regime were calculated long ago by Abrikosov and Khalatnikov~\cite{Abrikosov_rpp_1959}. They found that  $\eta_{\omega \to 0} = {\cal S}_\infty \tau_{\rm v}$, where $\tau_{\rm v}$ is a ``viscosity transport time'' of the order of (but not identical to) $\tau_{\rm ee}$ (here all quantities refer to a 3D system). The Abrikosov-Khalatnikov calculation was extended to the opposite regime of high frequency in Ref.~\onlinecite{Conti_prb_1999}. The main difficulty  in connecting these two regimes lies in the fact that e-e interactions play very different roles in the two cases.  In the low-frequency regime their main effect is to cut off an otherwise diverging shear viscosity: the corrected shear viscosity is proportional to $\tau_{\rm v}$, which is non-perturbative in the strength of e-e interactions. Conversely, in the high-frequency regime e-e interactions generate a non-zero value for the otherwise vanishing shear viscosity: this finite value can and has been calculated perturbatively. 

Our theoretical approach is based on ideas first presented in Ref.~\onlinecite{Conti_prb_1999}. 
We combine the perturbative information contained in $\eta_\infty$ and $\zeta_\infty$ with the calculation of $\tau_{\rm v}$ to generate non-perturbative interpolation formulas for $\eta_\omega$ and $\zeta_\omega$. The latter are approximately valid at all frequencies and consistently include many-body self-energy and vertex corrections.  The final formulas are~\cite{Conti_prb_1999}:
\begin{eqnarray} \label{eq:all_freq_viscosity}
\left\{
\begin{array}{l}
{\displaystyle \zeta_\omega = \zeta_\infty \frac{(\omega \tau_{\rm v})^2}{1 + (\omega \tau_{\rm v})^2}
}\vspace{0.2 cm}\\
{\displaystyle \eta_\omega = \frac{{\cal S}_\infty \tau_{\rm v} + \eta_\infty (\omega \tau_{\rm v})^2}{1 + (\omega \tau_{\rm v})^2}}
\end{array}
\right.~,
\end{eqnarray}  
where, in addition to the above-mentioned quantities $\eta_\infty$ and $\zeta_\infty$, we also see the high-frequency shear modulus $S_\infty$, which is to the shear viscosity what the ``Drude weight" is to the conductivity.  We note that the shear modulus is renormalized by e-e interactions: however, this effect is relatively small, and it is thus qualitatively correct to approximate ${\cal S}_\infty$ by its non-interacting value, which in the case of graphene is ${\cal S}_{\infty}^{(0)} = n \varepsilon_{\rm F}/4$, where $\varepsilon_{\rm F}$ is the Fermi energy.

From a mathematical point of view, the viscosities appear as coefficients in the expansion of the stress tensor $\tau_{\mu\nu}$ to first order in the spatial derivatives of oscillating velocity fields, $v_{\mu\nu} \equiv \frac{1}{2}\left(\partial_\mu v_\nu+\partial_\nu v_\mu\right)$.  This can also be viewed~\cite{Gao10} as the out-of-phase component of the response of $\tau_{\mu\nu}$ to an oscillating metric field $g_{\mu\nu}$. 
For a 2D isotropic fluid the expansion has the form
\be\label{eq:stress_definition}
\tau_{\mu\nu}(\omega)=(\zeta_\omega-\eta_\omega)  \big[\nabla\cdot{\bm v}(\omega)\big]\delta_{\mu\nu} +2\eta_\omega v_{\mu\nu}(\omega)~.
\ee
In a parabolic-band electron gas~\cite{Giuliani_and_Vignale}, and also in graphene in the Fermi liquid regime~\cite{kotov_rmp_2012}, the response of the stress tensor to the metric field is connected by  equations of motion to the non-local response of the current to a vector potential, i.e.~the coefficient of $q^2$ in the expansion of the non-local conductivity for small wave vectors $q$. This implies that the high-frequency viscosities can be extracted from the damping rate of plasmons, the high-frequency collective excitations of an electron liquid~\cite{Giuliani_and_Vignale,DiracplasmonsRPA}. 
On the other hand, no standard protocol exists at present to measure the low-frequency viscosities of electrons  in a solid-state host. Tomadin {\it et al.}~\cite{tomadin_prl_2014}  proposed a Corbino disk viscometer, which allows a determination of the  hydrodynamic shear viscosity $\eta_{\omega\to 0}$ from the dc potential difference that arises between the inner and the outer edge of the disk in response to an {\it oscillating} magnetic flux. More recently, it has been shown~\cite{torre_tobepublished_2015,bandurin_tobepublished_2015} that $\eta_{\omega\to 0}$ can also be extracted from purely-dc non-local transport measurements in ultra-clean multi-terminal Hall bar devices.

Our Article is organized as follows. In Sect.~\ref{sect:model} we introduce the tight-binding model of graphene, which we use~\cite{Principi_prb_2013,Principi_prbR_2013,Principi_prb_2014} to avoid broken gauge invariance due to the presence of a rigid ultraviolet cut-off  in the MDF low-energy theory. The MDF limit is indeed taken only at the very end of the calculation.  In Sect.~\ref{sect:viscosity} we derive the relativistic counterpart of the Navier-Stokes equation~\cite{Landau_6}, which describes the long-wavelength dynamics of quasiparticles in graphene in both the low- and high-frequency regimes. In this Section we also show the connection between the macroscopic bulk and shear viscosities and the longitudinal and transverse current-current response functions, which can be microscopically calculated from the usual Kubo formula~\cite{Giuliani_and_Vignale}. 
In Sect.~\ref{sect:RTA} we use a kinetic equation approach and the relaxation-time approximation to determine the interpolation formulas for the viscosities and the elastic moduli of graphene in terms of the high-frequency viscosity $\eta_\infty$ and a yet undetermined transport time $\tau$. These quantities are calculated in the remainder of the Article.
Since the longitudinal current-current response function of the 2D MDF liquid in a doped graphene sheet was calculated in Ref.~\onlinecite{Principi_prb_2013}, in Sect.~\ref{sect:high_frequency_viscosity} we focus on its transverse counterpart. We calculate it at the lowest non-vanishing order in the strength of e-e interactions, which is quantified by the graphene's fine structure constant~\cite{kotov_rmp_2012}
\be\label{eq:finestructure}
\alpha_{\rm ee} \equiv \frac{e^2}{\epsilon \hbar v_{\rm F}}~.
\ee
Our results for the high-frequencies bulk and shear viscosities are reported in Sect.~\ref{sect:high_frequency_viscosity}. We prove that the high-frequency bulk viscosity vanishes, while $\eta_\infty$ is finite. 
Sect.~\ref{sect:tau_v} is devoted to the calculation of the viscosity transport time $\tau_{\rm v}$. The approach we adopt is very similar to that used in Refs.~\onlinecite{Principi_arxiv_2014,Principi_arxiv_2015}, where the e-e contributions to the charge, spin, and thermal conductivities of the 2D MDF liquid in a doped graphene sheet were calculated. Therefore, only the main steps of the calculation are surveyed. We refer the reader interested in more details to Ref.~\onlinecite{Principi_arxiv_2015}.
Finally, In Sect.~\ref{sect:results} we show our result for the shear viscosity at finite frequency $\eta_\omega$.
Making use of Eq.~(\ref{eq:all_freq_viscosity}), we provide numerical results for the shear viscosity of the 2D MDF liquid in a doped graphene at all frequencies.  Appendix~\ref{sect:GRTA} presents a self-contained description of the generalized relaxation time approximation, leading to the formulas of Eq.~(\ref{eq:all_freq_viscosity}). Appendix~\ref{sect:SM_Upsilon_manipulation} contains several technical details of the calculation. In this Article we set, except when explicitly stated otherwise, $\hbar=1$.

\section{Model and basic definitions}
\label{sect:model}
Following Refs.~\onlinecite{Principi_prb_2013,Principi_prbR_2013,Principi_prb_2014}, we describe $\pi$-electrons in graphene by a one-orbital tight-binding (TB) model~\cite{castroneto_rmp_2009}. To keep the model as simple as possible, we set to zero all the hopping parameters but the nearest-neighbor one. The low-energy MDF limit will be taken only at the very end of the calculation, {\it after} carrying out all the necessary algebraic manipulations. By following this procedure we avoid problems associated with the introduction of a rigid ultraviolet cut-off, which breaks the gauge invariance~\cite{abedinpour_prb_2011} and is responsible for the appearance of anomalous commutators~\cite{abedinpour_prb_2011,sabio_prb_2008}.

The non-interacting Hamiltonian is
\begin{eqnarray} \label{eq:SM_non_int_H}
{\hat {\cal H}}_0 = \sum_{{\bm k} \in {\rm BZ}, \alpha, \beta} {\hat \psi}^\dagger_{{\bm k},\alpha} ({\bm f}_{{\bm k}} \cdot {\bm \sigma}_{\alpha\beta}) {\hat \psi}_{{\bm k},\beta}
~,
\end{eqnarray}
where the operator ${\hat \psi}^\dagger_{{\bm k},\alpha}$ (${\hat \psi}_{{\bm k},\alpha}$) creates (annihilates) an electron with Bloch momentum ${\bm k}$, which belongs to the sublattice~\cite{castroneto_rmp_2009} $\alpha = A,B$. The vector ${\bm f}_{\bm k}$ is defined as~\cite{castroneto_rmp_2009}
\begin{eqnarray} \label{eq:SM_f_vec}
{\bm f}_{{\bm k}} = -t \sum_{i=1}^3 \left(\Re e\left[e^{-i {\bm k}\cdot{\bm \delta}_i}\right], -\Im m\left[e^{-i {\bm k}\cdot{\bm \delta}_i}\right]\right)
~.
\end{eqnarray}
Here $t\sim 2.8~{\rm eV}$ is the nearest-neighbor tunneling amplitude, while ${\bm \delta}_i$ ($i=1,\ldots,3$) are the vectors which connect an atom to its three nearest neighbors, {\it i.e.} ${\bm \delta}_1 = a \sqrt{3} {\hat {\bm x}}/2 + a {\hat {\bm y}}/2$, ${\bm \delta}_2 = -a \sqrt{3} {\hat {\bm x}}/2 + a{\hat {\bm y}}/2$, and ${\bm \delta}_3 = - a{\hat {\bm y}}$. Here $a \sim 1.42$~\AA~ is the Carbon-Carbon distance in graphene. The sum over ${\bm k}$ in Eq.~(\ref{eq:SM_non_int_H}) is restricted to the first Brillouin zone (BZ) and the Pauli matrices $\sigma^i_{\alpha\beta}$ ($i=x,y,z$) operate on the sublattice degree of freedom. 

The TB problem posed by the Hamiltonian~(\ref{eq:SM_non_int_H}) can be easily solved analytically~\cite{castroneto_rmp_2009}. One finds the following eigenvalues $\varepsilon_{{\bm k},\lambda} = \lambda |{\bm f}_{\bm k}|$, with $\lambda=\pm$. These two bands touch at  two inequivalent points ($K$ and $K'$) in the hexagonal BZ. The low-energy MDF model is obtained from Eq.~(\ref{eq:SM_non_int_H}) by taking the limit $a\to 0$, while keeping the product $t a$ constant. In this limit ${\bm f}_{{\bm K}+{\bm k}} \to v_{\rm F} {\bm k}$, where $v_{\rm F} = 3 t a/2 \sim 10^6 ~{\rm m/s}$ is the density-independent Fermi velocity.

Introducing the field operator ${\hat c}^\dagger_{{\bm k},\lambda}$ (${\hat c}_{{\bm k},\lambda}$) as the creation (annihilation) operator in the eigenstate representation, 
Eq.~(\ref{eq:SM_non_int_H}) can be rewritten as 
\begin{equation}\label{eq:bandenergy}
{\hat {\cal H}}_0 = \sum_{{\bm k}, \lambda} \varepsilon_{{\bm k},\lambda} {\hat c}^\dagger_{{\bm k},\lambda} {\hat c}_{{\bm k},\lambda}~.
\end{equation}
In the same representation the Hamiltonian describing e-e interactions reads~\cite{Giuliani_and_Vignale}
\begin{eqnarray} \label{eq:SM_interaction_H}
{\hat {\cal H}}_{\rm ee} = \frac{1}{2} \sum_{{\bm q}} v_{\bm q} {\hat n}_{\bm q} {\hat n}_{-{\bm q}}
~,
\end{eqnarray}
where the density operator is
\begin{eqnarray} \label{eq:SM_density_op}
{\hat n}_{\bm q} &=& \sum_{{\bm k},\lambda,\lambda'} {\cal D}_{\lambda\lambda'}({\bm k}-{\bm q}/2, {\bm k}+{\bm q}/2) {\hat c}^\dagger_{{\bm k}-{\bm q}/2,\lambda} {\hat c}_{{\bm k}+{\bm q}/2,\lambda'}
~,
\nonumber\\
\end{eqnarray}
and $v_{\bm q}$ is the 2D {\it discrete} Fourier transform of the real-space Coulomb interaction, which is a periodic function of the reciprocal-lattice vectors, and reduces to $\sim 2\pi e^2/(\epsilon q)$ in the limit of $q\to 0$.  Finally, in Eq.~(\ref{eq:SM_density_op}) we have introduced the ``density vertex"
\begin{eqnarray} \label{eq:SM_D_element}
{\cal D}_{\lambda\lambda'} ({\bm k}, {\bm k}') =
\frac{e^{i(\theta_{\bm k}-\theta_{{\bm k}'})/2} + \lambda\lambda' e^{-i(\theta_{\bm k}-\theta_{{\bm k}'})/2}}{2}
\end{eqnarray}
with $\theta_{\bm k} = {\rm Arg}[f_{{\bm k},x}+ i f_{{\bm k},y}]$. Here $\{f_{{\bm k}, i}, i= x,y\}$ denotes the Cartesian component of the vector ${\bm f}_{\bm k}$. In the low-energy MDF limit, $\theta_{{\bm K}+{\bm k}} \to \varphi_{\bm k}$, where $\varphi_{\bm k}$ is the angle between ${\bm k}$ and the ${\hat {\bm x}}$ axis.

Note that in writing Eq.~(\ref{eq:SM_interaction_H}) we have neglected the one-body operator proportional to the total number of particles, which avoids self-interactions~\cite{Giuliani_and_Vignale}, since it has no effect on the calculations we will carry out below. The viscosities are indeed determined (at the lowest non-vanishing order in the strength of e-e interactions) by two-particle excitations only, which are generated by {\it two-body} operators.

We also introduce the current operator
\begin{eqnarray} \label{eq:current_op_definition}
{\hat j}_{{\bm q},\alpha} &=& \!\! \sum_{{\bm k},\beta} \sum_{\lambda,\lambda'} \frac{\partial f_{{\bm k},\alpha}}{\partial k_{\beta}} {\cal S}^{(\beta)}_{\lambda\lambda'}({\bm k}-{\bm q}/2, {\bm k}+{\bm q}/2)
\nonumber\\
&\times&
{\hat c}^\dagger_{{\bm k}-{\bm q}/2,\lambda} {\hat c}_{{\bm k}+{\bm q}/2,\lambda'}
~,
\end{eqnarray}
where the ``pseudospin-density" vertices are
\begin{eqnarray} \label{eq:SM_S_x_element}
{\cal S}^{(x)}_{\lambda\lambda'} ({\bm k}, {\bm k}') &=& \frac{\lambda' e^{i(\theta_{\bm k}+\theta_{{\bm k}'})/2} + \lambda e^{-i(\theta_{\bm k}+\theta_{{\bm k}'})/2}}{2}
\end{eqnarray}
and
\begin{eqnarray} \label{eq:SM_S_y_element}
{\cal S}^{(y)}_{\lambda\lambda'} ({\bm k}, {\bm k}') &=& \frac{\lambda' e^{i(\theta_{\bm k}+\theta_{{\bm k}'})/2} - \lambda e^{-i(\theta_{\bm k}+\theta_{{\bm k}'})/2}}{2i}~.
\end{eqnarray}

For the sake of definiteness, we assume the system to be $n$-doped, with an excess electron density $n$. Results for a $p$-doped system can be easily obtained by appealing to the particle-hole symmetry of the MDF model defined by Eqs.~(\ref{eq:bandenergy})-(\ref{eq:SM_interaction_H}). The Fermi wave vector is defined as $k_{\rm F} = \sqrt{4\pi n/N_{\rm f}}$, while $\varepsilon_{\rm F}= v_{\rm F} k_{\rm F}$ is the Fermi energy, and $N_{\rm f} =4$ is the number of fermion flavors in graphene. For future purposes we also define the matrix element of the ${\hat {\bm z}}$ component of the pseudospin between the states labeled by ${\bm k},\lambda$ and ${\bm k}',\lambda'$:
\begin{eqnarray} \label{eq:SM_S_z_element}
{\cal S}^{(z)}_{\lambda\lambda'} ({\bm k}, {\bm k}') &=& \frac{e^{i(\theta_{\bm k}-\theta_{{\bm k}'})/2} - \lambda \lambda' e^{-i(\theta_{\bm k}-\theta_{{\bm k}'})/2}}{2}
~.
\end{eqnarray}
\section{The bulk and shear viscosities --- general theory}
\label{sect:viscosity}
The bulk and shear viscosity are usually introduced~\cite{Landau_6} as phenomenological coefficients to describe the long-wavelength motion of a viscous fluid close to a quasi-equilibrium situation. The interactions between the elementary constituent of the fluid, although extremely complicated at the microscopic level, admit a rather simple description in terms of macroscopic coefficients. Their space and time average is in fact responsible for the friction between (macroscopic) fluid elements having different values of the momentum. The two viscosities then describe the forces between fluid elements that undergo either a shear or a compression-like long-wavelength deformation.

It is therefore possible, starting from a fluid-like description of the 2D MDF liquid in graphene, to derive the {\it macroscopic} response of the current to an external vector potential, in the linear response regime and in terms of hydrodynamic coefficients. Equating the coefficient of proportionality between the current and the vector potential to the microscopic current-current linear response functions of the system, one obtains a microscopic definition of the bulk and shear viscosities. It turns out that these can be calculated from the coefficients of the expansion to order $q^2/\omega^2$ of the current-current linear response functions (in the limit $q\to 0$). This approach was used in Ref.~\onlinecite{Conti_prb_1999} for the case of a parabolic-band ({\it i.e.} Galilean invariant) electron gas.

A more fundamental approach~\cite{Tao_prl_2009,Gao_prb_2010,Taylor_pra_2010} relies on the fact that it is possible to microscopically define the stress tensor operator $\tau_{\mu\nu}({\bm x},t)$, starting from the equation of motion of the momentum density, and to calculate the deviation of its average from the equilibrium value due to an applied strain. In the linear regime and at low frequencies, such variation is proportional to the applied strain. The coefficient of proportionality is the ``tensor of elasticity'', a rank-4 tensor whose imaginary part in a rotationally and time-reversal invariant system at $q=0$ can be characterized by two coefficients, the complex bulk and shear moduli. Note that in the linear regime the tensor of elasticity is equivalent to the stress-stress response function. Therefore, the knowledge of the latter response function constitutes a viable route to the calculation of the viscosity coefficients~\cite{Tao_prl_2009,Gao_prb_2010,Taylor_pra_2010}.

The connection between the two approaches is rather trivial in Galilean invariant systems. Indeed in such systems the current density is proportional to the momentum density, and therefore its time derivative is proportional to the divergence of the stress tensor. It is therefore possible to derive an equation of motion that relates the current-current response functions to the stress-stress response. From that, it is thus evident that the viscosity, which is proportional to the coefficient of the term of order $q^2/\omega^2$ in the expansion of the current-current response function, can also be calculated from the $q=0$ limit of the stress-stress response. 

In the 2D MDF liquid, however, the current and the momentum are not proportional to each other. The current is indeed an off-diagonal operator (in pseudospin space), which represents the hopping between the two inequivalent Carbon atoms in the unit cell. Conversely, the momentum operator is a diagonal one (as, e.g., the density operator). Therefore the two approaches could give, in principle, different results. Nevertheless, we find that in the Fermi liquid regime, {\it i.e.} when graphene is doped and the temperature $T \ll \varepsilon_{\rm F}/k_{\rm B}$ is sufficiently low ($k_{\rm B}$ is the Boltzmann constant), the two agree with each other. In the following subsections we  show in detail how the two approaches reconcile in the Fermi liquid regime. Here, however, we briefly discuss the problem in general terms.

It is quite easy to understand what goes wrong in the general case. While the second approach, {\it i.e.} the calculation from the stress-stress response, is completely general, the first one relies on the fact that it is possible to write a Navier-Stokes equation for the velocity as in the classical non-relativistic case. This statement is highly non-trivial. Indeed, in graphene the velocity operator, being proportional to the {\it unit vector} of the momentum, is not a conserved quantity. Therefore, it is not possible to write in general a Navier-Stokes equation for the macroscopic velocity in the same way as it is done in the Galilean invariant case. Since the quantities that are conserved by e-e interactions are the density, energy density, and momentum, one should in principle consider the Navier-Stokes equations for these three. In passing we note that, in a system with a linear energy dispersion, the momentum coincides, up to a proportionality constant,  with the non-interacting component of the energy current. 

However, as it was shown in Ref.~\onlinecite{Principi_arxiv_2015}, in the Fermi liquid regime the velocity {\it is} essentially a conserved quantity. This statement is rationalized as follows. At low temperature and in a doped system, the states that contribute to transport are those in a narrow region (of size $\sim k_{\rm B} T$) around the Fermi energy. Those states have momentum nearly identical to the Fermi momentum $k_{\rm F}$, and the velocity operator evaluated on them is simply proportional to the momentum (via the constant $k_{\rm F}$). Since the same states give the dominant contribution to the total momentum of the system, and the latter is conserved by e-e interactions, also the velocity is a conserved quantity~\cite{footnote_conductivity}. This paves the way to a hydrodynamic description of the velocity, with a Navier-Stokes equation~\cite{tomadin_prb_2013} that contains the same viscosities as the momentum one. It is indeed clear that the two equations must be proportional to each other via the cyclotron mass $m_{\rm c} = k_{\rm F}/v_{\rm F}$.

\subsection{The hydrodynamic approach}
The long-wavelength dynamics of the 2D MDF liquid in a doped graphene sheet is described by the relativistic Navier-Stokes equation in 2+1 dimensions~\cite{Landau_6}. In order to avoid confusion, we remark that the use of a relativistic equation is dictated by the low-energy linear-in-momentum band dispersion of quasiparticles, which, however, move with a Fermi velocity $v_{\rm F}$ that is much smaller than the speed of light $c$. For this reason we neglect retardation effects.

The Navier-Stokes equation for MDFs reads~\cite{tomadin_prb_2013}
\begin{equation} \label{eq:Navier_Stokes}
w u^\mu \partial_\mu u_\nu - (\delta^\mu_\nu - u_\nu u^\mu) \partial_\mu P + (\delta^\mu_\nu - u_\nu u^\mu) \partial_\rho \tau_\mu^\rho = F_\nu
~,
\end{equation}
where the space- and time-dependences of the various quantities are suppressed for brevity and the sum over repeated indices is understood.
Here $a^\mu$ is a contravariant vector, while $a_\mu = \eta_{\mu\nu}a^\nu$ is its covariant counterpart, and $\eta_{\mu\nu} = {\rm diag}(1,-1,-1,-1)$ is the metric tensor of flat spacetime. $\delta_\mu^\nu$ is the four-dimensional Kronecker delta.
In Eq.~(\ref{eq:Navier_Stokes}) $w \equiv w({\bm r},t)$ is the enthalpy density (which is equivalent to the Drude weight), $P\equiv P({\bm r},t)$ is the pressure, $u^\mu \equiv \gamma(v) (1, {\bm v}/v_{\rm F})$ is a three-component velocity vector, ${\bm v}={\bm v}({\bm r},t)$ is the 2D fluid-element velocity, and $\gamma(v) = (1-v^2/v_{\rm F}^2)^{-1/2}$. We assume that all the thermodynamic quantities in Eq.~(\ref{eq:Navier_Stokes}) have been regularized by subtracting the (divergent) non-interacting contribution due to the infinite sea of states in valence band. 
This implies that in the non-interacting limit they depend only on the excess carrier density in the conduction band. However, corrections due to e-e interactions between states in conduction and in valence band are fully taken into account.

Moreover, we assume that states in valence band do not contribute to transport, which is certainly a good approximation in the limit in which the wave vector and frequency of the external perturbation are small. The fluid velocity can thus be expressed, in the ``non-relativistic'' limit $|{\bm v}| \ll v_{\rm F}$, in terms of the macroscopic current ${\bm j}({\bm r},t)$ carried by quasiparticles in conduction band as ${\bm v}({\bm r},t)\equiv {\bm j}({\bm r},t)/n({\bm r},t)$. Here $n({\bm r},t)$ is the position- and time-dependent excess number density. In Eq.~(\ref{eq:Navier_Stokes}) we also defined the stress tensor $\tau_{\mu\nu} = \tau_{\mu\nu}({\bm r}, t)$, whose most general form, based on symmetry arguments, is reported in Eq.~(\ref{eq:stress_definition})~\cite{Landau_6}. Finally, in the Navier-Stokes Eq.~(\ref{eq:Navier_Stokes}) we introduced the driving force~\cite{Giuliani_and_Vignale} $F_\nu  \equiv F_\nu ({\bm r}, t)$. As explained above, we do not introduce any retardation effect in the driving term since the Fermi velocity is much smaller than the speed of light.

Linearizing Eq.~(\ref{eq:Navier_Stokes}), taking the ``non-relativistic'' limit $|{\bm v}| \ll v_{\rm F}$ [which implies $\gamma(v) \sim 1$], and considering only the spatial components we get~\cite{Landau_6,tomadin_prb_2013,Muller_prl_2009}
\begin{eqnarray} \label{eq:Navier_Stokes_linear}
&& \!\!\!\!\!\!\!\!
\frac{w}{v_{\rm F}^2} \partial_t  v_i + \left( \nabla_i + \frac{u_i}{v_{\rm F}^2} \partial_t \right) P - \nabla_j \Big\{\eta_0 \big[\nabla_j v_i + \nabla_i v_j 
\nonumber\\
&-& ({\bm \nabla}\cdot{\bm v}) \delta_{ij}\big] + \zeta_0 ({\bm \nabla}\cdot{\bm v}) \delta_{ij} \Big\} = n \frac{e}{c} \partial_t A_i
~,
\end{eqnarray}
where the Latin indices $i,j = x,y$ denote the Cartesian components of the vectors. Since we are interested in deriving the current response of the fluid to an external perturbation, in Eq.~(\ref{eq:Navier_Stokes_linear}) we expressed~\cite{Giuliani_and_Vignale} the driving field as the time derivative of a vector potential ${\bm A}\equiv {\bm A}({\bm r},t)$. We assume the density to be close to its equilibrium value (which only in this Section we denote as $n_{\rm eq}$), i.e.~$n({\bm r}, t) = n_{\rm eq} + \delta n({\bm r}, t)$ with $|\delta n({\bm r}, t)/n_{\rm eq}| \ll 1$. Moreover, in a linear-response fashion, also the current density ${\bm j}({\bm r}, t)$ and the vector potential ${\bm A}({\bm r},t)$ are assumed to be small (note that the current is zero at equilibrium). 

Using well-known thermodynamic relations we rewrite~\cite{Giuliani_and_Vignale}
\begin{equation} \label{eq:pression_bulk_mod_1}
\partial_t P = \frac{\partial P}{\partial n} \partial_t n = \frac{{\cal B}}{n_{\rm eq}} \partial_t n
~,
\end{equation}
and
\begin{equation} \label{eq:pression_bulk_mod_2}
{\bm \nabla} P = \frac{\partial P}{\partial n} {\bm \nabla} n = \frac{{\cal B}}{n_{\rm eq}} {\bm \nabla} n
~,
\end{equation}
where ${\cal B} = {\cal B}(n)$ is the bulk modulus~\cite{Giuliani_and_Vignale} of the 2D MDF liquid regularized, as explained above, by subtracting the non-interacting contribution of the valence band. In Eqs.~(\ref{eq:pression_bulk_mod_1}) and~(\ref{eq:pression_bulk_mod_2}) we suppressed the space- and time-dependences of the various quantities for brevity. These equations, together with the continuity equation $\partial_t n({\bm r},t) = -{\bm \nabla}\cdot {\bm j}({\bm r},t)$, allow us to rewrite the Fourier transform of Eq.~(\ref{eq:Navier_Stokes_linear}) as
\begin{equation} \label{eq:Navier_Stokes_linear_FT}
\frac{w}{v_{\rm F}^2} \omega  j_i - \frac{\cal B}{\omega} ({\bm q}\cdot{\bm j}) q_i + i \eta_0 q^2 j_i + i\zeta_0 ({\bm q}\cdot{\bm j})q_i = n_{\rm eq}^2 \frac{e}{c} \omega A_i
~.
\end{equation}
To obtain this equation we discarded a term proportional to the product $\delta n({\bm r},t) {\bm A}({\bm r}, t)$, which gives a contribution beyond the linear regime. Moreover we used that in this regime $\partial_t{\bm v}({\bm r},t) = \partial_t [{\bm j}({\bm r},t)/n({\bm r},t)] \simeq \partial_t {\bm j}({\bm r},t)/n_{\rm eq}$. A similar approximation was used for terms which contain spatial derivatives of ${\bm v}({\bm r},t)$.

We now observe~\cite{Giuliani_and_Vignale} that Eq.~(\ref{eq:Navier_Stokes_linear_FT}) is suitable to describe the long-wavelength dynamics of 2D MDFs in the collisional $\omega\tau_{\rm ee} \ll 1$ limit. In this regime~\cite{Giuliani_and_Vignale} the system is indeed expected to behave as a liquid, with a finite shear viscosity and a vanishing shear modulus. However, in the opposite, collisionless, limit ($\omega\tau_{\rm ee} \gg 1$) a solid-like behavior is expected to emerge, characterized by a small shear viscosity and a finite shear modulus~\cite{Giuliani_and_Vignale}. The system may thus be described by the equations of the elasticity theory~\cite{Landau_7}. A unified description  is achieved by introducing~\cite{Giuliani_and_Vignale} the complex bulk and shear moduli,  ${\tilde {\cal B}}_\omega = {\cal B}_\omega - i\omega \zeta_\omega$  and ${\tilde {\cal S}}_\omega = {\cal S}_\omega - i\omega \eta_\omega$ respectively, and by rewriting Eq.~(\ref{eq:Navier_Stokes_linear_FT}) as
\begin{eqnarray} \label{eq:Navier_Stokes_linear_FT_2}
\frac{w}{v_{\rm F}^2} \omega  j_i - \frac{{\tilde {\cal B}}_\omega}{\omega} ({\bm q}\cdot{\bm j}) q_i - \frac{{\tilde {\cal S}}_\omega}{\omega} q^2 j_i = n_{\rm eq}^2 \frac{e}{c} \omega A_i
~.
\end{eqnarray}
Note that, from the above discussion we expect $\Re e ({\tilde {\cal S}}_\omega) \to 0$ in the low-frequency regime. Separating the longitudinal and transverse components of Eq.~(\ref{eq:Navier_Stokes_linear_FT_2}), we finally get %
\begin{eqnarray} \label{eq:longitudinal_current}
{\bm j}_{\rm L}(q,\omega) &=& \frac{n_{\rm eq}^2 \omega^2}{w \omega^2/v_{\rm F}^2 - ({\tilde {\cal B}}_\omega+{\tilde {\cal S}}_\omega) q^2} \frac{e}{c} {\bm A}_{\rm L}(q,\omega) 
\nonumber\\
&\equiv& \chi_{\rm L}(q,\omega)\frac{e}{c} {\bm A}_{\rm L}(q,\omega)
~,
\end{eqnarray}
and
\begin{eqnarray} \label{eq:transverse_current}
{\bm j}_{\rm T}(q,\omega) &=& \frac{n_{\rm eq}^2 \omega^2}{w \omega^2/v_{\rm F}^2 - {\tilde {\cal S}}_\omega q^2} \frac{e}{c} {\bm A}_{\rm T}(q,\omega) 
\nonumber\\
&\equiv& \chi_{\rm T}(q,\omega)\frac{e}{c} {\bm A}_{\rm T}(q,\omega)
~.
\end{eqnarray}
The transverse component of any 2D vector ${\bm v}$ is defined as ${\bm v}_{\rm T} = {\bm v} - {\hat {\bm q}}({\hat {\bm q}}\cdot{\bm v})$, while its longitudinal component is $v_{\rm L} = {\hat {\bm q}}\cdot{\bm v}$. 
Eqs.~(\ref{eq:longitudinal_current}) and~(\ref{eq:transverse_current}) provide a macroscopic definition of the longitudinal [$\chi_{\rm L}(q,\omega)$] and transverse [$\chi_{\rm T}(q,\omega)$] components of the current-current linear response function. Indeed, in a rotationally-invariant system
\begin{eqnarray} \label{eq:current_current_decomposition}
\chi_{j_\ell j_k}({\bm q},\omega) &=&\chi_{\rm L}(q,\omega)\frac{q_\ell q_k}{q^2} \nonumber \\
&+& \chi_{\rm T}(q,\omega)\left(\delta_{\ell k} - \frac{q_\ell q_k}{q^2}\right)~,
\end{eqnarray}
where $\ell,k = x,y$ are Cartesian indices. It is easy to show that, if ${\bm q} = q {\hat {\bm x}}$, $\chi_{\rm L}(q,\omega) = \chi_{j_x j_x}({\bm q},\omega)$ and $\chi_{\rm T}(q,\omega) = \chi_{j_y j_y}({\bm q},\omega)$. These relations will be used in what follows.

Expanding Eqs.~(\ref{eq:longitudinal_current}) and~(\ref{eq:transverse_current}) to order $q^2/\omega^2$ we get
\begin{eqnarray} \label{eq:longitudinal_response_small_q}
\Im m[\chi_{\rm L}(q,\omega)] = -\left(\frac{v_{\rm F}^2}{w^{(0)}} n_{\rm eq}\right)^2 \frac{q^2}{\omega} (\zeta_\omega + \eta_\omega)
~,
\end{eqnarray}
and
\begin{eqnarray} \label{eq:transverse_response_small_q}
\Im m[\chi_{\rm T}(q,\omega)] = -\left(\frac{v_{\rm F}^2}{w^{(0)}} n_{\rm eq}\right)^2 \frac{q^2}{\omega} \eta_\omega
~,
\end{eqnarray}
where we have approximated the enthalpy density by its {\it non-interacting} value $w^{(0)} = n_{\rm eq} \varepsilon_{\rm F}$. This is justified because the relations~(\ref{eq:longitudinal_response_small_q})-(\ref{eq:transverse_response_small_q}) will be used  to estimate the {\it high-frequency} bulk and shear viscosities, $\zeta_\infty$ and $\eta_\infty$, respectively. In the high-frequency regime, the two viscosities can be calculated perturbatively, the first non-vanishing contribution being of second order in the dimensionless 
strength of e-e interactions, i.e.~$\alpha_{\rm ee}$ in Eq.~(\ref{eq:finestructure}). It is therefore legitimate, up to second order,  to neglect interaction corrections to $w$.

By inverting Eqs.~(\ref{eq:longitudinal_response_small_q}) and~(\ref{eq:transverse_response_small_q}) we can express the high-frequency bulk and shear viscosities of a 2D MDF liquid in terms of the current-current response functions. To second order in the strength of e-e interactions these hydrodynamic coefficients read~\cite{Giuliani_and_Vignale}
\begin{eqnarray} \label{eq:viscosities_def}
\eta_\infty &=& -\lim_{\omega\to 0} \lim_{q\to 0} \frac{\omega m_{\rm c}^2}{q^2} \Im m[\chi_{\rm T}(q,\omega)]~,
\nonumber\\
\zeta_\infty &=& -\lim_{\omega\to 0} \lim_{q\to 0} \frac{\omega m_{\rm c}^2}{q^2}\left\{\Im m[\chi_{\rm L}(q,\omega)] - \Im m[\chi_{\rm T}(q,\omega)]\right\}
~.
\nonumber\\
\end{eqnarray}
Note that in Eq.~(\ref{eq:viscosities_def}) the limit $\omega\to 0$ is taken {\it after} the limit $\omega\tau_{\rm ee} \gg 1$.

The current-current response functions on the right-hand side of Eq.~(\ref{eq:viscosities_def}) have a rigorous microscopic definition~\cite{Giuliani_and_Vignale} in terms of Kubo products, i.e.~\cite{proper_chi}
\begin{equation} \label{eq:chi_Kubo}
\chi_{AB}(\omega) = \frac{1}{S} \langle\langle {\hat A}; {\hat B} \rangle\rangle_\omega
~,
\end{equation}
where $S$ is the 2D electron system area, ${\hat A}$ and ${\hat B}$ are operators, and
\begin{eqnarray} \label{eq:SM_Kubo}
\langle\langle {\hat A}; {\hat B} \rangle\rangle_\omega \equiv -i \int_0^\infty dt e^{i(\omega+i\eta) t} \langle[{\hat A}(t),{\hat B}]\rangle
~.
\end{eqnarray}
Note that the average $\langle \ldots \rangle$ in Eq.~(\ref{eq:SM_Kubo}) is taken over the ground state of the interacting system. 

The longitudinal component $\chi_{\rm L}(q,\omega)$ of the current-current response function of 2D MDFs was calculated in Ref.~\onlinecite{Principi_prb_2013}. In this Article we evaluate the {\it transverse} current-current response function $\chi_{\rm T}(q,\omega)$ at second order in the strength of e-e interactions and in the limit $v_{\rm F} q \ll \omega \ll 2\varepsilon_{\rm F}$.

We note that, from Eq.~(\ref{eq:longitudinal_current}), it is possible to derive an expression for the non-local charge conductivity $\sigma(q,\omega)$. Replacing $\omega^2 \to \omega[\omega + i/\tau(q,\omega)]$ in its denominator to account for non-momentum-conserving dissipative effects, and neglecting the real parts of ${\tilde {\cal B}}_\omega$ and ${\tilde {\cal S}}_\omega$ (which are negligible in the limit $v_{\rm F} q\ll \omega$) we get
\be\label{eq:conductivity}
\sigma(q,\omega) &\equiv& \frac{\chi_{\rm L}(q,\omega)}{-i\omega}=
\frac{n/m_{\rm c}}{-i\omega + 1/\tau(q,\omega) + \nu_\omega q^2}~.
\ee
In Eq.~(\ref{eq:conductivity}) we defined the kinematic viscosity $\nu_\omega \equiv \eta_\omega/(n m_{\rm c})$, and we used the fact that $\zeta_\omega = 0$ (as will be demonstrated in what follows). Since Eq.~(\ref{eq:conductivity}) is valid to all orders in the perturbative expansion in the strength of the e-e coupling constant, the cyclotron mass $m_{\rm c}$ must be renormalized according to Landau theory of normal Fermi liquids~\cite{Giuliani_and_Vignale}.

\subsection{The microscopic approach}
\label{sect:microscopic_approach}
In order to get a macroscopic Navier-Stokes equation for a given quantity, it is necessary for the latter to be conserved by interactions at the microscopic level. For example the momentum density operator ${\hat {\bm p}}({\bm x},t)$, whose Fourier transform is microscopically defined as
\begin{eqnarray} 
{\hat {\bm p}}_{\bm q} = \sum_{{\bm k} \in {\rm BZ}, \alpha} {\bm k} {\hat \psi}^\dagger_{{\bm k}-{\bm q},\alpha} {\hat \psi}_{{\bm k},\alpha}~,
\end{eqnarray}
is a locally conserved quantity, and satisfies the following continuity equation
\begin{eqnarray} \label{eq:momentum_EOM}
\partial_t {\hat p}_j({\bm x}, t) = -\partial_i {\hat \tau}_{ij}({\bm x},t)~,
\end{eqnarray}
where $i,j = x,y$ are Cartesian indices and the summation over repeated indices is understood. Eq.~(\ref{eq:momentum_EOM}) defines the symmetric stress tensor operator ${\hat \tau}_{ij}({\bm x},t)$. 
In the non-interacting limit, the Fourier transform of ${\hat \tau}_{ij}({\bm x},t)$ is defined by
\begin{eqnarray} \label{eq:non_int_stress_tensor}
{\hat \tau}_{{\bm q},ij}^{(0)} = v_{\rm F}\sum_{{\bm k},\alpha,\beta} {\hat \psi}^\dagger_{{\bm k}-{\bm q},\alpha} \frac{k_i \sigma^j_{\alpha\beta} + k_j \sigma^i_{\alpha\beta}}{2} {\hat \psi}_{{\bm k},\beta}
~.
\end{eqnarray}

Let us now consider a time dependent deformation ${\bm u}({\bm r}, t)$ of the electron coordinates $\{{\bm r}_i, i=1,\ldots, N\}$, which are thus shifted as ${\bm r}_i \to {\bm r}_i - u({\bm r}_i,t)$. This deformation should not be confused with a similar one that could be applied to the lattice. To linear order, the metric tensor of the deformed system is
\begin{eqnarray}
g_{ij} &=& \delta_{ij} - (\partial_i u_j + \partial_j u_i)
\nonumber\\
&\equiv& 
\delta_{ij} - 2 u_{ij}
~,
\end{eqnarray}
where $u_{ij}$ is the strain tensor of the elasticity theory~\cite{Landau_7}. Note that $g^{ij} = \delta^{ij} + 2 u^{ij}$. Under this transformation the Hamiltonian becomes (to linear order in $u_{ij}$)
\begin{eqnarray} \label{eq:Hamiltonian_strain}
{\hat {\cal H}}' &=& {\hat {\cal H}} + 2\frac{\delta {\hat {\cal H}}}{\delta g^{ij}} \Bigg|_{g_{ij} = \delta_{ij}} u^{ij}
\nonumber\\
&=& 
{\hat {\cal H}} + {\hat \tau}_{ij} u^{ij}
~.
\end{eqnarray}
It can be shown~\cite{Birrell_and_Davies} that the non-interacting part of the stress tensor defined by Eq.~(\ref{eq:Hamiltonian_strain}) is identical to Eq.~(\ref{eq:non_int_stress_tensor}).  The variation of the Hamiltonian, ${\hat \tau}_{ij} u^{ij}$, induces a variation  in the expectation value of the stress tensor operator.   To linear order in $u^{ij}$ we get
\begin{eqnarray}
\delta \langle {\hat \tau}_{ij} \rangle ({\bm q},\omega) = Q_{ij,kl}({\bm q},\omega) u^{kl}({\bm q},\omega)
~,
\end{eqnarray}
which defines the tensor of elasticity $Q_{ij,kl}({\bm q},\omega)$. In a rotationally invariant system this tensor can be decomposed as
\begin{eqnarray}
\lim_{\omega\to 0} Q_{ij,kl}({\bm q}={\bm 0},\omega) &=& {\tilde {\cal B}}_\omega \delta_{ij} \delta_{kl} 
\nonumber\\
&+&
{\tilde {\cal S}}_\omega \left(\delta_{ik}\delta_{jl} + \delta_{il}\delta_{jk} - \frac{2}{d} \delta_{ij} \delta_{kl} \right)
~,
\nonumber\\
\end{eqnarray}
where $d=2$ is the dimensionality of the system. From the general theory of linear response~\cite{Giuliani_and_Vignale} and the form of the perturbation in Eq.~(\ref{eq:Hamiltonian_strain}), it is clear that the elasticity tensor can be calculated from the stress-stress response function, i.e.
\begin{eqnarray} \label{eq:elasticity_stress_stress}
Q_{ij,kl}({\bm q},\omega) = \frac{1}{S} \langle\langle {\hat \tau}_{ij}({\bm q}); {\hat \tau}_{kl}(-{\bm q}) \rangle\rangle_\omega
~.
\end{eqnarray}
In writing Eq.~(\ref{eq:elasticity_stress_stress}) we have neglected a ``contact'' term (analogous to the diamagnetic term of the current-current response function), which has been recently discussed e.g.~in Ref.~\onlinecite{Bradlyn_prb_2012}. However, this term is purely real and thus does not contribute to the bulk and shear viscosities, which are defined as
\begin{eqnarray} \label{eq:viscosities_stress_stress}
\eta_\omega &=& -\lim_{\omega\to 0} \frac{\Im m[\chi_{xy,xy}({\bm q}={\bm 0},\omega)]}{\omega}~,
\nonumber\\
\zeta_\omega &=& -\lim_{\omega\to 0} \Bigg\{
\frac{\Im m[\chi_{xx,xx}({\bm q}={\bm 0},\omega)]}{\omega} 
\nonumber\\
&-&
\frac{2(d-1)\Im m[\chi_{xy,xy}({\bm q}={\bm 0},\omega)]/d}{\omega}
\Bigg\}
~.
\nonumber\\
\end{eqnarray}
\subsection{Equivalence of the hydrodynamic and microscopic approaches in a Fermi liquid}
In a Galilean invariant system (in which the current operator is ${\hat {\bm j}}_{\bm q} = {\hat {\bm p}}_{\bm q}/m_{\rm c}$) it is rather easy to see that the hydrodynamic and microscopic approach give identical results. Thanks to Eq.~(\ref{eq:momentum_EOM}) and the following identity
\begin{eqnarray} \label{eq:derivative_Kubo_product}
\omega \langle\langle {\hat A}; {\hat B} \rangle\rangle_\omega &=& \langle\langle i\partial_t{\hat A}; {\hat B} \rangle\rangle_\omega + \langle[{\hat A}, {\hat B}]\rangle
\nonumber\\
&=&
\langle\langle {\hat A};- i\partial_t{\hat B} \rangle\rangle_\omega + \langle[{\hat A}, {\hat B}]\rangle
~,
\end{eqnarray}
we get (${\bm q} = q{\hat {\bm x}}$)
\begin{eqnarray} \label{eq:chi_stress_stress_current_current}
&& \Im m[\chi_{xy,xy}({\bm q}={\bm 0},\omega)] = \lim_{q\to 0} \frac{m_{\rm c}^2 \omega^2}{q^2} \Im m [\chi_{\rm T}(q,\omega)]~,
\nonumber\\
&&
\Im m[\chi_{xx,xx}({\bm q}={\bm 0},\omega)] = \lim_{q\to 0} \frac{m_{\rm c}^2 \omega^2}{q^2} \Im m [\chi_{\rm L}(q,\omega)]
~.
\nonumber\\
\end{eqnarray}
Here, we used twice Eq.~(\ref{eq:derivative_Kubo_product}) and the fact that the average of the commutator on the right-hand-side of that equation is purely real. From Eqs.~(\ref{eq:viscosities_stress_stress}) and~(\ref{eq:chi_stress_stress_current_current}) one immediately recovers Eq.~(\ref{eq:viscosities_def}). Notice that, since in a Galilean invariant system $m_{\rm c}$ is not renormalized by e-e interactions, Eq.~(\ref{eq:viscosities_def}) is valid to all orders in the strength of e-e interactions, and, therefore, at all frequencies (as long as $\omega \ll \varepsilon_{\rm F}$). 

The connection is much less straightforward in the case of the 2D MDF liquid in graphene. Indeed, in a system with a linear dispersion, the current operator is not proportional to the momentum-density operator. Therefore, Eq.~(\ref{eq:chi_stress_stress_current_current}) does not hold, in general. This fact  notwithstanding, we have found that in the Fermi liquid regime Eq.~(\ref{eq:chi_stress_stress_current_current}) still holds in an approximate sense and, therefore, to the level of accuracy we are interested in, the microscopic and hydrodynamic approaches coincide.
 The details of the argument are presented in Appendix~\ref{app:equivalence}, but the basic idea is quite simple.  First of all, it is clear that at sufficiently low frequency interband transitions are irrelevant and can be disregarded.  Then it is clear that, within a narrow band of energies around the Fermi level in the conduction band, there is no qualitative difference between the linear dispersion of MDFs and the {\it linearized} dispersion of ordinary massive fermions: the two dispersions are indistinguishable if the ordinary Schr\"{o}dinger fermions are assigned the cyclotron mass $m_{\rm c} = k_{\rm F}/v_{\rm F}$. We conclude that the equivalence of the microscopic and hydrodynamic approaches carries over to MDFs with the simple replacement of the effective mass by the cyclotron effective mass.

\section{Calculation of the viscosity in the relaxation time approximation}
\label{sect:RTA}
As we pointed out in the Introduction, e-e interactions enter the calculation of the viscosities quite differently in the high-frequency (collisionless) and low-frequency (collisional) regimes. In the high-frequency regime, the effect of the interaction is {\it perturbative}, meaning that there would be no viscosity without interactions creating a correlation between the motions of adjacent parts of the liquid. In the low-frequency regime, e-e interactions are {\it non-perturbative} as their primary role is to establish a finite mean free path $\ell_{\rm ee}$ for electrons: this mean free path would be infinite in the absence of interactions.   The problem is how to connect in a seamless way these two very different regimes of e-e scattering.  The relaxation time approximation, summarized in Eqs.~(\ref{eq:all_freq_viscosity}),  offers a simple and physically motivated way to achieve this connection.  The derivation of these formulas closely parallels the derivation given in Ref.~\onlinecite{Conti_prb_1999} for Galilean invariant systems. Only a minor adaptation is needed, namely the replacement of the bare electron mass $m$ by the cyclotron effective mass $m_{\rm c}$, as discussed in the previous Section.  We therefore refer the reader to Ref.~\onlinecite{Conti_prb_1999},  where a detailed derivation of  Eqs.~(\ref{eq:all_freq_viscosity}) is provided, and we focus in this Section on the calculation of the inputs for Eqs.~(\ref{eq:all_freq_viscosity}), i.e.~ the high-frequency viscosities and the corresponding relaxation times.  An alternative derivation of  Eqs.~(\ref{eq:all_freq_viscosity}) is presented in Appendix~\ref{sect:GRTA}.

\subsection{The high-frequency viscosities}
\label{sect:high_frequency_viscosity}
To evaluate the transverse current-current response function, we adopt the (Hamann-Overhauser~\cite{hamann_pr_1966} or Schrieffer-Wolff~\cite{schrieffer_pr_1966}) canonical transformation approach outlined in Refs.~\onlinecite{Principi_prb_2013,Principi_prbR_2013,Principi_prb_2014}. 

First of all, we introduce a unitary transformation generated by a Hermitian operator ${\hat F}$,
\begin{eqnarray} \label{eq:SM_F_tranf}
{\hat {\cal H}}' = e^{i {\hat F}} ({\hat {\cal H}}_0 + {\hat {\cal H}}_{\rm ee}) e^{-i {\hat F}}
~,
\end{eqnarray}
which cancels the e-e interaction from the transformed Hamiltonian, {\it i.e.} we require ${\hat {\cal H}}' \equiv {\hat {\cal H}}_0$ to {\it second order} in the dimensionless strength $\alpha_{\rm ee}$ of e-e interactions. The transformation ${\hat F} = {\hat \openone} + {\hat F}_1 + {\hat F}_2 +...$ is determined order-by-order in perturbation theory. Here ${\hat \openone}$ is the identity and ${\hat F}_n$ is the term of $n$-th order in the strength of e-e interactions. As shown in Refs.~\onlinecite{Principi_prb_2013,Principi_prbR_2013,Principi_prb_2014}, to calculate the imaginary part of the current-current response function in the limit $v_{\rm F} q \ll \omega \ll 2 \varepsilon_{\rm F}$ it is sufficient to determine ${\hat F}_1$, which satisfies the equality $[{\hat {\cal H}}_0, i {\hat F}_1] = {\hat {\cal H}}_{\rm ee}$.

After carrying out the transformation ${\hat F}$, the Kubo product in Eq.~(\ref{eq:SM_Kubo}) is reduced to the evaluation of a non-interacting response function $\propto \langle \langle {\hat A}', {\hat B}' \rangle \rangle_{0,\omega}$. The subscript ``0'' here means that the average $\langle \ldots \rangle$ is now performed over the ground state of the non-interacting system and that the time evolution is generated by ${\hat {\cal H}}_0$. However, the operators ${\hat A}' = e^{i {\hat F}} {\hat A} e^{-i {\hat F}}$ and ${\hat B}' = e^{i {\hat F}} {\hat B} e^{-i {\hat F}}$ are now dressed by e-e interactions. The key idea is to realize that the calculation of $\Im m [\chi_{j_\alpha j_\beta}({\bm q},\omega)]$ to {\it second} order in the strength of e-e interactions and in the limit $v_{\rm F} q \ll \omega \ll 2 \varepsilon_{\rm F}$ requires only the knowledge of the transformed current-density operator ${\hat {\bm j}}'_{{\bm q}}$ to {\it first} order~\cite{Principi_prb_2013,Principi_prbR_2013,Principi_prb_2014}, {\it i.e.} ${\hat {\bm j}}'_{1,{\bm q}} = {\hat {\bm j}}_{{\bm q}} + {\hat {\bm j}}_{1,{\bm q}}$, where
\be\label{eq:current_first_order}
{\hat {\bm j}}_{1,{\bm q}} = [i {\hat F}_1, {\hat {\bm j}}_{{\bm q}}]~.
\ee
Thus, to second order in the strength of e-e interactions, and for $v_{\rm F} q\ll \omega \ll 2\varepsilon_{\rm F}$, we get the following exact-eigenstate (Lehmann) representation~\cite{Giuliani_and_Vignale} of the current-current response function
\begin{eqnarray} \label{eq:SM_chi_jj_def_second}
\Im m[\chi_{j_\alpha j_\beta}({\bm q},\omega)] &=&
-\pi \sum_m 
\langle 0 | {\hat j}_{1,{\bm q},\alpha} |m\rangle \langle m | {\hat j}_{1,-{\bm q},\beta} |0\rangle 
\nonumber\\
&\times&
\delta(\omega-\omega_{m0})
~.
\nonumber\\
\end{eqnarray}
The calculation of ${\hat F}_1$ and ${\hat {\bm j}}_{1,{\bm q}}$ is carried out in Appendices~\ref{sect:SM_calculation_F1_j1} and~\ref{sect:SM_Upsilon_manipulation}. 

Here we quote the final formula for the two components (longitudinal and transverse) of the current-current response function, exact to second order in e-e interactions and in the large-$N_{\rm f}$ limit, which is
\begin{widetext}
\begin{eqnarray} \label{eq:chi_rhorho_mode_decoupling_def}
\Im m[\chi_{\ell}(q,\omega)]&=&
- \sum_{\alpha,\beta = x,y}\int \frac{d^2{\bm q}'}{(2\pi)^2} v_{{\bm q}'}^2
\int_0^\omega \frac{d\omega'}{\pi} \Big\{ \Gamma_\alpha^{(\ell)}({\bm q},{\bm q}') \Gamma_{\beta}^{(\ell)}(-{\bm q},-{\bm q}')
\Im m[\chi^{(0)}_{nn}(q',\omega')]\Im m[\chi^{(0)}_{j_\alpha j_\beta}({\bm q}',\omega-\omega')]
\nonumber\\
&+&
\Gamma_\alpha^{(\ell)}({\bm q},{\bm q}') \Gamma_\beta^{(\ell)}(-{\bm q},{\bm q}')
\Im m[\chi^{(0)}_{n j_\alpha}(-{\bm q}',\omega')]~\Im m[\chi^{(0)}_{n j_\beta}({\bm q}',\omega-\omega')]\Big\}~.
\end{eqnarray}
\end{widetext}
In this equation $\ell={\rm L},{\rm T}$ and $\chi^{(0)}_{nn}(q,\omega)$, $\chi^{(0)}_{j_\alpha j_\beta}({\bm q},\omega)$, and $\chi^{(0)}_{n j_\alpha}({\bm q},\omega)$ are the {\it non-interacting} density-density, current-current, and density-current response functions of a 2D gas of MDFs. The quantities $\{\Gamma_\alpha^{(\ell)}({\bm q},{\bm q}'); \alpha = x,y; \ell={\rm L},{\rm T}\}$ are defined as
\begin{eqnarray} \label{eq:Gamma_T_def}
\Gamma^{({\rm T})}_\alpha({\bm q},{\bm q}')
\!\! &=& \!\!
\frac{v_{\rm F} q}{\omega^2} 
\left[
\frac{q'_x q'_y}{q'^2} \frac{q'_\alpha}{k_{\rm F}} - 
\left(1 - \frac{q'^2}{4 k_{\rm F}^2} \right)
\right.
\nonumber\\
&\times&
\left.
\frac{q'_x \delta_{\alpha,y} + q'_y \delta_{\alpha, y}}{k_{\rm F}}
\right]
+
\frac{q'^2}{4 v_{\rm F} k_{\rm F}^3} \delta_{\alpha,x} 
~,
\end{eqnarray}
and
\begin{eqnarray} \label{eq:Gamma_L_def}
\Gamma_\alpha^{({\rm L})}({\bm q},{\bm q}')
\!\! &=& \!\!
\frac{v_{\rm F} q_x}{\omega^2} \Bigg[ \frac{q_y'^2}{q'^2} \frac{q'_\alpha}{k_{\rm F}}
-2 \frac{q'_x}{k_{\rm F}}
\Bigg( 1 - \frac{q'^2}{4 k_{\rm F}^2} \!\Bigg)
\delta_{\alpha,x} 
\Bigg]
\nonumber\\
&+&
\frac{q'^2}{4 v_{\rm F} k_{\rm F}^3} \delta_{\alpha,x} 
~.
\end{eqnarray}
We stress that the {\it imaginary} parts of the three linear-response functions $\chi^{(0)}_{nn}(q,\omega)$, $\chi^{(0)}_{j_\alpha j_\beta}({\bm q},\omega)$, and $\chi^{(0)}_{n j_\alpha}({\bm q},\omega)$ do not depend on any ultraviolet cut-off in the low-energy MDF limit. Moreover, since in the limit of $\omega\to 0$ the integral over $q'$ is naturally restricted to $0\leq q' \leq 2k_{\rm F}$, no ultraviolet regularization is needed in Eq.~(\ref{eq:chi_rhorho_mode_decoupling_def}). The only pathology of the integral in Eq.~(\ref{eq:chi_rhorho_mode_decoupling_def}) appears in the infrared $q' \to 0$ limit, due to the $1/q'$ singularity of the Coulomb potential $v_{{\bm q}'}$. This problem is cured by screening, as we will further discuss below.

We observe that, contrary to what happens in a parabolic-band electron gas, the matrix elements of Eqs.~(\ref{eq:Gamma_T_def}) and~(\ref{eq:Gamma_L_def}) do not vanish in the limit $q\to 0$. The terms that remain finite are due to broken Galilean invariance~\cite{abedinpour_prb_2011}, i.e.~due to the presence of the valence band and, as noted in Ref.~\onlinecite{Principi_prb_2013}, are responsible for a finite optical conductivity in the single-particle optical gap $\omega < 2 \varepsilon_{\rm F}$, which scales as $\sim \omega^2$.
Being a conductivity, it is conceptually wrong to include it in the viscosities, and therefore in what follows we neglect the terms of Eqs.~(\ref{eq:Gamma_T_def}) and~(\ref{eq:Gamma_L_def}) that do not vanish when $q\to 0$, {\it i.e.} the last terms in Eqs~(\ref{eq:Gamma_T_def})-(\ref{eq:Gamma_L_def}). By insisting in retaining these terms, we would (wrongly) get two diverging viscosities.

We stress that such a finite optical conductivity is an effect beyond the Fermi liquid theory, and that it is present only in the high-frequency limit. 
Note however that, contrary to the derivation of the low-frequency limit, the calculation of the high-frequency viscosities does not require any Fermi-liquid assumption. Therefore, the general discussion of Sect.~\ref{sect:viscosity} does not need to be amended to take care of the peculiarities of the high-frequency calculation.

Eq.~(\ref{eq:chi_rhorho_mode_decoupling_def}) is the main result of this Section. Note that in the large-$N_{\rm f}$ limit the second-order expression for the imaginary part of the current-current response function in Eq.~(\ref{eq:chi_rhorho_mode_decoupling_def}) has the appealing form of a convolution of two single-particle spectra~\cite{Nifosi_prb_1998}. The physical interpretation of Eq.~(\ref{eq:chi_rhorho_mode_decoupling_def}) is the following. At long wavelengths and to the lowest non-vanishing order of perturbation theory, the spectrum of the current-current correlator is dominated by the emission of two correlated electron-hole pairs. Each of the Kubo products on the right-hand side of Eq.~(\ref{eq:chi_rhorho_mode_decoupling_def}) describes the rate of generation of a single electron-hole pair. The spectral weight associated with the excitation of two particle-hole pairs with opposite momenta and given total energy $\omega$ is proportional to their convolution.

We now sketch how to make analytical progress in the evaluation of the current-current response function as from Eq.~(\ref{eq:chi_rhorho_mode_decoupling_def}). 

The integrals in Eq.~(\ref{eq:chi_rhorho_mode_decoupling_def}) can be carried out analytically with the help of known formulas for the response functions~\cite{DiracplasmonsRPA}. We first observe that in the low-energy MDF limit the system is translationally and rotationally invariant. The current-current response function $\chi^{(0)}_{j_\alpha j_\beta}({\bm q},\omega)$ is a rank-$2$ tensor that can be therefore decomposed according to Eq.~(\ref{eq:current_current_decomposition}). In the same way, the density-current response function can be seen to be equal to
\begin{equation} \label{eq:density_current_decomposition}
\Im m \chi^{(0)}_{j_\alpha n}({\bm q}',\omega) = \frac{q'_\alpha}{q'} \Im m \chi_{j_{\rm L}n}^{(0)}(q',\omega)
~,
\end{equation}
where $\chi_{j_{\rm L}n}^{(0)}(q',\omega)$ is the non-interacting longitudinal-current-density response function.  Eqs.~(\ref{eq:current_current_decomposition}) and~(\ref{eq:density_current_decomposition}) can be used to perform analytically the angular integration in Eq.~(\ref{eq:chi_rhorho_mode_decoupling_def}), which reads
\begin{eqnarray} \label{eq:chi_rhorho_mode_decoupling_ang_int}
&& \!\!\!\!\!\!\!\!
\Im m[\chi_{\ell}(q,\omega)] =
\int \frac{d^2{\bm q}'}{(2\pi)^2}
\int_0^\omega \frac{d\omega'}{\pi} v_{{\bm q}'}^2 \frac{q'^2}{k_{\rm F}^2}
\nonumber\\
&\times&
\Big\{
a_\ell \Im m[\chi^{(0)}_{nn}(q',\omega')]\Im m[\chi^{(0, {\rm T})}_{j j}(q',\omega-\omega')]
\nonumber\\
&+&
b_\ell \Im m[\chi^{(0)}_{nn}(q',\omega')]\Im m[\chi^{(0, {\rm L})}_{j j}(q',\omega-\omega')]
\nonumber\\
&+&
c_\ell
\Im m[\chi^{(0)}_{n j_{\rm L}}(q',\omega')]~\Im m[\chi^{(0)}_{n j_{\rm L}}(q',\omega-\omega')]\Big\}
~,
\nonumber\\
\end{eqnarray}
where
\begin{equation}
a_{\rm T} = a_{\rm L} = \frac{v_{\rm F}^2 q^2}{\omega^4}\frac{(q'^2/k_{\rm F}^2 - 4)^2}{32} 
~,
\end{equation}
\begin{eqnarray}
b_{\rm L} &=& \frac{v_{\rm F}^2 q^2}{\omega^4}\frac{3 q'^4 - 20 k_{\rm F}^2 q'^2 + 44 k_{\rm F}^4}{32 k_{\rm F}^4} 
~,
\nonumber\\
b_{\rm T} &=& \frac{v_{\rm F}^2 q^2}{\omega^4}\frac{(q'^2/k_{\rm F}^2-2)^2}{32} 
~,
\end{eqnarray}
and, finally,
\begin{eqnarray}
c_{\rm L} &=& \frac{v_{\rm F}^2 q^2}{\omega^4}\frac{3 q'^4 - 20 k_{\rm F}^2 q'^2 + 44 k_{\rm F}^4}{32 k_{\rm F}^4}
~,
\nonumber\\
c_{\rm T} &=& \frac{v_{\rm F}^2 q^2}{\omega^4}\frac{(q'^2/k_{\rm F}^2-2)^2}{32} 
~.
\end{eqnarray}
To cure infrared divergences associated with the long-range tail of the e-e interaction $v_{{\bm q}^\prime}$, we have evaluated the integral over ${\bm q}^\prime$ in Eq.~(\ref{eq:chi_rhorho_mode_decoupling_ang_int}) by replacing $v_{{\bm q}^\prime}$ with a statically-screened Thomas-Fermi interaction, i.e.~$v^{({\rm TF})}_{{\bm q}^\prime} = 2\pi e^2/[\epsilon(q' + q_{\rm TF})]$. Here $q_{\rm TF} = N_{\rm F} \alpha_{\rm ee} k_{\rm F}$ is the Thomas-Fermi screening wave vector~\cite{kotov_rmp_2012}.

In the limit $\omega \to 0$ we can expand the imaginary part of each response function 
$\chi^{(0)}_{nn}$, $\chi^{(0)}_{n j_{\rm L}}$, $\chi^{(0)}_{\rm L}$, and $\chi^{(0)}_{\rm T}$ in a power series of $\omega'$, $\omega -\omega'$ and retain only the leading order of this expansion. The leading contribution to $\Im m[\chi_{\ell}(q,\omega)]$ in Eq.~(\ref{eq:chi_rhorho_mode_decoupling_def}) in powers of $\omega$ in the limit $\omega \to 0$ can be extracted from the following asymptotic formulae:
\begin{equation}\label{q:smallomegaone}
\lim_{\omega' \to 0}\Im m [\chi_{nn}^{(0)}(q',\omega')] = - N_{\rm f}\frac{\sqrt{4 k_{\rm F}^2 - q'^2}}{2 q'} \frac{\omega'}{2\pi v_{\rm F}^2}
\end{equation}
and
\begin{equation}\label{q:smallomegatwo}
\lim_{\omega' \to 0}\Im m [\chi^{(0)}_{\rm T}(q',\omega')] = - N_{\rm f}\frac{2 k_{\rm F}}{q' \sqrt{4 k_{\rm F}^2 - q'^2}} \frac{\omega'}{2\pi v_{\rm F}}~.
\end{equation}
The imaginary parts of density-current and longitudinal current-current response functions scale with higher powers of $\omega'$. Indeed they can be derived from Eq.~(\ref{q:smallomegaone}) according to the formulae~\cite{Giuliani_and_Vignale} $\Im m [\chi_{j_{\rm L}n}^{(0)}(q',\omega')] = \omega' \Im m [\chi^{(0)}_{nn}(q',\omega')]/q'$ and $\Im m [\chi^{(0)}_{\rm L}(q',\omega')] = \omega'^2 \Im m [\chi^{(0)}_{nn}(q',\omega')]/q'^2$.

Since the last two lines of Eq.~(\ref{eq:chi_rhorho_mode_decoupling_ang_int}) give a subleading contribution, in the limit $v_{\rm F} q \ll \omega \ll 2\varepsilon_{\rm F}$, they can be disregarded. Plugging Eq.~(\ref{eq:chi_rhorho_mode_decoupling_ang_int}) back into Eq.~(\ref{eq:viscosities_def}) we finally get
\begin{eqnarray} \label{eq:viscosities_result}
\eta_\infty &=& n {\cal A}_{N_{\rm f}}(\alpha_{\rm ee}) 
~,
\nonumber\\
\zeta_\infty &=& 0
~,
\end{eqnarray}
where ${\cal A}_{N_{\rm f}}(\alpha_{\rm ee}) = 2N_{\rm f}\alpha_{\rm ee}^2 f(N_{\rm f}\alpha_{\rm ee})$, with
\begin{eqnarray}\label{eq:fofx}
f(x) &=& \frac{15 x^3 - 15 x^2  - 52 x + 42}{288 \pi} 
\nonumber\\
&-&
\frac{(5 x^4 - 24 x^2 + 16)}{96 \pi} {\rm arccoth}(1 + x)
~.
\end{eqnarray}
Note that ${\cal A}_{N_{\rm f}}(\alpha_{\rm ee})$ is defined with an extra factor of $2/N_{\rm f}$ with respect to Ref.~\onlinecite{Principi_prb_2013}, due to the factor $m^2_{\rm c}$ in Eq.~(\ref{eq:viscosities_def}), which has been absorbed into the definition of these functions. We underline that the dependence on the strength of e-e interactions beyond $\alpha_{\rm ee}^2$ encoded in $f(x)$ is due to the Thomas-Fermi screened interaction introduced above used to regularize the infrared behavior of the integrand in Eq.~(\ref{eq:chi_rhorho_mode_decoupling_ang_int}). 
As expected~\cite{Giuliani_and_Vignale}, the bulk viscosity is zero.

\begin{figure}[t]
\begin{center}
\begin{tabular}{c}
\includegraphics[width=0.99\columnwidth]{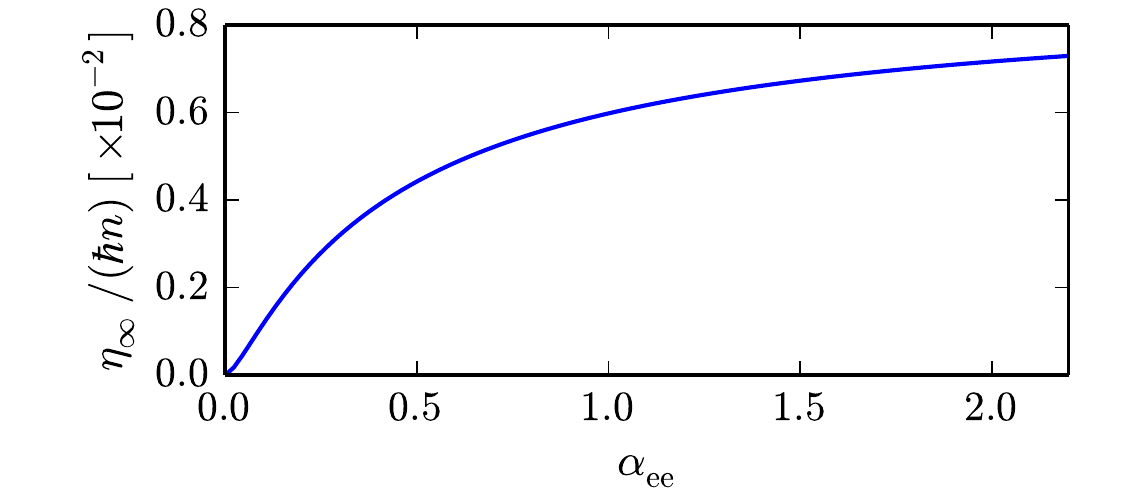}
\end{tabular}
\end{center}
\caption{(Color online) The high-frequency shear viscosity $\eta_\infty$ (in units of the excess carrier density $n$) of the 2D MDF liquid in a doped graphene sheet is plotted as a function of the dimensionless parameter $\alpha_{\rm ee}$, which defines the strength of e-e interactions. Note that the vertical axis must be multiplied by a factor $10^{-2}$.\label{fig:two}}
\end{figure}

In Fig.~\ref{fig:two} we show the high-frequency shear viscosity, as defined in Eq.~(\ref{eq:viscosities_result}), in units of the carrier density $n$, and as a function of the dimensionless parameter $\alpha_{\rm ee}$. In the limit of $\alpha_{\rm ee} \to 0$ we find that
\begin{equation} \label{eq:eta_infty_alpha0}
\eta_\infty \to - n \frac{N_{\rm f}}{6\pi} \alpha_{\rm ee}^2 \ln(\alpha_{\rm ee})
~.
\end{equation}
The extra logarithmic dependence on the coupling constant of Eq.~(\ref{eq:eta_infty_alpha0}) is due to the Thomas-Fermi screening used in the calculation. For $\alpha_{\rm ee} \gg 1$ the Thomas-Fermi screened interaction becomes independent of $\alpha_{\rm ee}$, and the high-frequency shear viscosity tends to $\eta_{\infty} \to n/(9 \pi N_{\rm f})$.

\begin{figure}[t]
\begin{center}
\begin{tabular}{c}
\includegraphics[width=0.99\columnwidth]{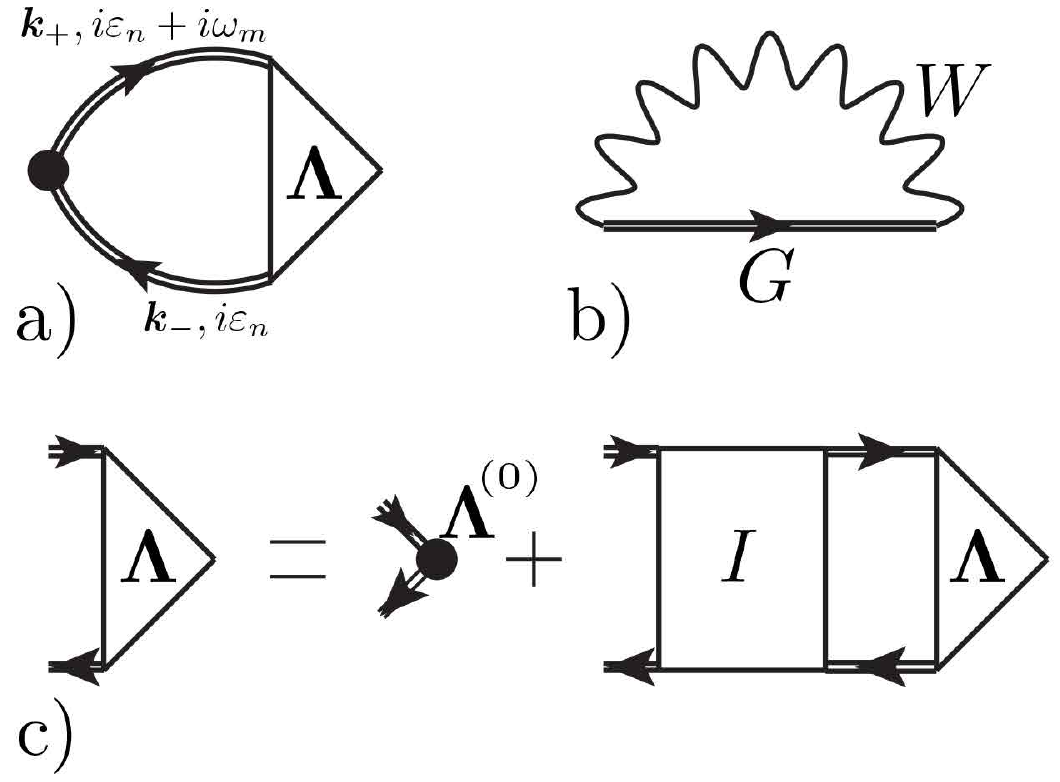}
\end{tabular}
\end{center}
\caption{
a) The diagrammatic representation of the current-current response function. The black filled circle represents the bare vertex $\Lambda^{(0,\alpha)}$ (we suppress the momentum-energy dependence for brevity), while the solid double lines are Green's functions dressed by the self-energy. In the large-$N_{\rm F}$ limit it corresponds to the GW self-energy, which is depicted in panel b). Wavy lines represent the RPA screened interaction $W$. Finally, the triangle represents the vertex function $\Lambda^{\beta}$ which is dressed by e-e interactions and satisfies the Bethe-Salpeter equation in panel c). Note that the form of the irreducible interaction $I$ is uniquely determined by the choice of the self-energy, provided that $\Lambda^\beta$ satisfies the Ward identities~\cite{Giuliani_and_Vignale} (see Fig.~\ref{fig:four}).\label{fig:three}}
\end{figure}

\begin{figure}[t]
\begin{center}
\begin{tabular}{c}
\includegraphics[width=0.99\columnwidth]{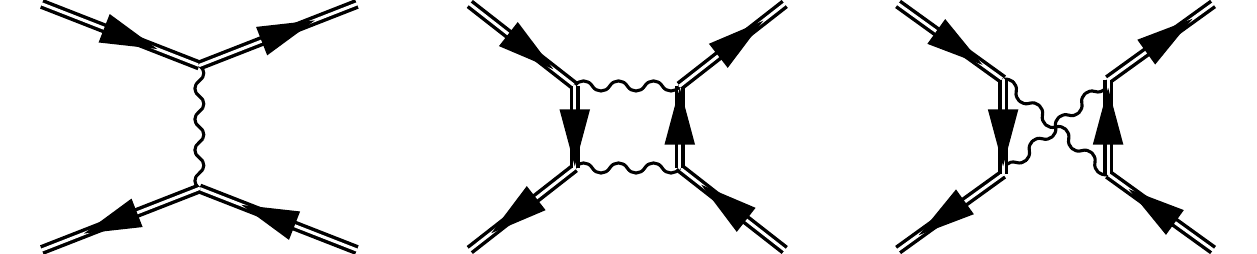}
\end{tabular}
\end{center}
\caption{Feynman diagrams that contribute to the irreducible interaction $I$ in Fig.~\ref{fig:three}.\label{fig:four}}
\end{figure}

\subsection{The viscosity transport time}
\label{sect:tau_v}
We now turn to the evaluation of the viscosity transport time $\tau_{\rm v}$, which enters Eq.~(\ref{eq:all_freq_viscosity}). This quantity can be estimated from a diagrammatic calculation of the low-frequency viscosity. Contrary to its high-frequency counterpart, the low-frequency viscosity is a non-perturbative quantity, and cannot be calculated by truncating the perturbative series of Feynman diagrams to any finite order. It is thus necessary to sum an infinite set of diagrams, to all orders in $\alpha_{\rm ee}$.  
This requires (i) that the Green's functions are dressed by self-energy insertions and (ii) that vertex corrections are included.

To bypass the tedious task of expanding the infinite series of diagrams for the current-current response function to order $q^2$, and then take the limits shown in Eq.~(\ref{eq:viscosities_def}), we use the definition of the shear viscosity given by Eq.~(\ref{eq:viscosities_stress_stress}), which we recall here:
\begin{eqnarray} \label{eq:eta_0_from_stress}
\eta_0 = -\lim_{\omega\to 0} \frac{\Im m[\chi_{xy,xy} ({\bm q}={\bm 0}, \omega)]}{\omega}
~.
\end{eqnarray}
The greatest advantage of using Eq.~(\ref{eq:eta_0_from_stress}) is that the hydrodynamic shear viscosity $\eta_0$ can be extracted from the response function $\chi_{xy,xy} ({\bm q}, \omega)$ evaluated at ${\bm q} = {\bm 0}$.
Fig.~\ref{fig:three}a) shows the general expression of the stress-stress linear response function we calculate in what follows. The double solid lines represent Green's functions dressed by the self-energy insertion shown in Fig.~\ref{fig:three}b). The wavy line represents the dynamically screened e-e interaction $W({\bm q},\omega)$. In the spirit of the large-$N_{\rm f}$ approximation, the self-energy can be approximated with its GW expression. This, in turn, implies that the screened interaction in Fig.~\ref{fig:three}b) contains the usual RPA dynamical dielectric function $\varepsilon({\bm q},\omega)$. Herein lies our large-$N_{\rm f}$ approximation.

The black filled circle on the left side of Fig.~\ref{fig:three}a) represents the bare stress-tensor vertex
\begin{eqnarray}
\Lambda_{\lambda\lambda'}^{(0,ij)}({\bm k},{\bm k}') = v_{\rm F} \frac{k_i {\cal S}^{(j)}_{\lambda\lambda'}({\bm k}_-, {\bm k}_+) + k_j {\cal S}^{(i)}_{\lambda\lambda'}({\bm k}_-,{\bm k}_+)}{2}~.
\nonumber\\
\end{eqnarray}
To account for vertex corrections, we need to dress one of the two vertices in Fig.~\ref{fig:three}a). The dressed vertex is marked as ``$\Lambda$'' and it is dictated to satisfy the self-consistent Bethe-Salpeter equation represented in Fig.~\ref{fig:three}c). The choice of the quasiparticle self-energy [Fig.~\ref{fig:three}b)] and the requirement of fulfilling the Ward identities uniquely determine the irreducible interaction $I$. The diagrams that contribute to $I$ are shown in Fig.~\ref{fig:four}.

The present calculation closely follows that reported in Ref.~\onlinecite{Principi_arxiv_2015}, the main difference being that here we have stress-tensor vertices {\it in lieu} of current vertices. Therefore, the calculation of Ref.~\onlinecite{Principi_arxiv_2015} should be adapted to the present case. These changes are explained in detail in Appendix~\ref{app:tau_v}. Here, we briefly summarize the procedure we adopted to solve the problem posed by the diagrams in Figs.~\ref{fig:three}-\ref{fig:four}.
 
As explained above, the choice of working with the stress-stress response function allows us to avoid the expansion of the Bethe-Salpeter equation for the {\it current} vertex to order $q^2$. We therefore set ${\bm q}= {\bm 0}$ from the very beginning of the present calculation. In what follows we focus on the imaginary part of the stress-stress response function, which is the quantity that controls the low-frequency viscosity [recall Eq.~(\ref{eq:eta_0_from_stress})].
Taking the low-frequency and -temperature limits, it becomes evident that only the states in a narrow region of size $\sim k_{\rm B} T$ around the Fermi surface give a significant contribution to the imaginary part of stress-stress response function. Moreover, in the limit $\omega, \tau_{\rm ee}^{-1} \ll 2 \varepsilon_{\rm F}$, whenever a product of two Green's functions with the same momentum argument appears, one of the two must be considered as retarded and the other as advanced (schematically, we get products of the form $G^{({\rm A})} G^{({\rm R})}$). Terms containing products of two retarded or two advanced Green's functions are responsible for subleading contributions in the limit $\varepsilon_{\rm F}\tau_{\rm ee} \gg 1$. Finally, in the spirit of the theory of normal Fermi liquids, each product $G^{({\rm A})} G^{({\rm R})}$ is approximated by a $\delta$-function, i.e.~the incoherent part of each Fermi-liquid Green's function is neglected and we retain only its singular part. This corresponds to the so-called ``quasiparticle approximation'', which is usually done in time-dependent perturbation theory to derive the Boltzmann transport equation from the Keldysh equation for the Green's function. Therefore, our theory describes the transport of momentum in the ``semiclassical'' regime, and neglects all quantum effects.

The set of approximations described above allows us to dramatically simplify the expression of the Bethe-Salpeter equation. The latter can be then solved {\it analytically} with the {\it Ansatz}~\cite{Yamada_PTP_1986} (suppressing all the frequency and momentum dependencies) $\Lambda = \gamma \Lambda^{(0)}$. In this way we require the dressed vertex to be proportional to the bare one. The Bethe-Salpeter equation becomes an algebraic equation, which can be solved analytically. We get
\begin{eqnarray} \label{eq:gamma_final_main}
\gamma(\omega) &=& \frac{\omega + i/\tau_{\rm ee}}{\omega + i/\tau_{\rm v}}
~.
\end{eqnarray}
The expression for $\tau_{\rm v}$ reads (see also Appendix~\ref{app:tau_v})
\begin{eqnarray} \label{eq:tau_v_final}
\frac{1}{\tau_{\rm v}}
&\simeq&
-\frac{2}{(2\pi)^2}\int_{-\infty}^{+\infty}d\omega~\frac{1- n_{\rm F}(\xi_{{\bm k},+} - \omega)}{1 - \exp(-\beta\omega)}\int_0^{+\infty}dq~q
\nonumber\\
&\times&
\left|\frac{v_q}{\varepsilon(q, \omega, T)}\right|^2\Im m[\chi^{(0)}(q,\omega,T)]A_{++}(k,q,\omega)
\nonumber\\
&\times&
4 \left[ 1 - \frac{q^2 - \omega^2/v^2_{\rm F}}{4k(k- \omega/v_{\rm F})} \right] \frac{q^2 - \omega^2/v^2_{\rm F}}{2k(k- \omega/v_{\rm F})}
~.
\end{eqnarray}
where
\begin{eqnarray}
\label{eq:final_angular_integral_MDF_intraband}
A_{++} &=& \frac{4(k - \omega/v_{\rm F})}{v_{\rm F}\sqrt{[(2 k - \omega/v_{\rm F})^2 - q^2](q^2 - \omega^2/v^2_{\rm F})}} 
\nonumber\\
&\times&
\left[ 1 - \frac{q^2 - \omega^2/v^2_{\rm F}}{4k(k- \omega/v_{\rm F})} \right] 
\nonumber\\
&\times&
\Theta\Big\{\big[(2 k - \omega/v_{\rm F})^2 - q^2\big](q^2 - \omega^2/v^2_{\rm F})\Big\}
~.
\nonumber\\
\end{eqnarray}
Eq.~(\ref{eq:tau_v_final}) describes the scattering of a quasiparticle with momentum ${\bm k} = k_{\rm F} {\hat {\bm x}}$ and energy equal to the Fermi energy into a quasiparticle with momentum $|{\bm k}+{\bm q}| = k_{\rm F}$. In this process the whole Fermi liquid is excited. The many-particle excitations created during the scattering event are encoded in the imaginary part of the density-density response function $\Im m[\chi^{(0)}_{nn}(q,\omega,T)]$.  The last line of Eq.~(\ref{eq:tau_v_final}) can be shown to be proportional to $1 - \cos^2(\varphi_{{\bm k}+{\bm q}})$ [see Appendix~\ref{app:tau_v}---to get Eq.~(\ref{eq:tau_v_final}) an integration over the angle of ${\bm q}$ has been performed]. At low temperature and in the limit of $\alpha_{\rm ee}\to 0$ Eq.~(\ref{eq:tau_v_final}) gives
\begin{eqnarray}\label{eq:low_T_approx}
\frac{1}{\tau_{\rm v}} &\to& - N_{\rm f} \frac{8 \pi}{3} 
\frac{(k_{\rm B} T)^2}{\hbar\varepsilon_{\rm F}}
\alpha_{\rm ee}^2 \ln(\alpha_{\rm ee})
~.
\end{eqnarray}

Note that the matrix element $1 - \cos^2(\varphi_{{\bm k}+{\bm q}})$ vanishes when ${\bm k}+{\bm q}$ is either parallel or antiparallel to ${\bm k}$ (recall that ${\bm k}$ is along the ${\hat {\bm x}}$ direction), and it is maximum for {\it transverse} excitations, for which ${\bm k}+{\bm q}$ is perpendicular to ${\bm k}$. Therefore it suppresses the contribution coming from the region of small transferred momenta $q\sim 0$, which is responsible for the logarithmic-in-temperatures correction to the quasiparticle lifetime. Indeed, at low temperature $1/\tau_{\rm ee} \propto - T^2 \ln(T)$, with a coefficient of proportionality which is {\it independent} of the coupling constant $\alpha_{\rm ee}$ (see Ref.~\onlinecite{polini_arxiv_2014} for more details). Therefore, because of the presence of the matrix element $1 - \cos^2(\varphi_{{\bm k}+{\bm q}})$, $1/\tau_{\rm v} \propto T^2$, and the coefficient of proportionality depends on $\alpha_{\rm ee}$.

Note also that, since the derivation of the viscosity transport time does not rely on the linear band dispersion of MDFs, a similar conclusion can be drawn for a Galilean invariant parabolic-band 2D electron gas. Also in the latter case there is no logarithmic-in-temperature correction to the low-temperature behavior of $\tau_{\rm v}$. Indeed, the matrix element suppresses both the regions $q\sim 0$ (forward scattering) and $q\sim 2 k_{\rm F}$ (perfect backscattering), which are responsible for the aforementioned correction~\cite{Giuliani_and_Vignale}.

\section{Results and Conclusions}
\label{sect:results}
%

\begin{figure}[t]
\begin{center}
\begin{tabular}{c}
\includegraphics[width=0.99\columnwidth]{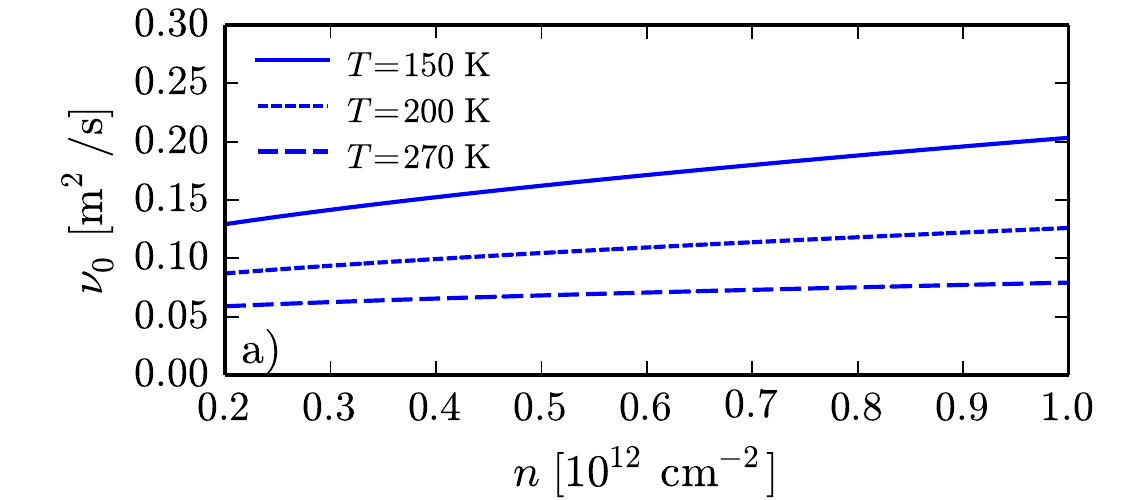}
\\
\includegraphics[width=0.99\columnwidth]{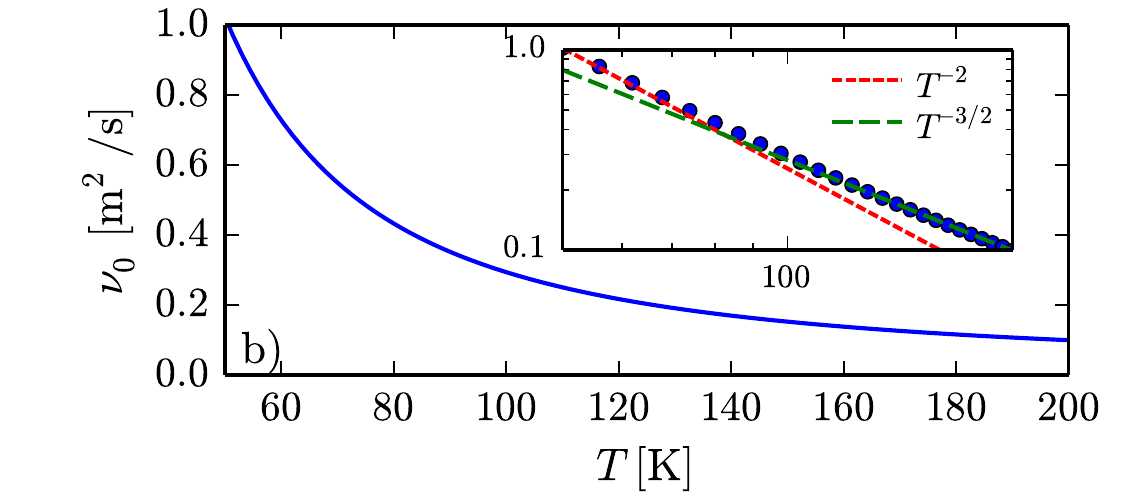}
\end{tabular}
\end{center}
\caption{(Color online) Panel a) shows the low-frequency kinematic viscosity $\nu_0 \equiv \eta_0/(n m_{\rm c})$ of a 2D MDF liquid (in units of ${\rm m}^2/{\rm s}$) as a function of the excess carrier density $n$ (in units of $10^{12}~{\rm cm}^{-2}$). The three curves refer to different values of the temperature, i.e.~$T = 150$, $200$,~and $270~{\rm K}$. In this plot $\alpha_{\rm ee}=0.5$. Note that in the range of densities shown in this plot the temperature is always smaller than $T_{\rm F} = \varepsilon_{\rm F}/k_{\rm B}$. This, in turn, implies that the low-temperature approximation ($T \ll T_{\rm F}$) for $1/\tau_{\rm v}$ in Eq.~(\ref{eq:tau_v_final}) is valid. Indeed, the minimum Fermi temperature in this plot, corresponding to $n= 0.2\times 10^{12}~{\rm cm}^{-2}$, is $T_{\rm F} \simeq 605~{\rm K}$. Therefore, in this plot $T/T_{\rm F} \lesssim 0.45$.
Panel b) The low-frequency kinematic viscosity is shown as a function of temperature $T$ (in units of ${\rm K}$) for an excess carrier density $n=0.4\times 10^{12}~{\rm cm}^{-2}$ (corresponding to $T_{\rm F} \simeq 855~{\rm K}$). Inset: a logarithmic plot of $\nu_0$ in the same range of temperatures as in the main panel. Note that $\nu_0$ grows like $T^{-2}$ for $T\ll T_{\rm F}$. A crossover to a different power law, $\sim T^{-3/2}$, is evident as temperature increases beyond the $T\ll T_{\rm F}$ regime.} \label{fig:five}
\end{figure}

In Fig.~\ref{fig:five}a) we illustrate the carrier density dependence of the low-frequency kinematic viscosity $\nu_0 \equiv \eta_0/(n m_{\rm c})$ of a 2D MDF liquid in a doped graphene sheet. We show three curves corresponding to different values of temperature $T$, i.e.~$T = 150$, $200$, and $270~{\rm K}$. Data in this plot have been obtained by setting 
$\alpha_{\rm ee}=0.5$. In Fig.~\ref{fig:five}b) we plot the same quantity but viewed as a function of temperature $T$ (in units of ${\rm K}$), for an excess carrier density $n=0.4\times 10^{12}~{\rm cm}^{-2}$. In the inset we also show a logarithmic plot of $\nu_0$, and we compare it with the power laws $T^{-2}$ and $T^{-3/2}$. While the kinematic viscosity goes as $T^{-2}$ for very small temperatures, we find that it goes as $T^{-3/2}$ for the larger temperatures shown in Fig.~\ref{fig:five}b). 
Note that the kinematic viscosity of the 2D MDF liquid in graphene is much higher than that of classical fluids~\cite{Landau_6}.

\begin{figure}[t]
\begin{center}
\begin{tabular}{c}
\includegraphics[width=0.99\columnwidth]{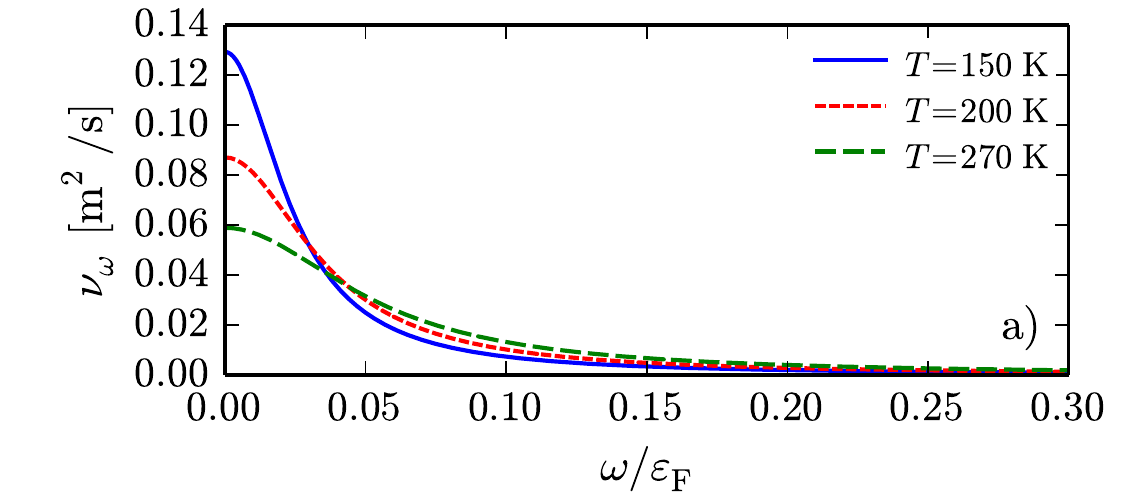}
\\
\includegraphics[width=0.99\columnwidth]{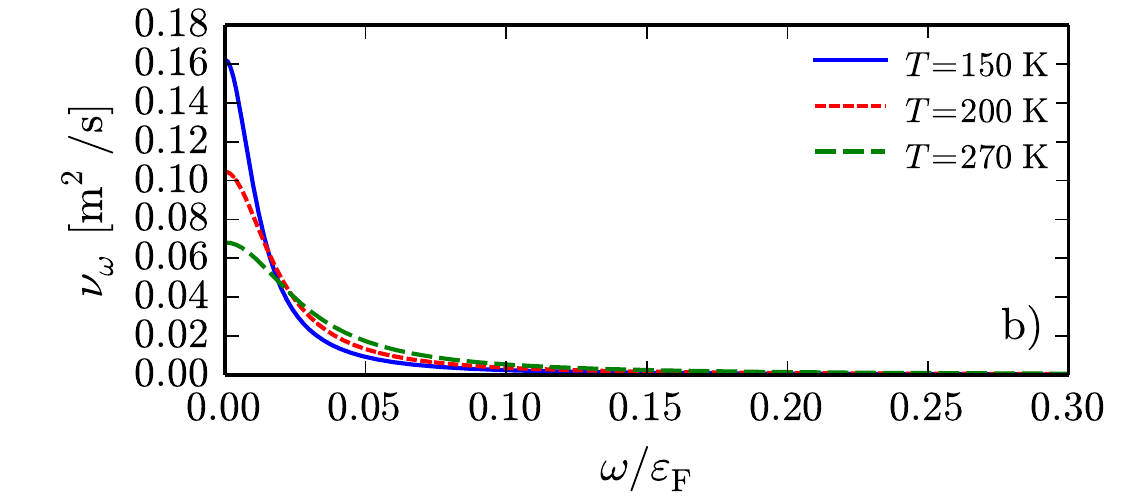}
\\
\includegraphics[width=0.99\columnwidth]{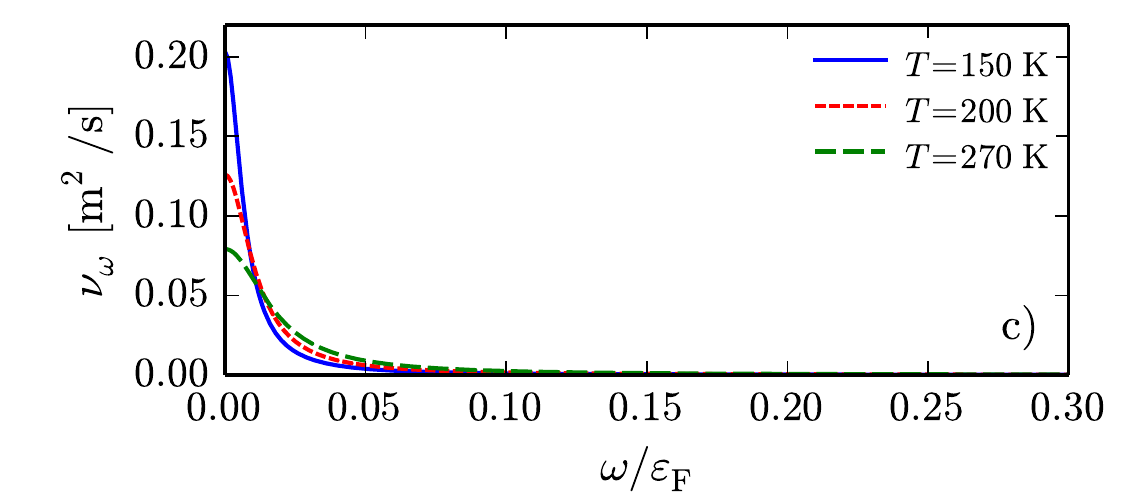}
\end{tabular}
\end{center}
\caption{(Color online) Panel a) shows the kinematic viscosity 
$\nu_\omega = \eta_\omega/(n m_{\rm c})$ (in units of ${\rm m}^2/{\rm s}$) of the 2D MDF liquid in a doped graphene sheet---Eq.~(\ref{eq:all_freq_viscosity})---as a function of $\omega$ (in units of the Fermi energy $\varepsilon_{\rm F}$). In this plot we set $n=0.2\times 10^{12}~{\rm cm}^{-2}$, $\alpha_{\rm ee}=0.5$, and we show results for three different temperatures: $T = 150$, $200$,~and $270~{\rm K}$. All the curves tend to the finite (albeit small) value at large $\omega$, whose magnitude can be extracted from Fig.~\ref{fig:two}. In this plot we subtracted the ``conductivity term'' proportional to ${\cal B}_{N_{\rm f}} (\alpha_{\rm ee})$ from the definition of $\eta_\infty$ given in Eq.~(\ref{eq:viscosities_result}).
Panel b) Same as in panel a) but for $n=0.5\times 10^{12}~{\rm cm}^{-2}$.
Panel c) Same as in panels a) and b) but for $n=1.0 \times 10^{12}~{\rm cm}^{-2}$.
\label{fig:six}}
\end{figure}

In Fig.~\ref{fig:six} we illustrate the frequency dependence of the kinematic viscosity $\nu_\omega \equiv \eta_\omega/(n m_{\rm c})$ of the 2D MDF liquid in a doped graphene sheet. The quantity $\eta_\omega$ is defined in Eq.~(\ref{eq:all_freq_viscosity}). We remind the reader that $\omega$ here represents the frequency of the external perturbation. In Fig.~\ref{fig:six}a) we fix the excess carrier density $n=0.2\times 10^{12}~{\rm cm}^{-2}$, the coupling constant $\alpha_{\rm ee}=0.5$, and we show three curves for $T = 150$, $200$,~and $270~{\rm K}$.  In Fig.~\ref{fig:six}b) and~c) we fix instead the carrier density at $n=0.5\times 10^{12}~{\rm cm}^{-2}$ and $n=10^{12}~{\rm cm}^{-2}$. Fig.~\ref{fig:six} shows that viscosity corrections to the lifetime of plasmons in the high-frequency (e.g.~mid infrared) regime are totally negligible. However, when the plasmon frequency is in the Terahertz (THz) spectral region [$\omega/\varepsilon_{\rm F} \lesssim 0.1$ in Figs.~\ref{fig:six}a)-c)] the viscosity of the electron liquid leads to corrections~\cite{Torre_prbR_2015} to the plasmon lifetime that may be comparable to those due to electron-impurity~\cite{Principi_prbR_2013} and electron-phonon scattering~\cite{Principi_prb_2014,Woessner_naturemater_2015}. Therefore, a careful comparison between measurements of the lifetime of THz plasmons in high-quality graphene samples and theoretical predictions can be used to extract the value of $\nu_\omega$.

In summary, we have calculated (i) the high-frequency bulk and shear viscosities---Eq.~(\ref{eq:viscosities_result})---and (ii) the viscosity transport time---Eq.~(\ref{eq:tau_v_final})---of the two-dimensional massless Dirac fermion liquid in a doped graphene sheet, as solely due to electron-electron interactions. As expected, the bulk viscosity vanishes. The shear viscosity is instead finite and is proportional to the excess carrier density of the doped system.  Note that, since the bulk viscosity vanishes, the high-frequency shear viscosity $\eta_\infty$ can be directly estimated by measuring the lifetime of Dirac plasmons~\cite{DiracplasmonsRPA} in high-quality encapsulated graphene sheets~\cite{Woessner_naturemater_2015}, by carrying out similar experiments in the Terahertz spectral range. Finally, from the knowledge of the high-frequency values of the shear viscosity and modulus, and of the viscosity transport time, we extracted the shear viscosity $\eta_\omega$ at all frequencies, using the interpolation formula given by Eq.~(\ref{eq:all_freq_viscosity}) and derived in Appendix~\ref{sect:GRTA}.

The low-frequency shear viscosity is equal to the high-frequency shear modulus ${\cal S}_\infty = n\varepsilon_{\rm F}/4$ multiplied by a ``viscosity transport time'' $\tau_{\rm v}$, which we microscopically calculated in this Article. We showed that $\tau_{\rm v}$ is both quantitatively {\it and} qualitatively different from the quasiparticle lifetime~\cite{polini_arxiv_2014,Li_prb_2013} $\tau_{\rm ee}$. Indeed, in the low temperature limit $1/\tau_{\rm ee} \propto -T^2 \ln(T/T_{\rm F})$, and the coefficient of proportionality is independent of the coupling constant $\alpha_{\rm ee}$ (as shown in detail in Ref.~\onlinecite{polini_arxiv_2014}). Conversely, we proved that $1/\tau_{\rm v} \propto T^2$ and that it depends on $\alpha_{\rm ee}$. 
Transverse excitations, in which a quasiparticle is scattered at $90^\circ$ with respect to its initial direction of motion, are responsible for the dominant contribution to the (shear) viscosity transport time. This is reflected by a matrix element in the expression of $1/\tau_{\rm v}$, which suppresses the contributions due to both the forward scattering and perfect backscattering of the quasiparticle ({\it i.e.} the perfectly longitudinal channels). The same matrix element is expected to show up in the expression for the viscosity transport time of a Galilean invariant (parabolic band) two-dimensional electron gas. Therefore, also in the latter case we expect $1/\tau_{\rm v} \propto T^2$ for $T\ll T_{\rm F}$.

\section{Acknowledgements}
A.P. and G.V. were supported by the BES Grant DE-FG02-05ER46203. M.C. acknowledges support by MIUR through the program ``FIRB - Futuro in Ricerca 2012" - Project HybridNanoDev (Grant No. RBFR1236VV). M.P. is supported by MIUR through the program ``Progetti Premiali 2012" - Project ABNANOTECH and by the European Community under the Graphene Flagship program (contract no. CNECT-ICT-604391). 

\appendix

\section{The equivalence of hydrodynamic and microscopic approaches in the Fermi liquid regime}
\label{app:equivalence}

In this Appendix, we demonstrate the equivalence of hydrodynamic and microscopic approaches for the 2D MDF liquid in a doped graphene sheet, in the low-temperature Fermi liquid regime. 

To this aim, let us focus on the non-interacting equation of motion for the current operator
\begin{eqnarray} \label{eq:app_EOM_current}
i \partial_t j_{{\bm q},i} &=& [{\hat {\cal H}}_0, j_{{\bm q},i}]
\nonumber\\
&=&
- v_{\rm F}^2 q_i {\hat n}_{\bm q} 
\nonumber\\
&-&
2 i v_{\rm F}^2 \sum_{{\bm k},\alpha\beta} {\hat \psi}^\dagger_{{\bm k}-{\bm q}/2,\alpha} ({\bm k}\times {\bm \sigma}_{\alpha\beta}) {\hat \psi}_{{\bm k}+{\bm q}/2,\beta}
~.
\nonumber\\
\end{eqnarray}
We remind the reader that we are interested in calculating the current-current response functions to order $q^2$, in order to take the limits of Eq.~(\ref{eq:viscosities_def}). This in turn implies that we can retain only the term of order $q$ of Eq.~(\ref{eq:app_EOM_current}), and replace the first term on its right-hand side with $q_i {\hat n}_{\bm q} \to q_i {\hat n}_{{\bm q}={\bm 0}}$. Since ${\hat n}_{{\bm q}={\bm 0}}$ is the total number of particle, it is exactly conserved by any microscopic process, and therefore does not contribute to any linear response function. Thus, we neglect it in what follows. We now use the fact that we are interested only in the states in conduction band around the Fermi surface. After a rotation of the second term on the right-hand side of Eq.~(\ref{eq:app_EOM_current}) to the chiral basis, we retain only the states that have chiral indices $\lambda=\lambda' =+$. We thus get
\begin{eqnarray} \label{eq:app_EOM_current_CB}
\partial_t j_{{\bm q},i} &=& - 2 v_{\rm F}^2 \sum_{{\bm k}} {\hat c}^\dagger_{{\bm k}-{\bm q}/2,+} ({\bm k}\times {\hat {\bm z}}) {\hat c}_{{\bm k}+{\bm q}/2,+}
\nonumber\\
&\times&
{\cal S}^{(z)}_{++}({\bm k}-{\bm q}/2,{\bm k}+{\bm q}/2)
~,
\end{eqnarray}
where ${\cal S}^{(z)}_{\lambda\lambda'} ({\bm k}, {\bm k}')$ is defined in Eq.~(\ref{eq:SM_S_z_element}). Therefore,
\begin{eqnarray} \label{eq:SM_S_z_element_pp}
{\cal S}^{(z)}_{++}({\bm k}-{\bm q}/2,{\bm k}+{\bm q}/2) &=& i \sin \left(\frac{ \theta_{{\bm k}-{\bm q}/2}-\theta_{{\bm k}+{\bm q}/2} }{2}\right)
\nonumber\\
&\to&
\frac{i}{2} \frac{{\bm q}\times {\bm k}}{k_{\rm F}^2} + {\cal O}(q^2)
~.
\end{eqnarray}
Here we used that, close to the ${\bm K}$-point of the Brillouin zone
\begin{eqnarray}
{\bm \nabla}_{\bm k} \theta_{\bm k} &=& {\bm \nabla}_{\bm k} \left[\arctan\left(\frac{k_y}{k_x}\right)\right]
\nonumber\\
&=&
\left( -\frac{k_y}{|{\bm k}|^2}, \frac{k_x}{|{\bm k}|^2} \right)
~.
\end{eqnarray}
Plugging Eq.~(\ref{eq:SM_S_z_element_pp}) into Eq.~(\ref{eq:app_EOM_current_CB}), after some simple algebra we can rewrite it to ${\cal O}(q^2)$ as
\begin{eqnarray} \label{eq:app_EOM_current_CB_2}
i \partial_t j_{{\bm q},i} &=& - \sum_{{\bm k}} {\hat c}^\dagger_{{\bm k},+} {\hat c}_{{\bm k},+}
\frac{v_{\rm F} q_j}{m_{\rm c} k_{\rm F}}
\left(
\begin{array}{cc}
k_x^2 & k_x k_y
\vspace{0.2cm}\\
k_x k_y & k_y^2
\end{array}
\right)_{ji}
\nonumber\\
&+&
v_{\rm F}^2 q_i
\sum_{{\bm k}} {\hat c}^\dagger_{{\bm k}-{\bm q}/2,+} {\hat c}_{{\bm k}+{\bm q}/2,+}
~.
\end{eqnarray}
The term on the second line is proportional to the total number of particles in the conduction band. At low frequencies, since interband processes are strongly suppressed, the number of particles in each band is conserved. Therefore, such term does not contribute to the response functions and can be neglected. On the other hand, we note that the matrix elements of the current, when projected at the Fermi surface, read ${\cal S}^{(i)}_{++}({\bm k},{\bm k}) = k_i/k_{\rm F}$. Therefore Eq.~(\ref{eq:app_EOM_current_CB_2}) can be rewritten as
\begin{eqnarray} \label{eq:app_EOM_current_CB_3}
i \partial_t j_{{\bm q},i} &=& - \sum_{{\bm k}} {\hat c}^\dagger_{{\bm k},+} {\hat c}_{{\bm k},+}
\frac{v_{\rm F} q_j}{m_{\rm c}}
\frac{k_i {\cal S}^{(j)}_{++}({\bm k},{\bm k}) + k_j {\cal S}^{(i)}_{++}({\bm k},{\bm k})}{2}
~.
\nonumber\\
\end{eqnarray}
Since (i) we always consider states around the Fermi surface, (ii) we do not allow interband transitions, and (iii) we use the equation of motion to order $q$, after a rotation back to the pseudospin basis we can rewrite Eq.~(\ref{eq:app_EOM_current_CB_3}) as 
\begin{eqnarray} \label{eq:app_EOM_current_CB_4}
i \partial_t j_{{\bm q},i} &=& - \frac{v_{\rm F} q_j}{m_{\rm c}} \sum_{{\bm k},\alpha,\beta} {\hat \psi}^\dagger_{{\bm k}-{\bm q}/2,\alpha} 
\frac{k_i \sigma^j_{\alpha\beta} + k_j \sigma^i_{\alpha\beta}}{2}
{\hat \psi}_{{\bm k}+{\bm q}/2,\beta}
~.
\nonumber\\
\end{eqnarray}
The quantity on the right-hand side is exactly the bare stress tensor operator of Eq.~(\ref{eq:non_int_stress_tensor}). Using Eq.~(\ref{eq:app_EOM_current_CB_4}) we can derive the analogous of Eq.~(\ref{eq:chi_stress_stress_current_current}) for an MDF liquid and prove the equivalence of the two approaches [although in the approximate sense explained after Eq.~(\ref{eq:app_EOM_current_CB_3})].

\section{The generalized relaxation time approximation}
\label{sect:GRTA}
The Generalized Relaxation Time Approximation (GRTA) offers a simple and powerful way to  interpolate between the collisionless and the collisional regimes of linear response functions.  This theory has its roots in Mermin's  work on the density-density response function of the electron gas~\cite{Mermin1970}, but it is applicable to a broader class of response functions and problems. The classic derivation is from the solution of the Boltzmann equation with a collision integral that describes relaxation towards a local equilibrium distribution function.

A local equilibrium distribution function  (LEDF) has the same form as the equilibrium distribution function (a function of constants of the motion) but is evaluated at the local values of the densities of the conserved quantities.  Thus, in the presence of impurity scattering, the LEDF depends on a local chemical potential and a local temperature, determined by the local particle density and energy density respectively.  Momentum is not conserved, and therefore the LEDF has zero average momentum.  This is the situation considered in Mermin's original paper.  On the other hand, if impurity scattering is absent or negligible on the time scale of particle-particle collisions, then the LEDF depends also on a local drift velocity, determined by the local momentum density.   

Regardless of whether electron-impurity collisions or electron-electron interactions dominate (we assume that it is always either one or the other) one can distinguish between (i)  a collisionless regime in which the frequency of the macroscopic motion is much higher than the collision rate, and  (ii) a collisional regime in which the frequency of the macroscopic motion is much lower than the collision rate.  

When electron-impurity collisions dominate, the  collisional regime is known as {\it diffusive regime}: there is a direct coupling between the current and the gradient of the density.  Similarly, when electron-electron collisions dominate, the  collisional regime is known as {\it hydrodynamic regime}: there is a direct connection between the stress tensor and the gradient of the velocity field.

In the diffusive regime a drift velocity can only arise from the deviation of the distribution function from the LEDF, whereas in the hydrodynamic regime a drift velocity is already present in the LEDF.   The diffusive regime is intrinsically limited  to drift velocities that are small relative to the Fermi velocity -- it is basically a linear theory of drift.  Whereas the hydrodynamic regime is suitable to describe large drift velocities, which cannot be treated in the linear approximation.

Let us formulate the general theory of the GRTA for a generic response function $\chi(q,\omega)$. We define the {\it  relaxation function} $K(q,\omega)$ as follows~\cite{Forster_book}:
\be \label{eq:GRTA_imp_chi_K_def}
\chi(q,\omega)\equiv\chi(q,0)\left[1+i\omega K(q,\omega)\right]
\ee
Since we will always be working at small $q$ ($q \ll k_F$), the $q$-dependence of $\chi(q,0)$ will be ignored: $\chi(q,0) \simeq \chi_0$. Here $\chi_0$ is (minus) the density of state of quasiparticles, and is thus renormalized by electron-electron interactions. 
The relaxation function can further be expressed in terms of  a generalized frequency- and wave vector-dependent relaxation time $T(q,\omega)$ for the quantity under consideration.  Its general structure is
\be\label{KStructure} \label{eq:GRTA_imp_K_T_def} 
K(q,\omega) = \frac{1}{-i\omega+\frac{1}{T(q,\omega)}}
\ee
This form guarantees that the finite frequency response function $\chi(q,\omega)$ vanishes if the generalized relaxation time $T$ is infinite, i.e., if the quantity under consideration is conserved. From Eqs.~(\ref{eq:GRTA_imp_chi_K_def})-(\ref{eq:GRTA_imp_K_T_def}) it is indeed immediate to get
\be\label{chi}
\chi(q,\omega)\equiv\frac{\chi_0}{1-i\omega T(q,\omega)}\,.
\ee\\
It is essential to realize that $T(q,\omega)$ is the lifetime of a collective mode (say a density fluctuation) and as such is conceptually different from the microscopic scattering times (quasiparticle lifetime and transport lifetime) upon which it may depend. 
In fact, we will see that the form of $T(q,\omega)$ depends crucially on the nature of the response function under study.

Let us now introduce the ``high-frequency" response function $\chi_\infty(q,\omega)$. The latter includes electron-electron interactions only to the extent of renormalizing the Fermi liquid parameters. Thus the quasiparticle lifetime and all the transport times, whether they be due to electron-electron, electron impurity, or electron-phonon collisions, are assumed to be infinite: we are in the collisionless regime.  The associated ``high-frequency" relaxation function is denoted by $K_\infty(q,\omega)$. We posit that the relation between  the full interacting relaxation function and the ``high-frequency"  one has the following form:
\be\label{Ansatz}
\frac{1}{K(q,\omega)}= \frac{1}{K_\infty\left(q,\omega+\frac{i}{\tau}\right)} -V_\tau(q,\omega)
\ee
where the replacement $\omega \to \omega + i/\tau$ accounts for self-energy insertions, and $V_\tau(q,\omega)$ for vertex corrections.  The two corrections $1/\tau$ and $V_\tau$ are not independent of each other, however.  They must be chosen consistently, in order to satisfy the conservation laws. In the following subsections we consider the case of the density response in both disorder and clean systems. Therefore, all the response functions that appear in what follows should be regarded as density-density ones (we suppress their indices for brevity).

\subsection{GRTA in a disordered system}
Let us first apply the theory developed so far to a system in which the dominant mechanism is the electron-impurity scattering. Since the only conserved quantity is the particle number, we now consider the relaxation function of density fluctuations.
At $q=0$ we have that the non-interacting relaxation function is $K_\infty^{-1}(0,\omega) =  -i\omega$.  Then we  simply choose $V_\tau(q,\omega)=1/\tau$ so that Eq.~(\ref{Ansatz}) yields $K^{-1}(0,\omega)=-i\omega$,
as it should be due to particle-number conservation. The inverse density-density response function, as calculated from Eq.~(\ref{eq:GRTA_imp_chi_K_def}), reads
\be\label{chiinv}
\frac{1}{\chi(q,\omega)}=\frac{1}{\chi_0} - \frac{i\omega}{\chi_0} \frac{K(q,\omega)}{1+i\omega K(q,\omega)}
\ee
Substituting in Eq.~(\ref{chiinv}) the {\it Ansatz}~(\ref{Ansatz}), we get
\be\label{chiinv2}
\frac{1}{\chi(q,\omega)}=\frac{1}{\chi_0} - \frac{i\omega}{\chi_0}\frac{K_\infty\left(q,\omega+\frac{i}{\tau}\right)}{1+i[\omega+i V_\tau(q,\omega)] K_\infty\left(q,\omega+\frac{i}{\tau}\right)}
~.
\nonumber\\
\ee
At the same time, taking the limit $\tau \to \infty$ in Eq.~(\ref{chiinv}), and replacing $\omega \to \omega + i/\tau$ we have
\be\label{chiinvinfty2}
\frac{1}{\chi_\infty\left(q,\omega+\frac{i}{\tau}\right)} &=& \frac{1}{\chi_0}
\nonumber\\
&-&
\frac{i\left(\omega+\frac{i}{\tau}\right)}{\chi_0}\frac{K_\infty\left(q,\omega+\frac{i}{\tau}\right)}{1+i\left(\omega+\frac{i}{\tau}\right) K_\infty\left(q,\omega+\frac{i}{\tau}\right)}
~.
\nonumber\\
\ee
Combining Eqs.~(\ref{chiinv2}) and~(\ref{chiinvinfty2}) we finally get
\be\label{chiinv2b}
&&
\frac{1}{\chi(q,\omega)}=\frac{1}{\chi_0} + \frac{\omega}{\left(\omega+\frac{i}{\tau}\right)\chi_0}\left(
\frac{\chi_0}{\chi_\infty\left(q,\omega+\frac{i}{\tau}\right)} -1\right)
\nonumber\\
&& \times
\left[1-
\frac{i\left[V_\tau(q,\omega) - \frac{1}{\tau}\right]}{\left(\omega+\frac{i}{\tau}\right)}\left(
\frac{\chi_0}{\chi_\infty\left(q,\omega+\frac{i}{\tau}\right)} -1\right)
\right]^{-1}
~.
\nonumber\\
\ee
Setting now $V_\tau(q,\omega)=1/\tau$ we get
\be\label{MERMIN}
\frac{1}{\chi(q,\omega)}
=\frac{\frac{i}{\tau}}{\left(\omega+\frac{i}{\tau}\right)}\frac{1}{\chi_0} +\frac{\omega}{\left(\omega+\frac{i}{\tau}\right)}\frac{1}{\chi_\infty\left(q,\omega+\frac{i}{\tau}\right)}
~,
\nonumber\\
\ee
which is precisely the Mermin's formula for the density-density response function~\cite{Mermin1970}.
In a Galilean-invariant electron gas in the diffusive regime
\be\label{DiffusiveResponse}
\frac{1}{\chi(q,\omega)}\simeq \frac{1}{\chi_0}\left(1- \frac{i\omega}{D_\tau q^2}\right)\,,
\ee
where $D_\tau = - {\cal D}_{\rm w} \tau/(\chi_0)$ is the diffusion constant, ${\cal D}_{\rm w}$ is the renormalized Drude weight ($n/m$ for the two-dimensional electron gas or $k_F v_F/\hbar$ for massless Dirac fermions), and we have used the $\omega \to 0$ diffusive form of the density-density response function $\chi_\infty(q,\omega+i/\tau) \simeq -{\cal D}_{\rm w}q^2\tau^2$.

However, it is well known that the relaxation time appearing in the diffusion constant, commonly denoted by $\tau_{tr}$,  is different from the self-energy relaxation time $\tau$ (quasiparticle lifetime).   The conservation of particle number, as implemented in the $q=0$ vertex correction, is not sufficient to produce the correct transport relaxation time.  
To capture the difference between $\tau$ and $\tau_{tr}$  we must allow for the $q$-dependence of the vertex correction in our Ansatz~(\ref{Ansatz}). Namely, we set
\be\label{NewChoice}
V_\tau(q,\omega)=\frac{1}{\tau} +\tilde D_\tau q^2\,.
\ee 
The coefficient $\tilde D_\tau$ has the dimensions of a diffusion constant and its value is  determined by the requirement that the final formula for $\chi(q,\omega)$ has a diffusive limit with the correct relaxation time, i.e., $\tau_{tr}$.
With the choice~(\ref{NewChoice}) for the vertex correction our  Eq.~(\ref{chiinv2b}) gives
\be\label{chiinv4}
\frac{1}{\chi(q,\omega)}
&\simeq&\frac{1}{\chi_0} - \frac{i\omega}{\chi_0} \frac{1}{(D_\tau-\tilde D_\tau)q^2}
~.
\ee
It is evident from this result that we obtain the correct  diffusive limit if and only if
\be
\tilde D_\tau = D_\tau\left(1-\frac{\tau_{\rm tr}}{\tau}\right)
\ee
Thus, the coefficient $\tilde D_\tau$ in the vertex correction is simply the difference between the ``naive" diffusion constant and the ``true" diffusion constant, in which $\tau$ is replaced by the transport lifetime $\tau_{\rm tr}$.

\subsection{GRTA in a clean system}
We now discuss the case of a clean Galilean-invariant system in the so-called ``hydrodynamic regime'', in which the main mechanism of quasiparticle relaxation is given by electron-electron interactions.
Replacing $\tau\to\tau_{\rm ee}$, from Eqs.~(\ref{MERMIN}) we get the ``naive" diffusion constant
\be\label{DNaive}
D_\tau = \frac{{\cal D}_{\rm w}}{\chi_0} \frac{1}{i\left(\omega+\frac{i}{\tau_{\rm ee}}\right)}
~.
\ee
However, since electron-electron collisions conserve the total momentum the {\it true} diffusion constant must be divergent in the limit of $\omega \to 0$, {\it i.e.} it must be
\be
D_{\rm tr} = \frac{{\cal D}_{\rm w}}{\chi_0} \frac{1}{i\omega}\,,
\ee
Therefore the role of $\tilde D_\tau$ is played by
\be\label{Dtilde}
\tilde D_\tau = D_\tau-D_{\rm tr} = -\frac{{\cal D}_{\rm w}}{\chi_0 \tau_{\rm ee}}\frac{1}{\omega\left(\omega+\frac{i}{\tau_{\rm ee}}\right)}\,.
\ee
Substituting this  in Eq.~(\ref{NewChoice}) for the $q$-dependent vertex correction, and then putting this vertex correction in Eq.~(\ref{chiinv2b}) we finally get
\be \label{eq:invchi4-ee}
&&
\frac{1}{\chi(q,\omega)} = \frac{1}{\chi_0} 
+
\frac{\omega}{\left(\omega+\frac{i}{\tau_{\rm ee}}\right)}\left(
\frac{1}{\chi_\infty\left(q,\omega+\frac{i}{\tau_{\rm ee}}\right)} -\frac{1}{\chi_0}\right)
\nonumber\\
&& \times
\left\{1+
\frac{i{\cal D}_{\rm w} q^2/(\omega\tau_{\rm ee})}{\left(\omega+\frac{i}{\tau_{\rm ee}}\right)^2}
\left[\frac{1}{\chi_\infty\left(q,\omega+\frac{i}{\tau_{\rm ee}}\right)} - \frac{1}{\chi_0}\right]
\right\}^{-1}
~,
\nonumber\\
\ee
We now introduce $1/{\bar \chi}_\infty(q,\omega)$ to be the $q^0$ coefficient of the expansion of $1/\chi_\infty(q,\omega)$, {\it i.e.}
\begin{eqnarray}
\frac{1}{\chi_\infty(q,\omega)} = \frac{\omega^2}{{\cal D}_{\rm w} q^2} + \frac{1}{{\bar \chi}_\infty(q,\omega)}
~.
\end{eqnarray}
This allows us to rewrite Eq.~(\ref{eq:invchi4-ee}) as
\be \label{eq:invchi4-ee_R}
&&
\frac{1}{\chi(q,\omega)} = \frac{1}{\chi_0} 
+
\left(\frac{\omega}{\omega+\frac{i}{\tau_{\rm ee}}}\right)^2 \left(
1 - \frac{\chi_\infty\left(q,\omega+\frac{i}{\tau_{\rm ee}}\right)}{\chi_0} \right)
\nonumber\\
&&
\times
\left(
\frac{1}{\chi_\infty\left(q,\omega+\frac{i}{\tau_{\rm ee}}\right)} - \frac{1}{{\bar \chi}_\infty\left(q,\omega+\frac{i}{\tau_{\rm ee}}\right)}\right)
\nonumber\\
&& \times
\left\{1-
\frac{\omega}{\omega + \frac{i}{\tau_{\rm ee}}}
\left[1+ \frac{i}{\omega\tau_{\rm ee}\chi_0}\chi_\infty\left(q,\omega+\frac{i}{\tau_{\rm ee}}\right)\right]
\right\}^{-1}
~,
\nonumber\\
\ee
Expanding Eq.~(\ref{eq:invchi4-ee_R}) for $q \ll \omega, 1/\tau_{\rm ee} \ll \varepsilon_{\rm F}$, and retaining all the terms up to the order $q^0$, we get
\be\label{chiinv4-final}
\frac{1}{\chi(q,\omega)}&=&\frac{1}{\chi_0} +\left(\frac{\omega}{\omega+\frac{i}{\tau_{\rm ee}}}\right)^2 \left[\frac{1}{\chi_\infty\left(q,\omega+\frac{i}{\tau_{\rm ee}}\right)}-\frac{1}{\chi_0}\right]
\nonumber\\
&-& \left(\frac{\omega}{\omega+\frac{i}{\tau_{\rm ee}}}\right)^2 \frac{1}{{\bar \chi}_\infty\left(q,\omega+\frac{i}{\tau_{\rm ee}}\right)}
\nonumber\\
&+&
\left(\frac{\omega}{\omega+\frac{i}{\tau_{\rm ee}}}\right)^3 \left[\frac{1}{{\bar \chi}_\infty\left(q,\omega+\frac{i}{\tau_{\rm ee}}\right)}+ \frac{i}{\omega \tau_{\rm ee} \chi_0}\right]
~.
\nonumber\\
\ee

To make a clearer connection with the solution of the Boltzmann equation in the relaxation time approximation we rewrite
\be
\left(\frac{\omega}{\omega+\frac{i}{\tau_{\rm ee}}}\right)^2 &=&\frac{\omega}{\omega+\frac{i}{\tau_{\rm ee}}}
-\frac{i \omega/\tau_{\rm ee}}{\left(\omega+\frac{i}{\tau_{\rm ee}}\right)^2}\nonumber\\
&=&\frac{\omega}{\omega+\frac{i}{\tau_{\rm ee}}}
-\frac{i \omega}{\tau_{\rm ee} {\cal D}_{\rm w} q^2} \frac{{\cal D}_{\rm w} q^2}{(\omega+i/\tau_{\rm ee})^2}
~.
\nonumber\\
\ee
The first term reproduces the Mermin formula, and the second term is the correction required to ensure momentum conservation, {\it i.e.}
\be\label{ChiFinal}
&&
\frac{1}{\chi(q,\omega)}
= \frac{\frac{i}{\tau_{\rm ee}}}{\left(\omega+\frac{i}{\tau_{\rm ee}}\right)}\frac{1}{\chi_0} +\frac{\omega}{\left(\omega+\frac{i}{\tau_{\rm ee}}\right)}\frac{1}{\chi_\infty\left(q,\omega+\frac{i}{\tau_{\rm ee}}\right)} 
\nonumber\\
&&-
i \frac{\omega}{\tau_{\rm ee} {\cal D}_{\rm w} q^2} 
\nonumber\\
&&- \left(\frac{\omega}{\omega+\frac{i}{\tau_{\rm ee}}}\right)^2 \left[\frac{\omega+\frac{i}{\tau_{\rm ee}}}{\omega} \frac{1}{{\bar \chi}_\infty\left(q,\omega+\frac{i}{\tau_{\rm ee}}\right)} - \frac{i}{\omega \tau_{\rm ee} \chi_0}\right]
\nonumber\\
&&+
\left(\frac{\omega}{\omega+\frac{i}{\tau_{\rm ee}}}\right)^3 \left[\frac{1}{{\bar \chi}_\infty\left(q,\omega+\frac{i}{\tau_{\rm ee}}\right)}+ \frac{i}{\omega \tau_{\rm ee} \chi_0}\right]
~.
\nonumber\\
\ee

To extract the hydrodynamic coefficients from Eq.~(\ref{ChiFinal}) we now observe that  
\be\label{ChiInfinity}
\frac{1}{\chi_\infty(q,\omega)} \simeq \frac{\omega^2}{{\cal D}_{\rm w}q^2}-\left[\frac{{\cal B}_\infty}{n^2}+\left(2-\frac{2}{d}\right)\frac{{\tilde {\cal S}}_\infty(\omega)}{n^2}   \right]
~,
\nonumber\\
\ee
where the bulk modulus ${\cal B}_\infty$ is real, frequency-independent, and related to $\chi_0$ by the well-known compressibility sum rule
\be\label{Chi0}
\chi_0=-\frac{n^2}{{\cal B}_\infty}
~,
\ee
while ${\tilde {\cal S}}_\infty(\omega)= {\cal S}_\infty-i \omega \eta_\infty$ is the complex shear modulus.
Plugging Eqs.~(\ref{ChiInfinity}) and (\ref{Chi0}) into Eq.~(\ref{ChiFinal}) we get, after some simple algebra,
\be\label{ChiFinalHydro}
&& \!\!\!\!\!\!\!\!
\frac{1}{\chi(q,\omega)}
\simeq  \frac{\omega^2}{{\cal D}_{\rm w}q^2} 
\nonumber\\
&-&
\left[\frac{{\cal B}_\infty}{n^2} +\left(2-\frac{2}{d}\right)\left(\frac{\omega}{\omega+\frac{i}{\tau_{\rm ee}}} \right)^3
\frac{{\tilde {\cal S}}_\infty(\omega+i/\tau_{\rm ee})}{n^2}\right]
~.
\nonumber\\
\ee  
These result does not matches with what we found in the main text of the paper. Indeed, the correct result should be
\be\label{ChiFinalHydro_correct}
\frac{1}{\chi(q,\omega)}
&\simeq&  \frac{\omega^2}{{\cal D}_{\rm w}q^2} 
\nonumber\\
&-&
\left[\frac{{\cal B}_\infty}{n^2} +\left(2-\frac{2}{d}\right)\frac{\omega}{\omega+i/\tau_{\rm v}}
\frac{{\tilde {\cal S}}_\infty(\omega)}{n^2}\right]
~.
\nonumber\\
\ee  
To get the correct result it is necessary to take into account the corrections to order $q^4$ to the vertex correction $V_\tau(q,\omega)$. We therefore set
\begin{eqnarray} \label{eq:vertex_q4}
V_\tau(q,\omega)=\frac{1}{\tau} +\tilde D_\tau q^2 + \frac{i \Xi (q,\omega)}{\chi_0} \frac{D_{\rm w}^2 q^4}{(\omega+i/\tau_{\rm ee})^3}
~,
\end{eqnarray}
where $\Xi(q,\omega)$ is to be determined. Plugging Eq.~(\ref{eq:vertex_q4}) into Eq.~(\ref{chiinv2b}) and expanding it up to order $q^0$, after some straightforward algebra we get
\be\label{ChiFinalHydro_vertex}
&& \!\!\!\!\!\!\!\!
\frac{1}{\chi(q,\omega)}
\simeq  \frac{\omega^2}{{\cal D}_{\rm w}q^2} 
\nonumber\\
&-&
\left[\frac{{\cal B}_\infty}{n^2} +\left(2-\frac{2}{d}\right)\left(\frac{\omega}{\omega+\frac{i}{\tau_{\rm ee}}} \right)^3
\frac{{\tilde {\cal S}}_\infty(\omega+i/\tau_{\rm ee})}{n^2}\right]
\nonumber\\
&+&
\left(\frac{\omega}{\omega+\frac{i}{\tau_{\rm ee}}} \right)^3 \Xi(q,\omega)
~.
\ee  
It is clear that it is possible to choose $\Xi(q,\omega)$ in such a way as to match Eq.~(\ref{ChiFinalHydro_vertex}) with Eq.~(\ref{ChiFinalHydro_correct}).

\section{Calculation of ${\hat F}_1$ and \texorpdfstring{${\hat {\bm j}}_{1,{\bm q}}$}{j_1}}
\label{sect:SM_calculation_F1_j1}
The operator ${\hat F}_1$ was calculated in Ref.~\onlinecite{Principi_prb_2013} and reads
\begin{eqnarray}
i {\hat F}_1 &\equiv& \frac{1}{2}\sum_{{\bm q}'} v_{{\bm q}'} \sum_{{\bm k},{\bm k}'} \sum_{\lambda,\lambda',\mu,\mu'} {\cal M}_{\lambda,\lambda',\mu,\mu'}({\bm k},{\bm k}',{\bm q}')
\nonumber\\
&\times&
c^\dagger_{{\bm k}_-,\lambda} c_{{\bm k}_+,\lambda'} c^\dagger_{{\bm k}'_+,\mu} c_{{\bm k}'_-,\mu'}
~.
\end{eqnarray}
where
\begin{equation} \label{eq:SM_M_def}
{\cal M}_{\lambda,\lambda',\mu,\mu'}({\bm k},{\bm k}',{\bm q}') = \frac{{\cal D}_{\lambda\lambda'}({\bm k}_-,{\bm k}_+) {\cal D}_{\mu\mu'}({\bm k}'_+,{\bm k}'_-)}{\varepsilon_{{\bm k}_-,\lambda} - \varepsilon_{{\bm k}_+,\lambda'} + \varepsilon_{{\bm k}'_+,\mu} - \varepsilon_{{\bm k}'_-,\mu'}}
~.
\end{equation}
To simplify the notation we have introduced ${\bm k}_\pm \equiv {\bm k} \pm {\bm q}'/2$ and ${\bm k}'_\pm \equiv {\bm k}' \pm {\bm q}'/2$.

The transverse component of the operator ${\hat {\bm j}}_{1,{\bm q}}$ is obtained from its definition of Eq.~(\ref{eq:current_first_order}). After some manipulations, along the lines of Ref.~\onlinecite{Principi_prb_2013}, we get
\begin{eqnarray} \label{eq:SM_j_1_element}
{\hat j}_{1,{\bm q},{\rm T}} =  \frac{1}{2}\sum_{{\bm q}'} v_{{\bm q}'}
\left[{\hat \Upsilon}^{({\rm T})}_{{\bm q}, {\bm q}'} {\hat n}_{-{\bm q}'} + {\hat n}_{{\bm q}'} {\hat \Upsilon}^{({\rm T})}_{{\bm q}, - {\bm q}'} \right]~,
\end{eqnarray}
where
\begin{eqnarray} \label{eq:SM_Upsilon_def}
{\hat \Upsilon}^{({\rm T})}_{{\bm q},{\bm q}'} = \sum_{{\bm k},\lambda,\lambda'} M_{\lambda,\lambda'} ({\bm k},{\bm q}',{\bm q}){\hat c}^\dagger_{{\bm k}_--{\bm q}/2,\lambda} {\hat c}_{{\bm k}_++{\bm q}/2,\lambda'} 
~,
\nonumber\\
\end{eqnarray}
with
\begin{eqnarray} \label{eq:SM_MatrEl_def}
&& \!\!\!\!\!\!\!\!\!\!
M_{\lambda,\lambda'} ({\bm k},{\bm q}',{\bm q})
\equiv
\nonumber\\
&& \!\!\!\!\!\!\!\!\!\!
\sum_{\rho} \!\!
\left[
\frac{\displaystyle {\cal D}_{\lambda\rho} \left({\bm k}_--\frac{{\bm q}}{2},{\bm k}_+-\frac{{\bm q}}{2}\right) {\cal S}^{(x)}_{\rho\lambda'}\left({\bm k}_+ -\frac{{\bm q}}{2}, {\bm k}_+ +\frac{{\bm q}}{2}\right)}{\omega + \varepsilon_{{\bm k}_++{\bm q}/2,\lambda'} - \varepsilon_{{\bm k}_+-{\bm q}/2,\rho}}
\right.
\nonumber\\
\!\! &-& \!\!
\left.
\frac{\displaystyle {\cal S}^{(x)}_{\lambda\rho}\left({\bm k}_- -\frac{{\bm q}}{2}, {\bm k}_- +\frac{{\bm q}}{2}\right) {\cal D}_{\rho\lambda'} \left({\bm k}_-+\frac{{\bm q}}{2},{\bm k}_++\frac{{\bm q}}{2}\right)}{\omega + \varepsilon_{{\bm k}_- +{\bm q}/2,\rho} - \varepsilon_{{\bm k}_--{\bm q}/2,\lambda}}
\right]
.
\nonumber\\
\end{eqnarray}
We stress that Eqs.~(\ref{eq:SM_j_1_element})-(\ref{eq:SM_MatrEl_def}) are valid only for their use in Eq.~(\ref{eq:SM_chi_jj_def_second}), and become exact in the low-energy MDF continuum limit, which will be taken momentarily. The same limit restores rotational invariance: we therefore have the freedom of fixing the direction of the wave vector ${\bm q}$ arbitrarily. In deriving these equations we have taken, without any loss of generality, ${\bm q} = q {\hat {\bm y}}$ which implies that ${\hat j}_{1,{\bm q},{\rm T}} = {\hat j}_{1,{\bm q},x}$.

Following the same steps outlined in Ref.~\onlinecite{Principi_prb_2013} it is possible to show that the operator in Eq.~(\ref{eq:SM_Upsilon_def}) can be written as (the details of this derivation are given in App.~\ref{sect:SM_Upsilon_manipulation})
\begin{eqnarray} \label{eq:SM_Upsilon_approx_3}
{\hat \Upsilon}^{({\rm T})}_{{\bm q}, {\bm q}'} 
\!\! &=& \!\!
\sum_{\alpha = x,y}
\left\{
\frac{v_{\rm F} q}{\omega^2 k_{\rm F}} 
\left[
\frac{q'_x q'_y}{q'^2} q'_\alpha - 
\left(1 - \frac{q'^2}{4 k_{\rm F}^2} \right)
\right. \right.
\nonumber\\
&\times&
\left. \left.
(q'_x \delta_{\alpha,y} + q'_y \delta_{\alpha, y})
\right]
+
\frac{q'^2}{4 v_{\rm F} k_{\rm F}^3} \delta_{\alpha,x} 
\right\}
{\hat j}_{{\bm q}',\alpha}
\nonumber\\
&\equiv&
\sum_{\alpha  = x,y} \Gamma^{({\rm T})}_\alpha({\bm q},{\bm q}') {\hat j}_{{\bm q}',\alpha}
~.
\end{eqnarray}
The main differences between Eq.~(\ref{eq:SM_Upsilon_approx_3}) and the corresponding expression that can be found for a 2DEG~\cite{Nifosi_prb_1998} are: (i) the factor $1-q'^2/(4 k_{\rm F}^2)$, which is due to the chirality of the MDF eigenstates and suppresses backscattering at the Fermi surface~\cite{Principi_prb_2013}, and (ii) the last term in curly brackets, which is finite even in the long-wavelength ${\bm q} \to {\bm 0}$ limit~\cite{Principi_prb_2013}. 

For a sake of completeness we recall~\cite{Principi_prb_2013} that the first-order correction to the {\it longitudinal} current operator is formally identical to Eq.~(\ref{eq:SM_j_1_element}), with the longitudinal counterpart [${\hat \Upsilon}^{({\rm L})}_{{\bm q}, {\bm q}'}$] of the operator ${\hat \Upsilon}^{({\rm T})}_{{\bm q}, {\bm q}'}$ taking its place. We recall that ${\hat \Upsilon}^{({\rm L})}_{{\bm q}, {\bm q}'}$ is also proportional to the untransformed current operator, {\it i.e.} ${\hat \Upsilon}^{({\rm L})}_{{\bm q}, {\bm q}'} = \sum_{\alpha  = x,y} \Gamma_\alpha^{({\rm L})}({\bm q},{\bm q}') {\hat j}_{{\bm q}',\alpha}$, where
\begin{eqnarray} \label{eq:longitudinal_Gamma}
\Gamma_\alpha^{({\rm L})}({\bm q},{\bm q}')
\!\! &=& \!\!
\frac{v_{\rm F} q_x}{\omega^2} \Bigg[ \frac{q_y'^2}{q'^2} \frac{q'_\alpha}{k_{\rm F}}
-2 \frac{q'_x}{k_{\rm F}}
\Bigg( 1 - \frac{q'^2}{4 k_{\rm F}^2} \!\Bigg)
\delta_{\alpha,x} 
\Bigg]
\nonumber\\
&+&
\frac{q'^2}{4 v_{\rm F} k_{\rm F}^3} \delta_{\alpha,x} 
~.
\end{eqnarray}
After the change of variables ${\bm q}' \to -{\bm q}'$ in the second term on the right-hand side of Eq.~(\ref{eq:SM_j_1_element}), the latter can be rewritten as ${\bm q}\cdot{\hat {\bm j}}_{1,{\bm q}} = \sum_{{\bm q}'} v_{{\bm q}'} {\hat \Upsilon}^{({\rm T})}_{{\bm q}, {\bm q}'} {\hat n}_{-{\bm q}'}$. A major simplification is suggested by the analysis of the Feynman graphs contributing to the non-interacting spectrum of ${\hat {\bm q}}\cdot {\hat {\bm j}}_{1,{\bm q}}$. As shown in Ref.~\onlinecite{Principi_prb_2013} the disconnected graphs contain {\it two} independent sums over the number $N_{\rm f}$ of fermion flavors, whereas the connected ones contain only one such sum. We conclude that the disconnected graphs dominate in the large-$N_{\rm f}$ limit. The final formula for the two components (longitudinal and transverse) of the current-current response function, which is exact to second order in e-e interactions and in the large-$N_{\rm f}$ limit, is
\begin{widetext}
\begin{eqnarray} \label{eq:chi_rhorho_mode_decoupling_def_app}
\Im m[\chi_{\ell}(q,\omega)]&=&
- \sum_{\alpha,\beta = x,y}\int \frac{d^2{\bm q}'}{(2\pi)^2} v_{{\bm q}'}^2
\int_0^\omega \frac{d\omega'}{\pi} \Big\{ \Gamma_\alpha^{(\ell)}({\bm q},{\bm q}') \Gamma_{\beta}^{(\ell)}(-{\bm q},-{\bm q}')
\Im m[\chi^{(0)}_{nn}(q',\omega')]\Im m[\chi^{(0)}_{j_\alpha j_\beta}({\bm q}',\omega-\omega')]
\nonumber\\
&+&
\Gamma_\alpha^{(\ell)}({\bm q},{\bm q}') \Gamma_\beta^{(\ell)}(-{\bm q},{\bm q}')
\Im m[\chi^{(0)}_{n j_\alpha}(-{\bm q}',\omega')]~\Im m[\chi^{(0)}_{n j_\beta}({\bm q}',\omega-\omega')]\Big\}
~,
\end{eqnarray}
\end{widetext}
In this equation $\chi^{(0)}_{nn}(q,\omega)$, $\chi^{(0)}_{j_\alpha j_\beta}({\bm q},\omega)$, and $\chi^{(0)}_{n j_\alpha}({\bm q},\omega)$ are the {\it non-interacting} density-density, current-current, and density-current response functions of a 2D gas of MDFs. The quantities $\{\Gamma_\alpha^{(\ell)}({\bm q},{\bm q}'); \alpha = x,y; \ell={\rm L},{\rm T}\}$ are defined in Eq.~(\ref{eq:SM_Upsilon_approx_3}) and~(\ref{eq:longitudinal_Gamma}). We stress that the {\it imaginary} parts of the three linear-response functions $\chi^{(0)}_{nn}(q,\omega)$, $\chi^{(0)}_{j_\alpha j_\beta}({\bm q},\omega)$, and $\chi^{(0)}_{n j_\alpha}({\bm q},\omega)$ do not depend on any ultraviolet cut-off in the low-energy MDF limit. Moreover, since in the limit of $\omega\to 0$ the integral over $q'$ is naturally restricted to $0\leq q' \leq 2k_{\rm F}$, no ultraviolet regularization is needed in Eq.~(\ref{eq:chi_rhorho_mode_decoupling_def_app}). The only pathology of the integral in Eq.~(\ref{eq:chi_rhorho_mode_decoupling_def_app}) appears in the infrared $q' \to 0$ limit, due to the $1/q'$ singularity of the Coulomb potential $v_{{\bm q}'}$. This problem is cured by screening, as discussed in the main text.

\section{Details of the manipulation of \texorpdfstring{${\hat \Upsilon}^{({\rm T})}_{{\bm q},{\bm q}'}$}{Upsilon}}
\label{sect:SM_Upsilon_manipulation}
In this appendix we approximate the expression of the operator ${\hat \Upsilon}^{({\rm T})}_{{\bm q},{\bm q}'}$ in Eqs.~(\ref{eq:SM_Upsilon_def})-(\ref{eq:SM_MatrEl_def}) by taking the limit $v_{\rm F} q \ll \omega \ll 2\varepsilon_{\rm F}$. We will try to slowly guide the reader through the many steps of this lengthy process.

To begin with, in the long-wavelength $q \to 0$ limit we can write
\begin{eqnarray} \label{first_term_denominator}
\frac{1}{\omega + \varepsilon_{{\bm k}_\pm+{\bm q}/2,\lambda} - \varepsilon_{{\bm k}_\pm-{\bm q}/2,\rho}} 
&\to&
\delta_{\lambda,\rho} \left[ \frac{1}{\omega} - \frac{q}{\omega^2} \frac{\partial \varepsilon_{{\bm k}_\pm}}{\partial k_y} \right] 
\nonumber\\
&+&
(1-\delta_{\lambda,\rho}) \frac{1}{\omega + 2\lambda\varepsilon_{\rm F}}
\nonumber\\
&+&
{\cal O}(q^2)~.
\end{eqnarray}

We then observe that in the regime of interest in this Paper, {\it i.e.} $v_{\rm F} q \ll \omega \ll 2\varepsilon_{\rm F}$, the particle-hole states created by the operator ${\hat \Upsilon}^{({\rm T})}_{{\bm q},{\bm q}'}$ are energetically close to the Fermi energy. The band indices $\lambda, \lambda'$ on the right-hand side of Eq.~(\ref{eq:SM_MatrEl_def}) are therefore constrained to take the values $\lambda = \lambda' = + 1$ (recall that $\varepsilon_{\rm F} > 0$). 

Note that the ``virtual state'' $\rho$, over which the sum on the right-hand side of Eq.~(\ref{eq:SM_MatrEl_def}) runs, can be either in conduction ($\rho = + 1$) or valence ($\rho = -1$) band, even though the states labeled by the band indices $\lambda$ and $\lambda'$ are bound to the Fermi surface. 

We first simplify Eqs.~(\ref{eq:SM_Upsilon_def})-(\ref{eq:SM_MatrEl_def}) by using Eq.~(\ref{first_term_denominator}). We are naturally led to define
\begin{widetext}
\begin{eqnarray} \label{eq:SM_M_intra_def}
M_{\rm intra}({\bm k}, {\bm q}', {\bm q}) &\equiv&\left.M_{++}({\bm k}, {\bm q}', {\bm q})\right|_{\rho = +1} = 
\cos\left(\frac{\theta_{{\bm k}_--{\bm q}/2} - \theta_{{\bm k}_+-{\bm q}/2}}{2}\right) \cos(\theta_{{\bm k}_+})
\left[ \frac{1}{\omega} - \frac{v_{\rm F} q}{\omega^2} \sin(\theta_{{\bm k}_+})\right]
\nonumber\\
&-&
\cos\left(\frac{\theta_{{\bm k}_-+{\bm q}/2} - \theta_{{\bm k}_++{\bm q}/2}}{2}\right) \cos(\theta_{{\bm k}_-})
\left[ \frac{1}{\omega} - \frac{v_{\rm F} q}{\omega^2} \sin(\theta_{{\bm k}_-}) \right]
+ {\cal O}(q^2)
~,
\end{eqnarray}
and
\begin{eqnarray} \label{eq:SM_M_inter_def}
M_{\rm inter}({\bm k}, {\bm q}', {\bm q}) &\equiv&\left.M_{++}({\bm k}, {\bm q}', {\bm q})\right|_{\rho = -1} = -\frac{1}{2\varepsilon_{\rm F}} \sin\left(\frac{\theta_{{\bm k}_--{\bm q}/2} - \theta_{{\bm k}_+-{\bm q}/2}}{2}\right) \sin\left(\frac{\theta_{{\bm k}_+ -{\bm q}/2} + \theta_{{\bm k}_+ +{\bm q}/2}}{2}\right)
\nonumber\\
&+&
\frac{1}{2\varepsilon_{\rm F}} \sin\left(\frac{\theta_{{\bm k}_-+{\bm q}/2} - \theta_{{\bm k}_++{\bm q}/2}}{2}\right) \sin\left(\frac{\theta_{{\bm k}_- -{\bm q}/2} + \theta_{{\bm k}_- +{\bm q}/2}}{2}\right)
 + {\cal O}(q^2)~.
\end{eqnarray}
\end{widetext}
so that in the limit $v_{\rm F} q \ll \omega \ll 2\varepsilon_{\rm F}$ we have
\begin{equation}
M_{++}({\bm k}, {\bm q}', {\bm q}) = M_{\rm intra}({\bm k}, {\bm q}', {\bm q})
+
M_{\rm inter}({\bm k}, {\bm q}', {\bm q})~.
\end{equation}
In writing Eq.~(\ref{eq:SM_M_inter_def}) we have taken the limit $\omega \to 0$ in the second term on the right-hand side of Eq.~(\ref{first_term_denominator}). Moreover, in obtaining Eq.~(\ref{eq:SM_M_intra_def}) we have used that
\begin{eqnarray}
\cos\left(\frac{\theta_{{\bm k}_\pm -{\bm q}/2} + \theta_{{\bm k}_\pm +{\bm q}/2}}{2}\right) = \cos(\theta_{{\bm k}_\pm}) + {\cal O} (q^2)
~,
\end{eqnarray}
and 
\begin{eqnarray}
\frac{\partial \varepsilon_{{\bm k}_\pm,\lambda}}{\partial k_y} \to \lambda v_{\rm F} \sin(\theta_{{\bm k}_\pm})
~.
\end{eqnarray}
The last equation becomes exact for ${\bm k}$ close to the ${\bm K}$ point of the BZ and therefore in the low-energy MDF limit.

Clearly we can carry out further approximations, relying on the fact that we are interested in the low-energy MDF limit. Eq.~(\ref{eq:SM_M_intra_def}) can be further simplified by noting that
\begin{eqnarray}
\cos\left(\frac{\theta_{{\bm k}_-\pm{\bm q}/2} - \theta_{{\bm k}_+\pm{\bm q}/2}}{2}\right)
\!\! &=& \!\! 
\cos\left(\frac{\theta_{{\bm k}_--{\bm q}/2} - \theta_{{\bm k}_++{\bm q}/2}}{2}\right)
\nonumber\\
\!\! &-& \!\! 
\frac{q}{2} \sin\left(\frac{\theta_{{\bm k}_-} - \theta_{{\bm k}_+}}{2}\right) \frac{\partial \theta_{{\bm k}_\mp}}{\partial k_y}
\nonumber\\
\!\! &+& \!\! 
{\cal O}(q^2)
~,
\end{eqnarray}
which leads to
\begin{eqnarray} \label{eq:SM_M_intra_2}
&& \!\!\!\!\!\!\!\!\!\!
M_{\rm intra}
=
\frac{\cos(\theta_{{\bm k}_+}) - \cos(\theta_{{\bm k}_-})}{\omega}
\cos\!\!\left(\!\frac{\theta_{{\bm k}_--{\bm q}/2} - \theta_{{\bm k}_++{\bm q}/2}}{2}\!\!\right)
\nonumber\\
&+&
\frac{v_{\rm F} q}{\omega^2} [\cos(\theta_{{\bm k}_-}) \sin(\theta_{{\bm k}_-}) - \cos(\theta_{{\bm k}_+})\sin(\theta_{{\bm k}_+})]
\nonumber\\
&\times&
\cos\left(\frac{\theta_{{\bm k}_-} - \theta_{{\bm k}_+}}{2}\right)
\nonumber\\
&+&
\frac{q}{2\omega} \frac{\partial [\sin(\theta_{{\bm k}_-}) - \sin(\theta_{{\bm k}_-})]}{\partial k_x}
\sin\left(\frac{\theta_{{\bm k}_-} - \theta_{{\bm k}_+}}{2}\right)
\nonumber\\
&+&
{\cal O}(q^2)
~.
\end{eqnarray}
In the first term on the right-hand side of Eq.~(\ref{eq:SM_M_intra_2}) we can approximate
\begin{eqnarray} \label{eq:SM_M_intra_first}
\cos(\theta_{{\bm k}_+}) - \cos(\theta_{{\bm k}_-}) \to \frac{q'_x}{k_{\rm F}}
~,
\end{eqnarray}
while the second term on the right-hand side of Eq.~(\ref{eq:SM_M_intra_2}) becomes
\begin{eqnarray} \label{eq:SM_M_intra_second}
&& \!\!\!\!\!\!\!\!
[\cos(\theta_{{\bm k}_-})\sin(\theta_{{\bm k}_-}) - \cos(\theta_{{\bm k}_+})\sin(\theta_{{\bm k}_+})] \cos\left( \!\! \frac{\theta_{{\bm k}_-} - \theta_{{\bm k}_+}}{2} \!\! \right)
\nonumber\\
&=& \!\!\!
- \frac{k_x q'_y + k_y q'_x}{k_{\rm F}^2} \cos\left(\frac{\theta_{{\bm k}_-} - \theta_{{\bm k}_+}}{2}\right)
\nonumber\\
&\to& \!\!\!
- \left[ \! \frac{q'_y}{k_{\rm F}} \frac{\cos(\theta_{{\bm k}_+}) + \cos(\theta_{{\bm k}_-})}{2} \! + \! \frac{q'_x}{k_{\rm F}} \frac{\sin(\theta_{{\bm k}_+}) + \sin(\theta_{{\bm k}_-})}{2} \! \right]
\nonumber\\
&\times& \!\!\!
\cos\left(\frac{\theta_{{\bm k}_-} - \theta_{{\bm k}_+}}{2}\right)
\nonumber\\
&=& \!\!\!
-\left(1-\frac{q'^2}{4 k_{\rm F}^2}\right)
\nonumber\\
&\times&
\left[ \frac{q'_x}{k_{\rm F}} \sin\left(\frac{\theta_{{\bm k}_-} + \theta_{{\bm k}_+}}{2}\right)+\frac{q'_y}{k_{\rm F}} \cos\left(\frac{\theta_{{\bm k}_-} + \theta_{{\bm k}_+}}{2}\right) \right]
~.
\nonumber\\
\end{eqnarray}
Finally, the derivative in the third term on the right-hand side of Eq.~(\ref{eq:SM_M_intra_2}) reduces to
\begin{eqnarray} \label{eq:SM_M_intra_third}
\frac{\partial [\sin(\theta_{{\bm k}_-}) - \sin(\theta_{{\bm k}_-})]}{\partial k_y}
&\to&
- \frac{\partial (q'_y/k_{\rm F})}{\partial k_y}= 0~.
\end{eqnarray}

Introducing Eq.~(\ref{eq:SM_M_intra_2}), approximated according to Eqs.~(\ref{eq:SM_M_intra_first})-(\ref{eq:SM_M_intra_third}), back into Eq.~(\ref{eq:SM_Upsilon_def}) we get the ``intra-band'' contribution to the operator ${\hat \Upsilon}^{({\rm T})}_{{\bm q},{\bm q}'}$, which reads
\begin{eqnarray} \label{eq:SM_Upsilon_intra_def}
{\hat \Upsilon}^{({\rm T},{\rm intra})}_{{\bm q},{\bm q}'} 
\!\!\! &=& \!\!\!
\left[\frac{v_{\rm F} q'_x}{k_{\rm F}\omega} {\hat n}_{{\bm q}+{\bm q}'}
- \frac{v_{\rm F} q}{\omega^2} \left(1 - \frac{q'^2}{4 k_{\rm F}^2} \right) 
\right.
\nonumber\\
&\times&
\left. \left(\frac{q'_x}{k_{\rm F}} {\hat j}_{{\bm q}',y} + \frac{q'_y}{k_{\rm F}} {\hat j}_{{\bm q}',x}\right)
\right]
+{\cal O}(q^2)
~.
\end{eqnarray}
We remind the reader that, without loss of generality, we have taken ${\bm q} = q {\hat {\bm y}}$. 
Here we used that ${\hat {\bm j}}_{\bm q} = v_{\rm F} {\hat {\bm \sigma}}_{\bm q}$ close to the ${\bm K}$ point of the BZ.

We now consider Eq.~(\ref{eq:SM_M_inter_def}). Steps similar to what summarized above for the intra-band contribution to ${\hat \Upsilon}^{({\rm T})}_{{\bm q}, {\bm q}'}$ yield
\begin{eqnarray} \label{eq:SM_MatrEl_inter_def}
M_{\rm inter} &=&
\frac{1}{2\varepsilon_{\rm F}}
[\sin(\theta_{{\bm k}_-}) - \sin(\theta_{{\bm k}_+})] \sin\left(\frac{\theta_{{\bm k}_-} - \theta_{{\bm k}_+}}{2}\right)
\nonumber\\
&=&
\frac{1}{\varepsilon_{\rm F}} \sin^2\left(\frac{\theta_{{\bm k}_-} - \theta_{{\bm k}_+}}{2}\right) \cos\left(\frac{\theta_{{\bm k}_-} + \theta_{{\bm k}_+}}{2}\right)
\nonumber\\
&=&
\frac{q'^2}{4 v_{\rm F} k_{\rm F}^3} {\cal S}^{(x)}_{\lambda\lambda'}({\bm k}_-,{\bm k}_+)
~.
\end{eqnarray}
Once Eq.~(\ref{eq:SM_MatrEl_inter_def}) is introduced into Eq.~(\ref{eq:SM_Upsilon_def}), it gives the ``inter-band'' contribution to the operator ${\hat \Upsilon}^{({\rm T})}_{{\bm q},{\bm q}'}$, {\it i.e.}
\begin{eqnarray} \label{eq:SM_Upsilon_inter_def}
{\hat \Upsilon}^{({\rm T},{\rm inter})}_{{\bm q},{\bm q}'} &=&
\frac{q'^2}{4 v_{\rm F} k_{\rm F}^3} j_{{\bm q}',x}
+{\cal O}(q^2)
~.
\end{eqnarray}
Again, we used the fact that ${\hat {\bm j}}_{\bm q} = v_{\rm F} {\hat {\bm \sigma}}_{\bm q}$ close to the ${\bm K}$ point of the BZ.

Summing Eqs.~(\ref{eq:SM_Upsilon_intra_def}) and~(\ref{eq:SM_Upsilon_inter_def}) we finally get
\begin{eqnarray} \label{eq:SM_Upsilon_approx}
{\hat \Upsilon}^{({\rm T})}_{{\bm q},{\bm q}'} &=& 
\frac{v_{\rm F} q'_x}{k_{\rm F}\omega} {\hat n}_{{\bm q}+{\bm q}'}
+ {\hat \Upsilon}'_{{\bm q},{\bm q}'}
\end{eqnarray}
with
\begin{eqnarray}
{\hat \Upsilon}'_{{\bm q},{\bm q}'} &=& - \frac{v_{\rm F} q}{\omega^2} \left(1 - \frac{q'^2}{4 k_{\rm F}^2} \right) 
\left(\frac{q'_x}{k_{\rm F}} {\hat j}_{{\bm q}',y} + \frac{q'_y}{k_{\rm F}} {\hat j}_{{\bm q}',x}\right)
\nonumber\\
&+&
\frac{q'^2}{4 v_{\rm F} k_{\rm F}^3} {\hat j}_{{\bm q}',x}
~.
\nonumber\\
\end{eqnarray}

The first term on the right-hand side of Eq.~(\ref{eq:SM_Upsilon_approx}) can be further manipulated. Indeed, when it is introduced in Eq.~(\ref{eq:SM_j_1_element}) it gives
\begin{eqnarray} \label{eq:SM_rhorho_manipulation}
&& \!\!\!\!\!\!\!\!
\frac{1}{2\omega k_{\rm F}}\sum_{{\bm q}'} v_{{\bm q}'} 
\left[q'_x{\hat n}_{{\bm q} + {\bm q}'} {\hat n}_{-{\bm q}'} - q'_x {\hat n}_{{\bm q}'} {\hat n}_{{\bm q} - {\bm q}'} \right]
\nonumber\\
&=&
\frac{v_{\rm F} }{2\omega k_{\rm F}} \sum_{{\bm q}'} {\hat n}_{{\bm q} + {\bm q}'} {\hat n}_{-{\bm q}'}
[ q'_x v_{{\bm q}'} - q'_x v_{{\bm q} + {\bm q}'}]
\nonumber\\
&\to&
\frac{v_{\rm F} q}{2\omega k_{\rm F}}\sum_{{\bm q}'} \frac{q_x' q_y'}{q'^2} v_{{\bm q}'} {\hat n}_{{\bm q}'} {\hat n}_{-{\bm q}'}
+ {\cal O}(q^2)
~.
\end{eqnarray}
Here, we performed the shift ${\bm q}' \to {\bm q}+{\bm q}'$ in the term proportional to ${\hat n}_{{\bm q}'} {\hat n}_{{\bm q} - {\bm q}'}$, using that ${\bm q} = q{\hat {\bm y}}$, and we took the small-${\bm q}$ limit in the last line of Eq.~(\ref{eq:SM_rhorho_manipulation}). Finally, using the continuity equation
\begin{eqnarray}
\omega {\hat n}_{{\bm q}'} {\hat n}_{-{\bm q}'} = - {\bm q}'\cdot {\hat {\bm j}}_{{\bm q}'} {\hat n}_{-{\bm q}'} + {\hat n}_{{\bm q}'} {\bm q}'\cdot {\hat {\bm j}}_{-{\bm q}'}
~,
\end{eqnarray}
we can rewrite Eq.~(\ref{eq:SM_Upsilon_def}) in the form of Eq.~(\ref{eq:SM_Upsilon_approx_3}).

\section{The viscosity transport time}
\label{app:tau_v}
We now guide the reader through the all-order diagrammatic calculation of the low-frequency viscosity, from which we extract the corresponding transport time $\tau_{\rm v}$. We remind the reader that the low-frequency shear viscosity is given by
\begin{eqnarray} \label{eq:app_eta_0_from_stress}
\eta_0 = -\lim_{\omega\to 0} \frac{\Im m[\chi_{xy,xy} ({\bm q}={\bm 0}, \omega)]}{\omega}
~,
\end{eqnarray}
where $\chi_{\alpha\beta,\mu\nu} ({\bm q}, \omega)$ is the stress-stress response function. Figs.~\ref{fig:three}-\ref{fig:four} summarize the all-order microscopic calculation of $\chi_{\alpha\beta,\mu\nu} ({\bm q}, \omega)$, which is given by [Fig.~\ref{fig:three}a)]
\begin{eqnarray} \label{eq:chi_ss_def}
\chi_{\alpha\beta,\mu\nu} ({\bm q}, i\omega_m) &=& -k_{\rm B} T \sum_{{\bm k}, \varepsilon_n, \lambda, \lambda'} G_\lambda({\bm k}_-, i\varepsilon_n) 
\nonumber\\
&\times&
\Lambda^{(0,\alpha\beta)}_{\lambda,\lambda'}({\bm k}_-,{\bm k}_+) G_{\lambda'}({\bm k}_+, i\varepsilon_n + i\omega_m) 
\nonumber\\
&\times&
\Lambda^{(\mu\nu)}_{\lambda'\lambda} ({\bm k}_+, i\varepsilon_n + i\omega_m,{\bm k}_-,i\varepsilon_n)
~.
\end{eqnarray}
Here $G_\lambda({\bm k},i\varepsilon)$ is the Green's function (represented by a double solid line in Figs.~\ref{fig:three}-\ref{fig:four}) dressed by the self-energy insertion of Fig.~\ref{fig:three}b), ${\bm k}_\pm = {\bm k}\pm{\bm q}/2$, $\varepsilon_n$ ($\omega_m$) is a fermionic (bosonic) Matsubara frequency, and $\lambda$ and $\lambda'$ are band indices. In Eq.~(\ref{eq:chi_ss_def}) we defined the bare vertex (represented as a solid dot in Fig.~\ref{fig:three})
\begin{eqnarray}
\Lambda_{\lambda\lambda'}^{(0,\alpha\beta)}({\bm k},{\bm k}') = v_{\rm F} \frac{k_\alpha {\cal S}^{(\beta)}_{\lambda\lambda'}({\bm k}_-, {\bm k}_+) + k_\beta {\cal S}^{(\alpha)}_{\lambda\lambda'}({\bm k}_-,{\bm k}_+)}{2}
~.
\nonumber\\
\end{eqnarray}
Finally, $\Lambda^{(\mu\nu)}_{\lambda\lambda'} ({\bm k}, i\varepsilon,{\bm k}',i\varepsilon')$ is the dressed vertex function (represented as a triangle in Figs.~\ref{fig:three}-\ref{fig:four}), which satisfies the self-consistent Bethe-Salpeter equation of Fig.~\ref{fig:three}c). The choice of the quasiparticle self-energy, and the requirement of fulfilling the Ward identities, uniquely determine the form of the Bethe-Salpeter equation, {\it i.e.} the irreducible interaction $I$. Fig.~\ref{fig:four} shows the three contributions to the irreducible interactions. In formulas, the Bethe-Salpeter equation is
\begin{eqnarray} \label{eq:Lambda_def}
&&
\Lambda_{\lambda'\lambda}^{(\alpha\beta)} ({\bm k}_+, i\varepsilon_n + i\omega_m,{\bm k}_-,i\varepsilon_n) = \Lambda^{(0,\alpha\beta)}_{\lambda'\lambda} ({\bm k}_+,{\bm k}_-)
\nonumber\\
&&
+ \sum_{i=1}^3 \Lambda^{(i,\alpha\beta)}_{\lambda'\lambda} ({\bm k}_+, i\varepsilon_n + i\omega_m,{\bm k}_-,i\varepsilon_n)
~,
\end{eqnarray}
where the three $\big\{ \Lambda^{(i,\alpha\beta)}_{\lambda'\lambda} ({\bm k}_+, i\varepsilon_n + i\omega_m,{\bm k}_-,i\varepsilon_n), i=1,\dots,3 \big\}$ correspond to the three diagrams in fig.~\ref{fig:four}, and read
\begin{eqnarray} \label{eq:Lambda_1}
&&
\Lambda^{(1,\alpha\beta)}_{\lambda'\lambda} ({\bm k}_+, i\varepsilon_n + i\omega_m,{\bm k}_-,i\varepsilon_n) =
k_{\rm B} T  \sum_{{\bm k}',\varepsilon_{n'}} 
\nonumber\\
&&
\times
W^{(1)}_{\lambda\lambda'\mu\mu'}({\bm k}',{\bm k},i\varepsilon_{n'}-i\varepsilon_n) G_{\mu'}({\bm k}'_+,i\varepsilon_{n'}+i\omega_m)
\nonumber\\
&&
\times
\Lambda^{(\alpha\beta)}_{\mu'\mu} ({\bm k}'_+, i\varepsilon_{n'} + i\omega_m,{\bm k}'_-,i\varepsilon_{n'})
G_{\mu}({\bm k}'_-,i\varepsilon_{n'})
~,
\end{eqnarray}
and
\begin{eqnarray} \label{eq:Lambda_2}
&&
\Lambda^{(2,\alpha\beta)}_{\lambda'\lambda} ({\bm k}_+, i\varepsilon_n + i\omega_m,{\bm k}_-,i\varepsilon_n) =
k_{\rm B} T  \sum_{{\bm k}',\varepsilon_{n'}} 
\nonumber\\
&&
\times
W^{(2)}_{\lambda\lambda'\mu\mu'}({\bm k}',{\bm k},i\varepsilon_{n'}-i\varepsilon_n) G_{\mu'}({\bm k}'_+,i\varepsilon_{n'}+i\omega_m)
\nonumber\\
&&
\times
\Lambda^{(\alpha\beta)}_{\mu'\mu} ({\bm k}'_+, i\varepsilon_{n'} + i\omega_m,{\bm k}'_-,i\varepsilon_{n'})
G_{\mu}({\bm k}'_-,i\varepsilon_{n'})
~,
\end{eqnarray}
and finally
\begin{eqnarray} \label{eq:Lambda_3}
&&
\Lambda^{(3,\alpha\beta)}_{\lambda'\lambda} ({\bm k}_+, i\varepsilon_n + i\omega_m,{\bm k}_-,i\varepsilon_n) = k_{\rm B} T  \sum_{{\bm k}',\varepsilon_{n'}}
\nonumber\\
&&
\times
W^{(3)}_{\lambda\lambda'\mu\mu'}({\bm k}', {\bm k}, i\varepsilon_{n'}+i\varepsilon_{n} +i\omega_m)
G_{\mu'}({\bm k}'_+,i\varepsilon_{n'}+i\omega_m) 
\nonumber\\
&&
\times
\Lambda^{(\alpha\beta)}_{\mu'\mu} ({\bm k}'_+, i\varepsilon_{n'} + i\omega_m,{\bm k}'_-,i\varepsilon_{n'})
G_\mu({\bm k}'_-,i\varepsilon_{n'})
~.
\end{eqnarray}
Here 
\begin{eqnarray}
&&
W^{(1)}_{\lambda\lambda'\mu\mu'}({\bm k}',{\bm k},i\omega_m) \equiv W({\bm k}-{\bm k}',i\omega_m) {\cal D}_{\lambda'\mu'}({\bm k}_+,{\bm k}'_+) 
\nonumber\\
&&
\times 
{\cal D}_{\mu\lambda}({\bm k}'_-,{\bm k}_-)
~,
\end{eqnarray}
where $W({\bm q},i\omega_m)$ is the screened interaction, represented as a wavy line in Figs.~\ref{fig:three}-\ref{fig:four}. In the large-$N_{\rm f}$ limit this is given by
\begin{equation} \label{eq:W_RPA}
W({\bm q},i\Omega_{m}) = \frac{v_{\bm q}}{1-v_{\bm q} \chi_{nn}({\bm q},i\Omega_m)}
~,
\end{equation}
where $\chi_{nn}({\bm q},\omega)$ is the proper density-density response function~\cite{Giuliani_and_Vignale} of graphene, {\it i.e.}~\cite{Principi_arxiv_2015}
\begin{eqnarray} \label{eq:chi_nn}
\chi_{nn}({\bm q},i\omega_m) &=& N_{\rm f} k_{\rm B} T \sum_{{\bm q}',\varepsilon_n} \sum_{\lambda'',\mu''} G_{\lambda''}({\bm q}',i\varepsilon_n)
\nonumber\\
&\times&
G_{\mu''}({\bm q'}+{\bm q}, i\varepsilon_n+i\omega_m)
\nonumber\\
&\times&
{\cal D}_{\lambda''\mu''}({\bm q}',{\bm q}'+{\bm q}) {\cal D}_{\mu''\lambda''}({\bm q}'+{\bm q},{\bm q}') 
~.
\nonumber\\
\end{eqnarray}
Moreover,
\begin{eqnarray} \label{eq:W_1_2}
&&
W^{(2)}_{\lambda\lambda'\mu\mu'} ({\bm k}',{\bm k},i\varepsilon_{n'}-i\varepsilon_n) = - k_{\rm B} T \!\! \sum_{{\bm q}',\omega_{m'}} \sum_{\lambda'',\mu''} W({\bm q}',i\omega_{m'}) 
\nonumber\\
&&
\times
W({\bm q}'-{\bm q},i\omega_{m'}-i\omega_m) 
{\cal D}_{\lambda'\lambda''}({\bm k}_+,{\bm k}_+-{\bm q}')
\nonumber\\
&&
\times
{\cal D}_{\lambda''\lambda}({\bm k}_+-{\bm q}',{\bm k}_-)
{\cal D}_{\mu\mu''}({\bm k}'_-,{\bm k}'_+-{\bm q}')
\nonumber\\
&&
\times
{\cal D}_{\mu''\mu'}({\bm k}'_+-{\bm q}',{\bm k}'_+)
G_{\lambda''}({\bm k}_+-{\bm q}',i\varepsilon_n+i\omega_m -i\omega_{m'})
\nonumber\\
&&
\times
G_{\mu''}({\bm k}'_+-{\bm q}',i\varepsilon_{n'}+i\omega_m -i\omega_{m'})
~,
\end{eqnarray}
and
\begin{eqnarray} \label{eq:W_3}
&&
W^{(3)}_{\lambda\lambda'\mu\mu'} ({\bm k}',{\bm k},i\varepsilon_{n'}+i\varepsilon_n+i\omega_m) = -k_{\rm B} T \sum_{{\bm q}',\omega_{m'}}\sum_{\lambda'',\mu''} 
\nonumber\\
&&
\times
W({\bm q}',i\omega_{m'}) 
W({\bm q}'-{\bm q},i\omega_{m'}-i\omega_m) 
{\cal D}_{\lambda\lambda''}({\bm k}_-,{\bm k}_-+{\bm q}')
\nonumber\\
&&
\times
{\cal D}_{\lambda''\lambda'}({\bm k}_-+{\bm q}',{\bm k}_+)
{\cal D}_{\mu\mu''}({\bm k}'_-,{\bm k}'_+-{\bm q}')
\nonumber\\
&&
\times
{\cal D}_{\mu''\mu'}({\bm k}'_+-{\bm q}',{\bm k}'_+)
G_{\lambda''}({\bm k}_-+{\bm q}',i\varepsilon_n +i\omega_{m'})
\nonumber\\
&&
\times
G_{\mu''}({\bm k}'_+-{\bm q}',i\varepsilon_{n'}+i\omega_m -i\omega_{m'})
~.
\end{eqnarray}
In Eqs.~(\ref{eq:Lambda_1})-(\ref{eq:W_3}) we suppressed the ${\bm q}$-dependence of $W^{(2)} ({\bm k}',{\bm k},i\varepsilon_{n'}-i\varepsilon_n)$ and $W^{(3)} ({\bm k}',{\bm k},i\varepsilon_{n'}+i\varepsilon_n+i\omega_m)$. 

In what follows we first analytically continue Eqs.~(\ref{eq:chi_ss_def})-(\ref{eq:W_3}), and then we consider the limit of ${\bm q}= 0$, small frequency and low temperature. These limits allows us to exactly solve the Bethe-Salpeter equation. This information is then used to calculate the low-frequency viscosity as shown by Eq.~(\ref{eq:app_eta_0_from_stress}).

\subsection{Analytical continuation to real frequencies}
The analytical continuation of equations similar to Eqs.~(\ref{eq:chi_ss_def})-(\ref{eq:W_3}) was performed in Ref.~\onlinecite{Principi_arxiv_2015}. Therefore, we summarize here only the main results. After the analytical continuation, the integral on the right-hand side of Eq.~(\ref{eq:chi_ss_def}) contains products of two advanced Green's functions (schematically $G^{(A)} G^{(A)}$), two retarded ones ($G^{(R)} G^{(R)}$), or one advanced and one retarded ($G^{(A)} G^{(R)}$). The first two combinations have both poles on the same side of the complex plane. Therefore, in the limit $\varepsilon_{\rm F} \tau_{\rm ee} \gg 1$ they give a negligible contribution, and we can retain only the combination $G^{(A)} G^{(R)}$. We thus get
\begin{eqnarray} \label{eq:R_chi_jj_final_omega_finite}
&&
\chi_{\alpha\beta,\mu\nu} ({\bm q}, \omega) = \sum_{{\bm k}, \lambda, \lambda'} \int \frac{d\varepsilon}{2\pi} \big[n_{\rm F} (\varepsilon+\omega) - n_{\rm F} (\varepsilon)\big]
\nonumber\\
&&
\times
G^{({\rm A})}_\lambda({\bm k}_-, \varepsilon) 
\Lambda^{(0,\alpha\beta)}_{\lambda,\lambda'}({\bm k}_-,{\bm k}_+) G^{({\rm R})}_{\lambda'}({\bm k}_+, \varepsilon + \omega)
\nonumber\\
&&
\times
\Lambda^{(\mu\nu)}_{\lambda'\lambda} ({\bm k}_+, \varepsilon_+ + \omega,{\bm k}_-,\varepsilon_-)
~,
\nonumber\\
\end{eqnarray}
where $\varepsilon_\pm = \varepsilon \pm i 0^+$, and $G^{({\rm R})}_{\lambda}({\bm k}, \varepsilon) \equiv G_{\lambda}({\bm k}, \varepsilon_+)$ [$G^{({\rm A})}_{\lambda}({\bm k}, \varepsilon) \equiv G_{\lambda}({\bm k}, \varepsilon_-)$] is the retarded [advanced] Green's function.
The vertex function satisfies the following Bethe-Salpeter equation
\begin{eqnarray} \label{eq:R_Lambda_def}
&&
\Lambda_{\lambda'\lambda}^{(\alpha\beta)} ({\bm k}_+, \varepsilon_+ + \omega,{\bm k}_-,\varepsilon_-) = \Lambda^{(0,\alpha\beta)}_{\lambda'\lambda} ({\bm k}_+,{\bm k}_-)
\nonumber\\
&&
+ \sum_{i=1}^3 \Lambda^{(i,\alpha\beta)}_{\lambda'\lambda} ({\bm k}_+, \varepsilon_+ + \omega,{\bm k}_-,\varepsilon_-)
~,
\end{eqnarray}
where
\begin{eqnarray} \label{eq:R_Lambda_1_f_final}
&&
\Lambda^{(1,\alpha\beta)}_{\lambda'\lambda} ({\bm k}_+, \varepsilon_++\omega,{\bm k}_-,\varepsilon_-) =
-\sum_{{\bm k}',\mu,\mu'} \int \frac{d\varepsilon'}{2\pi i} 
\nonumber\\
&&
\times
\big[n_{\rm F}(\varepsilon') + n_{\rm B}(\varepsilon'-\varepsilon)\big]
\nonumber\\
&&
\times
\Big[ W^{(1)}_{\lambda\lambda'\mu\mu'}({\bm k}',{\bm k},\varepsilon'_--\varepsilon)  - W^{(1)}_{\lambda\lambda'\mu\mu'}({\bm k}',{\bm k},\varepsilon'_+-\varepsilon) \Big]
\nonumber\\
&&
\times
G^{({\rm R})}_{\mu'}({\bm k}'_+,\varepsilon'+\omega) \Lambda^{(\alpha\beta)}_{\mu'\mu} ({\bm k}'_+, \varepsilon'_++\omega, {\bm k}'_-,\varepsilon'_-)
\nonumber\\
&&
\times
G^{({\rm A})}_\mu({\bm k}'_-,\varepsilon')
~,
\nonumber\\
\end{eqnarray}
and
\begin{eqnarray} \label{eq:R_Lambda_2_f_final}
&&
\Lambda^{(2,\alpha\beta)}_{\lambda'\lambda} ({\bm k}_+, \varepsilon_++\omega,{\bm k}_-,\varepsilon_-) =
-\sum_{{\bm k}',\mu,\mu'} \int \frac{d\varepsilon'}{2\pi i} 
\nonumber\\
&&
\times
\big[n_{\rm F}(\varepsilon') + n_{\rm B}(\varepsilon'-\varepsilon)\big]
\nonumber\\
&&
\times
\Big[ W^{(2)}_{\lambda\lambda'\mu\mu'}({\bm k}',{\bm k},\varepsilon'_--\varepsilon)  - W^{(2)}_{\lambda\lambda'\mu\mu'}({\bm k}',{\bm k},\varepsilon'_+-\varepsilon) \Big]
\nonumber\\
&&
\times
G^{({\rm R})}_{\mu'}({\bm k}'_+,\varepsilon'+\omega) \Lambda^{(\alpha\beta)}_{\mu'\mu} ({\bm k}'_+, \varepsilon'_++\omega, {\bm k}'_-,\varepsilon'_-)
\nonumber\\
&&
\times
G^{({\rm A})}_\mu({\bm k}'_-,\varepsilon')
~,
\nonumber\\
\end{eqnarray}
and finally
\begin{eqnarray} \label{eq:R_Lambda_3_f_final}
&&
\Lambda^{(3,\alpha\beta)}_{\lambda'\lambda} ({\bm k}_+, \varepsilon_++\omega,{\bm k}_-,\varepsilon_-) =
-\sum_{{\bm k}',\mu,\mu'} \int \frac{d\varepsilon'}{2\pi i} 
\nonumber\\
&&
\times
\big[n_{\rm F}(\varepsilon') + n_{\rm B}(\varepsilon' + \varepsilon)\big]
\nonumber\\
&&
\times
\Big[W^{(3)}_{\lambda\lambda'\mu\mu'}({\bm k}', {\bm k}, \varepsilon'_-+\varepsilon) - W^{(3)}_{\lambda\lambda'\mu\mu'}({\bm k}', {\bm k}, \varepsilon'_++\varepsilon) \Big]
\nonumber\\
&&
\times
G^{({\rm R})}_{\mu'}({\bm k}'_+,\varepsilon'+\omega) \Lambda^{(\alpha\beta)}_{\mu'\mu} ({\bm k}'_+, \varepsilon_++\omega,{\bm k}'_-,\varepsilon'_-)
\nonumber\\
&&
\times
G^{({\rm A})}_\mu({\bm k}'_-,\varepsilon')
~.
\nonumber\\
\end{eqnarray}
In these equations we defined the Fermi and Bose distributions $n_{\rm F}(\varepsilon) = [e^{\varepsilon/(k_{\rm B} T)} + 1]^{-1}$ and $n_{\rm B}(\varepsilon) = [e^{\varepsilon/(k_{\rm B} T)} - 1]^{-1}$, and the potentials
\begin{eqnarray} \label{eq:R_W_1_A-R}
&&
W^{(1)}_{\lambda\lambda'\mu\mu'}(\varepsilon_-'-\varepsilon) - W^{(1)}_{\lambda\lambda'\mu\mu'}(\varepsilon_+'-\varepsilon) =
-4 \sum_{{\bm q}',\lambda'',\mu''}
\nonumber\\
&&
\times
|W({\bm k}-{\bm k}',\varepsilon'-\varepsilon)|^2
 \int \frac{d\omega'}{2\pi i} \big[n_{\rm F} (\omega'+\varepsilon') - n_{\rm F} (\omega'+\varepsilon)\big] 
\nonumber\\
&&
\times
\Im m \Big[G^{({\rm R})}_{\lambda''}({\bm q}'-{\bm k},\omega'+\varepsilon)\Big] \Im m\Big[G^{({\rm R})}_{\mu''}({\bm q'}-{\bm k}', \omega'+\varepsilon')\Big]
\nonumber\\
&&
\times
{\cal D}_{\lambda'\mu'}({\bm k}_+,{\bm k}'_+) {\cal D}_{\mu\lambda}({\bm k}'_-,{\bm k}_-)
\nonumber\\
&&
\times
{\cal D}_{\lambda''\mu''}({\bm q}'-{\bm k},{\bm q}'-{\bm k}') {\cal D}_{\mu''\lambda''}({\bm q}'-{\bm k}',{\bm q}'-{\bm k}) 
~,
\nonumber\\
\end{eqnarray}
and
\begin{widetext}
\begin{eqnarray} \label{eq:R_W_2_A-R}
&&
W^{(2)}_{\lambda\lambda'\mu\mu'} (\varepsilon'_--\varepsilon) - W^{(2)}_{\lambda\lambda'\mu\mu'} (\varepsilon'_+-\varepsilon) =
\sum_{{\bm q}',\lambda'',\mu''} \int \frac{d\omega'}{2\pi i} 
W({\bm q}',\omega'_+) W({\bm q}',\omega'_--\omega) 
\big[n_{\rm F}(\omega'-\varepsilon-\omega) - n_{\rm F}(\omega'-\varepsilon'-\omega)\big]
\nonumber\\
&&\times
\Big[ G^{({\rm R})}_{\lambda''}({\bm k}_+-{\bm q}',\varepsilon+\omega-\omega') - G^{({\rm A})}_{\lambda''}({\bm k}_+-{\bm q}',\varepsilon+\omega-\omega') \Big]
\Big[ G^{({\rm R})}_{\mu''}({\bm k}'_+-{\bm q}',\varepsilon'+\omega -\omega') - G^{({\rm A})}_{\mu''}({\bm k}'_+-{\bm q}',\varepsilon'+\omega -\omega') \Big]
\nonumber\\
&&\times
{\cal D}_{\lambda'\lambda''}({\bm k}_+,{\bm k}_+-{\bm q}')
{\cal D}_{\lambda''\lambda}({\bm k}_+-{\bm q}',{\bm k}_-)
{\cal D}_{\mu\mu''}({\bm k}'_-,{\bm k}'_+-{\bm q}')
{\cal D}_{\mu''\mu'}({\bm k}'_+-{\bm q}',{\bm k}'_+)
~,
\end{eqnarray}
and finally
\begin{eqnarray} \label{eq:R_W_3_A-R}
&&
W^{(3)}_{\lambda\lambda'\mu\mu'} (\varepsilon'_-+\varepsilon+\omega) - W^{(3)}_{\lambda\lambda'\mu\mu'} (\varepsilon'_++\varepsilon+\omega) =
-\sum_{{\bm q}',\lambda'',\mu''} 
\int \frac{d\omega'}{2\pi i}
W({\bm q}',\omega'_+) W({\bm q}',\omega'_--\omega)
\big[n_{\rm F}(\omega'+\varepsilon) - n_{\rm F}(\omega'-\varepsilon')\big]
\nonumber\\
&&\times
\Big[ G^{({\rm R})}_{\lambda''}({\bm k}_- +{\bm q}',\varepsilon+\omega') - G^{({\rm A})}_{\lambda''}({\bm k}_- +{\bm q}',\varepsilon+\omega') \Big]
\Big[ G^{({\rm R})}_{\mu''}({\bm k}'_+ -{\bm q}',\varepsilon' +\omega- \omega') - G^{({\rm A})}_{\mu''}({\bm k}'_+ -{\bm q}',\varepsilon'+\omega - \omega') \Big]
\nonumber\\
&&\times
{\cal D}_{\lambda\lambda''}({\bm k}_-,{\bm k}_-+{\bm q}')
{\cal D}_{\lambda''\lambda'}({\bm k}_-+{\bm q}',{\bm k}_+)
{\cal D}_{\mu\mu''}({\bm k}'_-,{\bm k}'_+-{\bm q}')
{\cal D}_{\mu''\mu'}({\bm k}'_+-{\bm q}',{\bm k}'_+)
~.
\end{eqnarray}
\end{widetext}

\subsection{The Bethe-Salpeter equation}
\label{sect:BSE}
Setting $q=0$ and taking the limit $\omega \to 0$, Eq.~(\ref{eq:R_chi_jj_final_omega_finite}) becomes
\begin{eqnarray} \label{eq:R_chi_jj_final_omega0}
&&
\chi_{\alpha\beta,\mu\nu} ({\bm q} = {\bm 0}, \omega) =\frac{2 i \omega}{\omega + i/\tau_{\rm ee}}
\sum_{{\bm k}, \lambda} \int \frac{d\varepsilon}{2\pi i} \left(-\frac{\partial n_{\rm F} (\varepsilon)}{\partial \varepsilon} \right)
\nonumber\\
&&
\times
\Im m\big[ G^{({\rm R})}_{\lambda}({\bm k}, \varepsilon) \big]
\Lambda^{(0,\alpha\beta)}_{\lambda,\lambda}({\bm k},{\bm k}) \Lambda^{(\mu\nu)}_{\lambda\lambda} ({\bm k}, \varepsilon_+ + \omega,{\bm k},\varepsilon_-)
~.
\nonumber\\
\end{eqnarray}
Here we used that
\begin{eqnarray} \label{eq:R_GR_GA_product}
G^{({\rm A})}_\lambda({\bm k}, \varepsilon) G^{({\rm R})}_{\lambda'}({\bm k}, \varepsilon+\omega) &\simeq&
-\frac{2 i \delta_{\lambda\lambda'} }{\omega + i/\tau_{\rm ee}}
\Im m\big[ G^{({\rm R})}_{\lambda'}({\bm k}, \varepsilon) \big]
~.
\nonumber\\
\end{eqnarray}
In writing Eq.~(\ref{eq:R_GR_GA_product}) we retained only the singular part of the product of the two Green's functions, {\it i.e.} the quasiparticle pole, and we neglected the regular part. Herein relies our Fermi-liquid approximation. In so doing we are able to significantly simplify the following expressions, especially that of the Bethe-Salpeter equation, which can then be solved analytically. This allows us to determine the viscosity transport time. However, by neglecting the regular part of the product~(\ref{eq:R_GR_GA_product}), we miss the Fermi-liquid renormalization of the shear modulus $S_\infty$. Note that, while the transport time is a non-perturbative quantity which diverges in the limit $\alpha_{\rm ee}\to 0$, the renormalization of $S_\infty$ can be calculated perturbatively. Therefore our approach captures the most significant effect of e-e interactions in the regime in which they are not too strong ({\it i.e.} $\alpha_{\rm ee} \lesssim 1$).

We observe that, at low temperature, the derivative of the Fermi function is strongly peaked at $\varepsilon \sim 0$. This in turn implies that we can set $\varepsilon = 0$ in $\Im m\big[ G^{({\rm R})}_{\lambda}({\bm k}, \varepsilon) \big]$ on the right-hand side of Eq.~(\ref{eq:R_chi_jj_final_omega0}). The latter is then strongly peaked at the Fermi surface, and allows us to set $k = k_{\rm F}$ and $\lambda = +$ in the rest of the integrand. As shown in Ref.~\onlinecite{Principi_arxiv_2015}, we cannot set $\varepsilon = 0$ in the factor $\Lambda^{(\mu\nu)}_{\lambda\lambda} ({\bm k}, \varepsilon_+,{\bm k},\varepsilon_-)$. This happens because the latter contains Fermi and Bose distributions which combine with the factor $-\partial n_{\rm F} (\varepsilon)/(\partial \varepsilon)$ on the right-hand side of Eq.~(\ref{eq:R_chi_jj_final_omega0}) to produce the correct vertex correction,. 

After the analytical continuation Eqs.~(\ref{eq:R_Lambda_1_f_final})-(\ref{eq:R_W_3_A-R}) read
\begin{widetext}
\begin{eqnarray} \label{eq:R_Lambda_1_together}
&&
\Lambda^{(1,\alpha\beta)}_{\lambda\lambda} ({\bm k}, \varepsilon_++\omega,{\bm k},\varepsilon_-) =
- \frac{8 i}{\omega + i/\tau_{\rm ee}}
\sum_{{\bm k}',{\bm q}'} \sum_{\mu, \mu'',\lambda''} \int \frac{d\varepsilon'}{2\pi i} 
\int \frac{d\omega'}{2\pi i}
\big[n_{\rm F}(\varepsilon') + n_{\rm B}(\varepsilon'-\varepsilon)\big]
\big[n_{\rm F} (\omega'+\varepsilon') - n_{\rm F} (\omega'+\varepsilon)\big] 
\nonumber\\
&&
\times
|W({\bm k}-{\bm k}',\varepsilon')|^2
\Im m\big[ G^{({\rm R})}_{\mu}({\bm k}', \varepsilon') \big]
\Im m \Big[G^{({\rm R})}_{\lambda''}({\bm q}'-{\bm k},\omega')\Big]
\Im m\Big[G^{({\rm R})}_{\mu''}({\bm q'}-{\bm k}', \omega'+\varepsilon')\Big]
{\cal D}_{\lambda\mu}({\bm k},{\bm k}') {\cal D}_{\mu\lambda}({\bm k}',{\bm k})
\nonumber\\
&&
\times
{\cal D}_{\lambda''\mu''}({\bm q}'-{\bm k},{\bm q}'-{\bm k}') {\cal D}_{\mu''\lambda''}({\bm q}'-{\bm k}',{\bm q}'-{\bm k}) 
\Lambda^{(\alpha\beta)}_{\mu\mu} ({\bm k}', \varepsilon'_++\omega, {\bm k}',\varepsilon'_-)
~,
\end{eqnarray}
and
\begin{eqnarray} \label{eq:R_Lambda_2_together}
&&
\Lambda^{(2,\alpha\beta)}_{\lambda\lambda} ({\bm k}, \varepsilon_++\omega,{\bm k},\varepsilon_-) =
-\frac{8 i}{\omega + i/\tau_{\rm ee}} \sum_{{\bm k}',{\bm q}'} \sum_{\mu,\lambda'',\mu''}  \int \frac{d\varepsilon'}{2\pi i}
\int \frac{d\omega'}{2\pi i} 
\big[n_{\rm F}(\varepsilon') + n_{\rm B}(\varepsilon'-\varepsilon)\big]
\big[n_{\rm F}(\omega'-\varepsilon) - n_{\rm F}(\omega'-\varepsilon')\big]
\nonumber\\
&&
\times
|W({\bm q}',\omega')|^2 
\Im m\big[ G^{({\rm R})}_{\mu}({\bm k}', \varepsilon') \big]
\Im m\big[ G^{({\rm R})}_{\lambda''}({\bm k}-{\bm q}',-\omega') \big]
\Im m\big[ G^{({\rm R})}_{\mu''}({\bm k}'-{\bm q}',\varepsilon' -\omega') \Big]
{\cal D}_{\lambda\lambda''}({\bm k},{\bm k}-{\bm q}')
\nonumber\\
&&
\times
{\cal D}_{\lambda''\lambda}({\bm k}-{\bm q}',{\bm k})
{\cal D}_{\mu\mu''}({\bm k}',{\bm k}'-{\bm q}')
{\cal D}_{\mu''\mu}({\bm k}'-{\bm q}',{\bm k}')
\Lambda^{(\alpha\beta)}_{\mu\mu} ({\bm k}', \varepsilon'_++\omega, {\bm k}',\varepsilon'_-)
~,
\end{eqnarray}
and finally
\begin{eqnarray} \label{eq:R_Lambda_3_together}
&&
\Lambda^{(3,\alpha\beta)}_{\lambda\lambda} ({\bm k}, \varepsilon_++\omega,{\bm k},\varepsilon_-) =
\frac{8 i }{\omega + i/\tau_{\rm ee}} \sum_{{\bm k}',{\bm q}'} \sum_{\mu,\lambda'',\mu''} \int \frac{d\varepsilon'}{2\pi i} 
\int \frac{d\omega'}{2\pi i} 
\big[n_{\rm F}(\varepsilon') + n_{\rm B}(\varepsilon'+\varepsilon)\big]
\big[n_{\rm F}(\omega'+\varepsilon) - n_{\rm F}(\omega'-\varepsilon')\big]
\nonumber\\
&&
\times
|W({\bm q}',\omega')|^2
\Im m \big[ G^{({\rm R})}_{\mu}({\bm k}',\varepsilon')]
\Im m \big[ G^{({\rm R})}_{\lambda''}({\bm k} +{\bm q}',\omega') \big]
\Im m \big[ G^{({\rm R})}_{\mu''}({\bm k}' -{\bm q}',\varepsilon'- \omega') \big]
{\cal D}_{\lambda\lambda''}({\bm k},{\bm k}+{\bm q}')
\nonumber\\
&&
\times
{\cal D}_{\lambda''\lambda}({\bm k}+{\bm q}',{\bm k})
{\cal D}_{\mu\mu''}({\bm k}',{\bm k}'-{\bm q}')
{\cal D}_{\mu''\mu}({\bm k}'-{\bm q}',{\bm k}')
\Lambda^{(\alpha\beta)}_{\mu\mu} ({\bm k}', \varepsilon'_++\omega,{\bm k}',\varepsilon'_-)
~.
\end{eqnarray}
\end{widetext}
In these equations the limit $\omega\to 0$ is understood.

We now recall that Eqs.~(\ref{eq:R_Lambda_1_together})-(\ref{eq:R_Lambda_3_together}) must be introduced into Eq.~(\ref{eq:R_chi_jj_final_omega0}) and integrated over $\varepsilon$ with the weighting factor $\partial n_{\rm F}(\varepsilon)/(\partial \varepsilon)$. To perform such integration we use the fact that
\begin{eqnarray} \label{eq:nB_nF_approx}
{\cal N} &=&
\frac{\partial n_{\rm F}(\varepsilon)}{\partial \varepsilon} \big[n_{\rm F}(\varepsilon') + n_{\rm B}(\varepsilon'-\varepsilon)\big] 
\nonumber\\
&\times&
\big[ n_{\rm F}(\varepsilon''+\varepsilon) - n_{\rm F}(\varepsilon''+\varepsilon') \big] 
\nonumber\\
&=&
\frac{\partial n_{\rm B}(\varepsilon'')}{\partial \varepsilon''} \big[ n_{\rm F}(\varepsilon+\varepsilon'') - n_{\rm F}(\varepsilon) \big] 
\nonumber\\
&\times&
\big[ n_{\rm F}(\varepsilon'+\varepsilon'') - n_{\rm F}(\varepsilon') \big]
\nonumber\\
&\to&
\varepsilon''^2 \frac{\partial n_{\rm B}(\varepsilon'')}{\partial \varepsilon''} \frac{\partial n_{\rm F}(\varepsilon)}{\partial \varepsilon} \frac{\partial n_{\rm F}(\varepsilon')}{\partial \varepsilon'} 
~.
\end{eqnarray}
In evaluating an integral of the form 
\begin{equation}
{\cal I} = \int_{-\infty}^{\infty} d\varepsilon'' \frac{\partial n_{\rm B}(\varepsilon'')}{\partial \varepsilon''} \varepsilon''^2 f(\varepsilon'')
~,
\end{equation}
where $f(\varepsilon'')$ is some smooth function of its argument, we exploit the fact that the weighting function $\varepsilon''^2\partial n_{\rm B}(\varepsilon'')/\partial \varepsilon''$  is strongly peaked at $\varepsilon''=0$ and its width scales with $k_{\rm B}^2 T^2/\varepsilon_{\rm F}$.  This does not mean, however, that one can simply replace $f(\varepsilon'')$ by $f(0)$.  Such a crude approximation may introduce spurious divergences~\cite{Principi_arxiv_2015}. To take this into account we approximate
\begin{equation} \label{eq:energy_approx}
{\cal I} = -\frac{2 \pi^2 (k_{\rm B} T)^2}{3} f({\bar \varepsilon}) + {\cal O}(T^4)\,,
\end{equation}
where~\cite{Principi_arxiv_2015} ${\bar \varepsilon} = \zeta k_{\rm B} T$ is approximated with half the variance of the distribution $\varepsilon^2 \partial n_{\rm B}(\varepsilon)/(\partial\varepsilon)$. Therefore, $\zeta= \pi/\sqrt{5}$. With this approximation we can rewrite
\begin{eqnarray} \label{eq:R_Lambda_1_together_2}
&&
\Lambda^{(1,\alpha\beta)}_{\lambda\lambda} ({\bm k}, \omega_+,{\bm k},0^-) =
-\frac{4 i (k_{\rm B} T)^2}{3(\omega + i/\tau_{\rm ee})}
\sum_{{\bm k}',{\bm q}'} \sum_{\mu, \mu'',\lambda''}
\nonumber\\
&&
\times
|W({\bm k}-{\bm k}',0)|^2
\Im m\big[ G^{({\rm R})}_{\mu}({\bm k}',0) \big]
\Im m \Big[G^{({\rm R})}_{\lambda''}({\bm q}'-{\bm k},0)\Big]
\nonumber\\
&&
\times
\Im m\Big[G^{({\rm R})}_{\mu''}({\bm q'}-{\bm k}', 0)\Big]
{\cal D}_{\lambda\mu}({\bm k},{\bm k}') {\cal D}_{\mu\lambda}({\bm k}',{\bm k})
\nonumber\\
&&
\times
{\cal D}_{\lambda''\mu''}({\bm q}'-{\bm k},{\bm q}'-{\bm k}') {\cal D}_{\mu''\lambda''}({\bm q}'-{\bm k}',{\bm q}'-{\bm k}) 
\nonumber\\
&&
\times
\Lambda^{(\alpha\beta)}_{\mu\mu} ({\bm k}', \omega_+, {\bm k}',0^-)
~,
\end{eqnarray}
and
\begin{eqnarray} \label{eq:R_Lambda_2_together_2}
&&
\Lambda^{(2,\alpha\beta)}_{\lambda\lambda} ({\bm k}, \omega_+,{\bm k},0^-) =
-\frac{4 i (k_{\rm B} T)^2}{3(\omega + i/\tau_{\rm ee})} \sum_{{\bm k}',{\bm q}'} \sum_{\mu,\lambda'',\mu''}  
\nonumber\\
&&
\times
|W({\bm q}',0)|^2 
\Im m\big[ G^{({\rm R})}_{\mu}({\bm k}', 0) \big]
\Im m\big[ G^{({\rm R})}_{\lambda''}({\bm k}-{\bm q}',0) \big]
\nonumber\\
&&
\times
\Im m\big[ G^{({\rm R})}_{\mu''}({\bm k}'-{\bm q}',0) \Big]
{\cal D}_{\lambda\lambda''}({\bm k},{\bm k}-{\bm q}')
{\cal D}_{\lambda''\lambda}({\bm k}-{\bm q}',{\bm k})
\nonumber\\
&&
\times
{\cal D}_{\mu\mu''}({\bm k}',{\bm k}'-{\bm q}')
{\cal D}_{\mu''\mu}({\bm k}'-{\bm q}',{\bm k}')
\nonumber\\
&&
\times
\Lambda^{(\alpha\beta)}_{\mu\mu} ({\bm k}', \omega_+, {\bm k}',0^-)
~,
\end{eqnarray}
and finally
\begin{eqnarray} \label{eq:R_Lambda_3_together_2}
&&
\Lambda^{(3,\alpha\beta)}_{\lambda\lambda} ({\bm k}, \omega_+,{\bm k},0^-) =
\frac{4 i (k_{\rm B} T)^2}{3(\omega + i/\tau_{\rm ee})} \sum_{{\bm k}',{\bm q}'} \sum_{\mu,\lambda'',\mu''}  
\nonumber\\
&&
\times
|W({\bm q}',0)|^2
\Im m \big[ G^{({\rm R})}_{\mu}({\bm k}',0)]
\Im m \big[ G^{({\rm R})}_{\lambda''}({\bm k} +{\bm q}',0) \big]
\nonumber\\
&&
\times
\Im m \big[ G^{({\rm R})}_{\mu''}({\bm k}' -{\bm q}',0) \big]
{\cal D}_{\lambda\lambda''}({\bm k},{\bm k}+{\bm q}')
{\cal D}_{\lambda''\lambda}({\bm k}+{\bm q}',{\bm k})
\nonumber\\
&&
\times
{\cal D}_{\mu\mu''}({\bm k}',{\bm k}'-{\bm q}')
{\cal D}_{\mu''\mu}({\bm k}'-{\bm q}',{\bm k}')
\nonumber\\
&&
\times
\Lambda^{(\alpha\beta)}_{\mu\mu} ({\bm k}', \omega_+,{\bm k}',0^-)
~.
\end{eqnarray}
Shifting ${\bm k'}\to {\bm k}-{\bm q}''$ and ${\bm q}' \to {\bm k}-{\bm k}''$ in Eq.~(\ref{eq:R_Lambda_2_together_2}), we obtain
\begin{eqnarray} \label{eq:R_Lambda_2_together_shift}
&&
\Lambda^{(2,\alpha\beta)}_{\lambda\lambda} ({\bm k}, \omega_+,{\bm k},0^-) =
-\frac{4 i (k_{\rm B} T)^2}{3(\omega + i/\tau_{\rm ee})} \sum_{{\bm k}'',{\bm q}''} \sum_{\mu,\lambda'',\mu''}  
\nonumber\\
&&
\times
|W({\bm k} - {\bm k}'',0)|^2 
\Im m\big[ G^{({\rm R})}_{\mu}({\bm k}-{\bm q}'', 0) \big]
\Im m\big[ G^{({\rm R})}_{\lambda''}({\bm k}'',0) \big]
\nonumber\\
&&
\times
\Im m\big[ G^{({\rm R})}_{\mu''}({\bm k}''-{\bm q}'',0) \Big]
{\cal D}_{\lambda\lambda''}({\bm k},{\bm k}'')
{\cal D}_{\lambda''\lambda}({\bm k}'',{\bm k})
\nonumber\\
&&
\times
{\cal D}_{\mu\mu''}({\bm k}-{\bm q}'',{\bm k}''-{\bm q}'')
{\cal D}_{\mu''\mu}({\bm k}''-{\bm q}'',{\bm k}-{\bm q}'')
\nonumber\\
&&
\times
\Lambda^{(\alpha\beta)}_{\mu\mu} ({\bm k}-{\bm q}'', \omega_+, {\bm k}-{\bm q}'',0^-)
~,
\end{eqnarray}
while shifting ${\bm k}'\to {\bm k}''-{\bm q}''$ and ${\bm q}'\to {\bm k}''-{\bm k}$ in Eq.~(\ref{eq:R_Lambda_3_together_2}), we get
\begin{eqnarray} \label{eq:R_Lambda_3_together_shift}
&&
\Lambda^{(3,\alpha\beta)}_{\lambda\lambda} ({\bm k}, \omega_+,{\bm k},0^-) =
\frac{4 i (k_{\rm B} T)^2}{3(\omega + i/\tau_{\rm ee})} \sum_{{\bm k}',{\bm q}'} \sum_{\mu,\lambda'',\mu''}  
\nonumber\\
&&
\times
|W({\bm k}''-{\bm k},0)|^2
\Im m \big[ G^{({\rm R})}_{\mu}({\bm k}''-{\bm q}'',0)]
\Im m \big[ G^{({\rm R})}_{\lambda''}({\bm k}'',0) \big]
\nonumber\\
&&
\times
\Im m \big[ G^{({\rm R})}_{\mu''}({\bm k}-{\bm q}'',0) \big]
{\cal D}_{\lambda\lambda''}({\bm k},{\bm k}'')
{\cal D}_{\lambda''\lambda}({\bm k}'',{\bm k})
\nonumber\\
&&
\times
{\cal D}_{\mu\mu''}({\bm k}''-{\bm q}'',{\bm k}-{\bm q}'')
{\cal D}_{\mu''\mu}({\bm k}-{\bm q}'',{\bm k}''-{\bm q}'')
\nonumber\\
&&
\times
\Lambda^{(\alpha\beta)}_{\mu\mu} ({\bm k}''-{\bm q}'', \omega_+,{\bm k}''-{\bm q}'',0^-)
~.
\end{eqnarray}
We now observe that the products of the three Green's functions on the right-hand sides of Eqs.~(\ref{eq:R_Lambda_1_together_2})-(\ref{eq:R_Lambda_3_together_2}) constrains (i) $\mu=\mu''=\lambda''=+$ and (ii) the momenta of their arguments to be at the Fermi surface.
After renaming dummy momentum variables and noting that ${\cal D}_{\lambda\lambda'}(-{\bm k},-{\bm k}')={\cal D}_{\lambda\lambda'}({\bm k},{\bm k}')$, we finally get
\begin{eqnarray} \label{eq:Bethe_Salpeter_final}
&&
\Lambda_{++}^{(\alpha\beta)} ({\bm k}, \omega^+, {\bm k}, 0^-) = 
\Lambda^{(0,\alpha\beta)}_{++} ({\bm k},{\bm k})
-\frac{4 i (k_{\rm B} T)^2}{3(\omega + i/\tau_{\rm ee})}
\nonumber\\
&&
\times
\sum_{{\bm k}',{\bm q}'} \sum_{\mu, \mu'',\lambda''}
|W({\bm k}-{\bm k}',0)|^2
\Im m\big[ G^{({\rm R})}_{\mu}({\bm k}',0) \big]
\nonumber\\
&&
\times
\Im m \Big[G^{({\rm R})}_{\lambda''}({\bm q}'-{\bm k},0)\Big]
\Im m\Big[G^{({\rm R})}_{\mu''}({\bm q'}-{\bm k}', 0)\Big]
\nonumber\\
&&
\times
{\cal D}_{\lambda\mu}({\bm k},{\bm k}') {\cal D}_{\mu\lambda}({\bm k}',{\bm k})
{\cal D}_{\lambda''\mu''}({\bm q}'-{\bm k},{\bm q}'-{\bm k}')
\nonumber\\
&&
\times
{\cal D}_{\mu''\lambda''}({\bm q}'-{\bm k}',{\bm q}'-{\bm k}) 
\big[\Lambda^{(\alpha\beta)}_{++} ({\bm k}', \omega_+, {\bm k}',0^-) 
\nonumber\\
&&
+\Lambda^{(\alpha\beta)}_{++} ({\bm k}-{\bm q}', \omega_+, {\bm k}-{\bm q}',0^-) 
\nonumber\\
&&
- \Lambda^{(\alpha\beta)}_{++} ({\bm k}'-{\bm q}', \omega_+,{\bm k}'-{\bm q}',0^-)
\big]
~.
\end{eqnarray}
We solve Eq.~(\ref{eq:Bethe_Salpeter_final}) with the following {\it Ansatz}:
\begin{equation} \label{eq:Lambda_ansatz}
\Lambda^{(\alpha\beta)}_{++} ({\bm k}, \omega_+, {\bm k},0^-) = \gamma(\omega) \Lambda^{(0,\alpha\beta)}_{++} ({\bm k},{\bm k}) 
~,
\end{equation}
which reduces the self-consistent equation to an algebraic one. We then project the result along $\Lambda^{(0,\alpha\beta)}_{++} ({\bm k},{\bm k})$, {\it i.e.} we multiply it by $\Lambda^{(0,\beta\gamma)}_{++} ({\bm k},{\bm k})$ and we take the trace over the greek indices. We get
\begin{eqnarray} \label{eq:Bethe_Salpeter_final_2}
&& \gamma(\omega) = 1
- \gamma(\omega) \frac{4i (k_{\rm B} T)^2}{3(\omega + i/\tau_{\rm ee})}
\sum_{{\bm k}',{\bm q}'} \sum_{\mu, \mu'',\lambda''}
|W({\bm k}-{\bm k}',0)|^2
\nonumber\\
&&
\times
\Im m\big[ G^{({\rm R})}_{\mu}({\bm k}',0) \big]
\Im m \Big[G^{({\rm R})}_{\lambda''}({\bm q}'-{\bm k},0)\Big]
\nonumber\\
&&
\times
\Im m\Big[G^{({\rm R})}_{\mu''}({\bm q'}-{\bm k}', 0)\Big]
{\cal D}_{\lambda\mu}({\bm k},{\bm k}') {\cal D}_{\mu\lambda}({\bm k}',{\bm k})
\nonumber\\
&&
\times
{\cal D}_{\lambda''\mu''}({\bm q}'-{\bm k},{\bm q}'-{\bm k}') 
{\cal D}_{\mu''\lambda''}({\bm q}'-{\bm k}',{\bm q}'-{\bm k}) 
\nonumber\\
&&
\times
\big[\cos^2(\varphi_{{\bm k}'}) + \cos^2(\varphi_{{\bm k}-{\bm q}'}) - \cos^2(\varphi_{{\bm k}'-{\bm q}'})\big]
~,
\end{eqnarray}
Here we used that $k=k'=k_{\rm F}$ and that the Green's functions of Eq.~(\ref{eq:Bethe_Salpeter_final}) constrain their momentum arguments to be at the Fermi surface. We also assumed ${\bm k} = k{\hat {\bm x}}$. Eq.~(\ref{eq:Bethe_Salpeter_final_2}) can be rewritten as
\begin{eqnarray} \label{eq:Bethe_Salpeter_final_3}
\gamma(\omega) &=& 1
+ \gamma(\omega) \frac{i/\tau_{\rm ee} - i/\tau_{\rm v}}{\omega + i/\tau_{\rm ee}}
~,
\end{eqnarray}
where
\begin{eqnarray} \label{eq:tau_visc}
\frac{1}{\tau_{\rm v}} &=& 
 \frac{4(k_{\rm B} T)^2}{3}
\sum_{{\bm k}',{\bm q}'} \sum_{\mu, \mu'',\lambda''}
|W({\bm k}-{\bm k}',0)|^2
\nonumber\\
&&
\times
\Im m\big[ G^{({\rm R})}_{\mu}({\bm k}',0) \big]
\Im m \Big[G^{({\rm R})}_{\lambda''}({\bm q}'-{\bm k},0)\Big]
\nonumber\\
&&
\times
\Im m\Big[G^{({\rm R})}_{\mu''}({\bm q'}-{\bm k}', 0)\Big]
{\cal D}_{\lambda\mu}({\bm k},{\bm k}') {\cal D}_{\mu\lambda}({\bm k}',{\bm k})
\nonumber\\
&&
\times
{\cal D}_{\lambda''\mu''}({\bm q}'-{\bm k},{\bm q}'-{\bm k}')
{\cal D}_{\mu''\lambda''}({\bm q}'-{\bm k}',{\bm q}'-{\bm k}) 
\nonumber\\
&&
\times
\big[\cos^2(\varphi_{{\bm k}'}) + \cos^2(\varphi_{{\bm k}-{\bm q}'}) - \cos^2(\varphi_{{\bm k}'-{\bm q}'})-1\big]
~.
\nonumber\\
\end{eqnarray}
Eq.~(\ref{eq:Bethe_Salpeter_final_3}) is readily solved by
\begin{eqnarray} \label{eq:gamma_final}
\gamma(\omega) &=& \frac{\omega + i/\tau_{\rm ee}}{\omega + i/\tau_{\rm v}}
~,
\end{eqnarray}
which contains all the information about the vertex corrections.

Finally, putting Eqs.~(\ref{eq:Lambda_ansatz}) and~(\ref{eq:gamma_final}) back into Eq.~(\ref{eq:R_chi_jj_final_omega_finite}) and Eq.~(\ref{eq:app_eta_0_from_stress}), and taking the limit $\omega\to 0$, we get
\begin{eqnarray} \label{eq:R_chi_jj_final}
\eta_0
&=& \frac{\tau_{\rm v}}{\tau_{\rm ee}} \sum_{{\bm k}, \lambda, \lambda'} \int \frac{d\varepsilon}{2\pi}\frac{\partial n_{\rm F} (\varepsilon)}{\partial \varepsilon}
\Im m\Big[ G^{({\rm A})}_\lambda({\bm k}, \varepsilon) \Lambda^{(0,xy)}_{\lambda,\lambda'}({\bm k},{\bm k}) 
\nonumber\\
&&
\times
G^{({\rm R})}_{\lambda'}({\bm k},\varepsilon) \Lambda^{(0,xy)}_{\lambda'\lambda} ({\bm k},{\bm k}) \Big]
\nonumber\\
&=&
\frac{\tau_{\rm v}}{2 \pi \tau_{\rm ee}} \sum_{{\bm k}} 
\Im m\Big[ G^{({\rm A})}_+ ({\bm k}, 0) G^{({\rm R})}_+({\bm k},0) \Big] \Big[\Lambda^{(0,xy)}_{++}({\bm k},{\bm k}) \Big]^2
\nonumber\\
&=&
\frac{1}{4} n \varepsilon_{\rm F} \tau_{\rm v}
~.
\end{eqnarray}
One immediately recognizes ${\cal S}_0 = n \varepsilon_{\rm F}/4$ as the non-interacting bulk modulus (which is not renormalized by e-e interactions, as discussed before) and $\tau_{\rm v}$ as the viscosity transport time.

\subsection{Transformation of Eq.~(\ref{eq:tau_visc}) to a computationally efficient formula}
We start from the definition of Eq.~(\ref{eq:tau_visc}), and we shift ${\bm k}'\to {\bm k} - {\bm q}$, ${\bm q}'\to {\bm k}''+{\bm k}$. We get
\begin{eqnarray}
\frac{1}{\tau_{\rm v}} &=& 
\frac{4 (k_{\rm B} T)^2}{3}
\sum_{{\bm q},{\bm q}'} \sum_{\mu, \mu'',\lambda''}
|W({\bm q},0)|^2
\Im m\big[ G^{({\rm R})}_{\mu}({\bm k} - {\bm q},0) \big]
\nonumber\\
&\times&
\Im m \Big[G^{({\rm R})}_{\lambda''}({\bm k}'',0)\Big]
\Im m\Big[G^{({\rm R})}_{\mu''}({\bm k}''+ {\bm q}, 0)\Big]
\nonumber\\
&\times&
{\cal D}_{\lambda\mu}({\bm k},{\bm k}-{\bm q}) {\cal D}_{\mu\lambda}({\bm k}-{\bm q},{\bm k})
{\cal D}_{\lambda''\mu''}({\bm k}'',{\bm k}''+{\bm q})
\nonumber\\
&\times&
{\cal D}_{\mu''\lambda''}({\bm k}''+{\bm q},{\bm k}'') 
\nonumber\\
&\times&
\big[\cos^2(\varphi_{{\bm k}-{\bm q}}) + \cos^2(\varphi_{{\bm k}''}) - \cos^2(\varphi_{{\bm k}''+{\bm q}})-1\big]
~.
\nonumber\\
\end{eqnarray}
Now we use the fact that, if the scattering occurs at the Fermi surface, ${\bm k}''+{\bm q}$ (${\bm k}''$) is opposite to ${\bm k}$ (${\bm k}+{\bm q}$) (see also Ref.~\onlinecite{Principi_arxiv_2015}). Therefore
\begin{eqnarray} \label{eq:tau_visc_final_sempl}
\frac{1}{\tau_{\rm v}} &=& 
- \frac{8(k_{\rm B} T)^2}{3}
\sum_{{\bm q},{\bm q}'} \sum_{\mu, \mu'',\lambda''}
|W({\bm q},0)|^2
\Im m \Big[G^{({\rm R})}_{\lambda''}({\bm k}'',0)\Big]
\nonumber\\
&\times&
\Im m\big[ G^{({\rm R})}_{\mu}({\bm k} - {\bm q},0) \big]
\Im m\Big[G^{({\rm R})}_{\mu''}({\bm k}''+ {\bm q}, 0)\Big]
\nonumber\\
&\times&
{\cal D}_{\lambda\mu}({\bm k},{\bm k}-{\bm q}) {\cal D}_{\mu\lambda}({\bm k}-{\bm q},{\bm k})
{\cal D}_{\lambda''\mu''}({\bm k}'',{\bm k}''+{\bm q})
\nonumber\\
&\times&
{\cal D}_{\mu''\lambda''}({\bm k}''+{\bm q},{\bm k}'') 
[1 - \cos^2(\varphi_{{\bm k}+{\bm q}})]
~.
\end{eqnarray}
Note that the expression of Eq.~(\ref{eq:tau_visc_final_sempl}), if one exclude the matrix element in its last line, coincides with the quasiparticle lifetime for $k_{\rm B}T\ll \varepsilon_{\rm F}$ reported in Refs.~\onlinecite{Principi_arxiv_2014,Principi_arxiv_2015}. The expression of the quasiparticle lifetime valid at all temperatures was also given in Ref.~\onlinecite{polini_arxiv_2014}.
At low temperature (a limit that almost always holds in graphene), we can thus rewrite $1/\tau_{\rm v}$ using the formulas provided in Ref.~\onlinecite{polini_arxiv_2014}, amending them by multiplying the integrand with the matrix element $1 - \cos^2(\varphi_{{\bm k}+{\bm q}})$. Since that was derived in a regime in which intraband transitions are responsible for the dominant contribution, in what follows we approximate the formulas of Ref.~\onlinecite{polini_arxiv_2014} by neglecting contributions due to interband processes. This approximation is valid in the low-temperature regime. We get
\begin{eqnarray} \label{eq:app_tau_v_final}
\frac{1}{\tau_{\rm v}}
&\simeq&
\frac{4}{(2\pi)^2}\int_{-\infty}^{\infty} d\xi ~\frac{\partial n_{\rm F}(\xi)}{\partial \xi} \int_{-\infty}^{+\infty}d\omega~\frac{1- n_{\rm F}(\xi - \omega)}{1 - \exp(-\beta\omega)}
\nonumber\\
&\times&
\int_0^{+\infty}dq~q
\left|\frac{v_q}{\varepsilon(q, \omega, T)}\right|^2\Im m[\chi^{(0)}_{nn}(q,\omega,T)]
\nonumber\\
&\times&
A_{++}(k_{\rm F},q,\omega) \left[ 1 - \frac{q^2 - \omega^2/v^2_{\rm F}}{4k_{\rm F}(k_{\rm F}- \omega/v_{\rm F})} \right] 
\nonumber\\
&\times&
\frac{4(q^2 - \omega^2/v^2_{\rm F})}{2k_{\rm F}(k_{\rm F}- \omega/v_{\rm F})}
~.
\end{eqnarray}
where
\begin{eqnarray}
\label{eq:app_final_angular_integral_MDF_intraband}
A_{++} &=& \frac{4(k - \omega/v_{\rm F})}{v_{\rm F}\sqrt{[(2 k - \omega/v_{\rm F})^2 - q^2](q^2 - \omega^2/v^2_{\rm F})}} 
\nonumber\\
&\times&
\left[ 1 - \frac{q^2 - \omega^2/v^2_{\rm F}}{4k(k- \omega/v_{\rm F})} \right] 
\nonumber\\
&\times&
\Theta\Big\{\big[(2 k - \omega/v_{\rm F})^2 - q^2\big](q^2 - \omega^2/v^2_{\rm F})\Big\}
~,
\nonumber\\
\end{eqnarray}
and we used that
\begin{eqnarray}
\cos(\varphi_{{\bm k}+{\bm q}}) \to 1 - \frac{q^2 - \omega^2/v^2_{\rm F}}{2k(k- \omega/v_{\rm F})}
~. 
\end{eqnarray}
Eq.~(\ref{eq:app_tau_v_final}) has a form similar to the quasiparticle contribution to the decay rate given in Ref.~\onlinecite{polini_arxiv_2014}. We did not write the quasihole contribution, but we included it with an extra factor of two that multiplies the whole expression. Indeed, in the low-temperature limit the quasiparticle and quasihole contributions are identical~\cite{polini_arxiv_2014}.
The integration over $\varepsilon$ in Eq.~(\ref{eq:app_tau_v_final}), with the weighting factor $\partial n_{\rm F}(\xi)/(\partial \xi)$, is the same integration we performed in the Bethe-Salpeter equation [see the discussion after Eq.~(\ref{eq:R_Lambda_3_together})]. Its origin has to be found in the fact that, in the limit of low temperature, the energy-dependent self-energy can be replaced by its {\it average} over the thermally excited states. Note that, since in the low-temperature limit the function $\partial n_{\rm F}(\xi)/(\partial \xi)$ is strongly peaked at $\xi = 0$ (which in turn implies $k=k_{\rm F}$), we set $k=k_{\rm F}$ everywhere in the integrand on the right-hand side of Eq.~(\ref{eq:app_tau_v_final}), {\it except} in the Fermi function on its first line, which strongly depends on $\xi$.

Following Refs.~\onlinecite{Principi_arxiv_2014,Principi_arxiv_2015,polini_arxiv_2014}, we can now calculate Eq.~(\ref{eq:app_tau_v_final}).
As noted in the main text, the matrix element $1-\cos^2(\varphi_{{\bm k}+{\bm q}})$ completely suppresses the divergence due to the small-$q$ region. Following the same manipulations performed in Ref.~\onlinecite{polini_arxiv_2014} we get
\begin{eqnarray} \label{eq:app_tau_v_intermediate_1}
\frac{1}{\tau_{\rm v}} &=& 32 N(0) \alpha_{\rm ee}^2 \int_{-\infty}^{\infty} d\xi \frac{\partial n_{\rm F}(\xi)}{\partial \xi}\int_{-\infty}^{+\infty} d\omega ~\omega 
n_{\rm B}(-\omega) 
\nonumber\\
&\times&
\int_{|\omega|/v_{\rm F}}^{2 k_{\rm F} - \omega/v_{\rm F}} \frac{dq}{q} \frac{n_{\rm F}(\xi-\omega)}{\displaystyle \left(1 + \frac{q_{\rm TF}}{q} \right)^2 + \frac{q_{\rm TF}^2}{q^2} \frac{\omega^2}{v_{\rm F}^2} \frac{1-q^2/(4 k_{\rm F}^2)}{q^2 - \omega^2/v_{\rm F}^2} } 
\nonumber\\
&\times&
\frac{1-q^2/(4 k_{\rm F}^2)}{q^2 - \omega^2/v_{\rm F}^2} 
\left[ 1 - \frac{q^2 - \omega^2/v^2_{\rm F}}{4k_{\rm F}(k_{\rm F}- \omega/v_{\rm F})} \right] 
\nonumber\\
&\times&
\frac{q^2 - \omega^2/v^2_{\rm F}}{2k(k_{\rm F}- \omega/v_{\rm F})}
~.
\end{eqnarray}
Here $N(0) = N_{\rm f} \varepsilon_{\rm F}/[2\pi (\hbar v_{\rm F})^2]$ is the density-of-states at the Fermi energy.
Since the ratio in the integrand on the first line of Eq.~(\ref{eq:app_tau_v_intermediate_1}) is strongly peaked around $\xi,\omega = 0$, we take the limit of $\xi,\omega\to 0$ in the rest of the integrand. Contrary to the case of the quasiparticle lifetime this does not lead to any divergence. Indeed, the integrand is regular for $q\to 0, 2k_{\rm F}$. The integration is then straightforward and the result reads (restoring $\hbar$)
\begin{eqnarray}
\frac{1}{\tau_{\rm v}} &=& 
-\frac{\pi N_{\rm f}}{9} \frac{(k_{\rm B} T)^2}{\hbar\varepsilon_{\rm F}}\alpha_{\rm ee}^2 
\big[
 42 - 52 N_{\rm f} \alpha_{\rm ee} - 15 N_{\rm f}^2 \alpha_{\rm ee}^2 
 \nonumber\\
 &+& 15 N_{\rm f}^3 \alpha_{\rm ee}^3 
- 3 (16 - 24 N_{\rm f}^2 \alpha_{\rm ee}^2 + 5 N_{\rm f}^4 \alpha_{\rm ee}^4) 
\nonumber\\
&\times&
{\rm coth}^{-1}(1 + N_{\rm f} \alpha_{\rm ee})
\big]
~.
\nonumber\\
\end{eqnarray}
To get this equation we used that $q_{\rm TF} = N_{\rm f} \alpha_{\rm ee} k_{\rm F}$, and that in the low temperature limit
\begin{equation}
-\int_{-\infty}^{\infty} \!\! d\xi \frac{\partial n_{\rm F}(\xi)}{\partial \xi}\int_{-\infty}^{+\infty} \!\! d\omega \omega \frac{1-n_{\rm F}(\xi-\omega)}{1-\exp(-\beta\omega)} 
\to 
\frac{3\pi^2}{16}
(k_{\rm B} T)^2
~.
\end{equation}
In the limit of $\alpha_{\rm ee} \to 0$ we get
\begin{eqnarray}
\frac{1}{\tau_{\rm v}} &\to& - N_{\rm f} \frac{8 \pi}{3} 
\frac{(k_{\rm B} T)^2}{\hbar\varepsilon_{\rm F}}
\alpha_{\rm ee}^2 \ln(\alpha_{\rm ee})
~.
\end{eqnarray}

\begin{thebibliography}{77}
\bibitem{footnote_conductivity}
	Therefore, e-e interactions do not contribute to the charge conductivity, 
	which indeed diverges in the absence of electron-impurity and electron-phonon scattering.
\end{thebibliography}
\end{document}